\documentclass[a4paper,11pt]{article}
\pdfoutput=1

\usepackage{jheppub}

\usepackage[T1]{fontenc}
\usepackage{array}

\usepackage{amssymb}
\usepackage{amsmath}
\usepackage{graphicx}

\usepackage{longtable}
\newcolumntype{C}{>{$}c<{$}}
\usepackage{pdflscape}

\usepackage{chngpage}

\usepackage[table,dvipsnames]{xcolor}
\usepackage{diagbox}

\usepackage{multirow}

\mathchardef\ordinarycolon\mathcode`\:
\mathcode`\:=\string"8000
\begingroup \catcode`\:=\active
  \gdef:{\mathrel{\mathop\ordinarycolon}}
\endgroup

\usepackage{caption}
\usepackage{subcaption}

\newcommand{\GitHub}{{\tt \href{https://GitHub.com/julius-julius/qsc}{GitHub}} }

\newcommand{\la}[1]{\label{#1}}
\newcommand{\eq}[1]{(\ref{#1})}

\def\[{\left[}
\def\]{\right]}
\def\({\left(}
\def\){\right)}

\newcommand\beqa{\begin{eqnarray}}
\newcommand\eeqa{\end{eqnarray}}

\newcommand{\bP}{{\bf P}}
\newcommand{\bQ}{{\bf Q}}
\newcommand{\Qai}{Q_{a|i}}

\newcommand{\nstates}{219\;}

\newcommand{\beq}{\begin{eqnarray}}
\newcommand{\eeq}{\end{eqnarray}}

\newcommand{\be}{\begin{equation}}
\newcommand{\ee}{\end{equation}}

\newcommand{\bm}{\begin{multline}}
\newcommand{\fm}{\end{multline}}

\title{\boldmath Fast QSC Solver: tool for systematic study of ${\cal N}=4$ Super-Yang--Mills spectrum}

\author{Nikolay Gromov$^{a}$}
\author{\'Arp\'ad Heged\H us$^{b}$}
\author{Julius Julius$^{a}$}
\author{Nika Sokolova$^a$}

\affiliation{
 $^a$ Department of Mathematics, King's College London, Strand WC2R 2LS, United Kingdom
 \\
 $^b$
 Institute for Particle and Nuclear Physics, HUN-REN Wigner Research Centre for Physics,
 H-1525 Budapest 114, P.O.B. 49, Hungary
}
 \emailAdd{nikolay.gromov@kcl.ac.uk}
\emailAdd{hegedus.arpad@wigner.hun-ren.hu}
 \emailAdd{julius.julius@kcl.ac.uk}
 \emailAdd{nika.sokolova@kcl.ac.uk}

\abstract{
Integrability methods give us access to a number of observables in the planar ${\cal N}=4$ SYM. Among them, the Quantum Spectral Curve (QSC) governs the spectrum of anomalous dimensions. Low lying states were successfully studied in the past using the QSC.
However, with the increased demand for a systematic study of a large number of states for various applications, there is a clear need for a fast QSC solver which can easily access a large number of excited states. Here, we fill this gap by developing a new algorithm and applied it to study all $\nstates$ states with the bare dimension $\Delta_0\leq 6$ in a wide range of couplings.

The new algorithm has an improved performance at weak coupling and allows to glue numerics smoothly the available perturbative data, resolving the previous obstruction. Further $\sim 8$-fold efficiency gain comes from C++ implementation over the best available {\tt Mathematica} implementation.
We have made the code and the data to be available via a \GitHub repository.

The method is generalisable for non-local observables as well as for other theories such as deformations of ${\cal N}=4$ SYM and ABJM. It may find applications in the separation of variables and bootstrability approaches to the correlation functions. Some applications to correlators at strong coupling are also presented.

}

\begin{document}
\maketitle
\flushbottom

\section{Introduction}
\label{sec:intro}
The spectrum of planar ${\cal N}=4$ Super-Yang-Mills (SYM) theory is the most studied non-protected observable. Yet only a handful of states are available in the literature~\cite{Gromov:2009zb,Basso:2011rs, Gromov:2011de, Roiban:2011fe, Vallilo:2011fj,Gromov:2011bz,Frolov:2012zv,Gromov:2014bva,Hegedus:2016eop,Marboe:2017dmb,Marboe:2018ugv}. Recently, new methods became available which utilise a combination of the conformal bootstrap techniques with the insights from integrability to obtain observables beyond the spectrum~\cite{Cavaglia:2021bnz,Cavaglia:2022qpg,Caron-Huot:2022sdy,Cavaglia:2022yvv}. The information about the non-perturbative spectrum combined with the conformal bootstrap method is already shown to give very accurate bounds on the structure constants in some examples~\cite{Cavaglia:2021bnz, Cavaglia:2022qpg}.

The non-perturbative spectrum can only be studied numerically, except for some limiting cases, such as in the near-BPS regime~\cite{Basso:2011rs, Gromov:2011de,Gromov:2014bva,Alfimov:2020obh} or in the limit of infinitely long operators, using the Asymptotic Bethe Ansatz (ABA)~\cite{Beisert:2005fw,Beisert:2006ez}. There is very little chance of an exact analytic expression for all states in the theory, which is already quite clear from the cumbersome structure of the weak coupling expansion for the states which are tractable analytically~\cite{Marboe:2017dmb,Marboe:2018ugv}.

The numerical methods to tackle the non-perturbative spectrum of short operators was pioneered in~\cite{Gromov:2009zb} on the example of the simplest non-trivial Konishi operator, and was based on the infinite set of integral equations called the Thermodynamic Bethe Ansatz (TBA). The TBA based approach has a number of problems: the linear convergence rate, high computational cost, and finally going beyond the simplest set of operators was proven to be complicated~\cite{Quinn:2012cb}.
With the discovery of the Quantum Spectral Curve (QSC) \cite{Gromov:2013pga, Gromov:2014caa}, which is a finite system of equations, both analytic techniques~\cite{Leurent:2013mr,Marboe:2014gma,Agmon:2017xes,Gromov:2015vua,Marboe:2017dmb,Marboe:2018ugv} and numerical methods~\cite{Gromov:2015wca,Hegedus:2016eop} were developed to go well beyond the state of the art of what was possible at the level of TBA.

In this paper we take the previous developments of the numerical approaches to the non-perturbative spectrum to an industrial level. By adding new ideas to improve its performance at weak coupling, where the perturbative data is readily available~\cite{Marboe:2018ugv}, to start up the numerics for a huge number of states, and by using fast C++ implementation of our new algorithm. We computed all \nstates states with bare dimension $\Delta_0 \leq 6$ on one PC (with two multi-core CPUs) in a timescale of a few months, which also included the time for adjusting our algorithm.

We make the data and the code available via \GitHub keeping in mind that it can be modified to be applicable for a variety of situations: this may include the
Regge complex-spin trajectories~\cite{Alfimov:2018cms,Alfimov:2020obh}, boundary problems such as Wilson defects~\cite{Gromov:2015dfa,Gromov:2016rrp,Cavaglia:2018lxi,Giombi:2018hsx,Grabner:2020nis,Cavaglia:2021bnz,Julius:2021uka,spec1DCFT}, other theories such as ABJM~\cite{Bombardelli:2017vhk,Bombardelli:2018bqz, Correa:2023lsm,Lee:2017mhh,Lee:2018jvn,Lee:2019oml} or $\rm AdS_3$ -- where the QSC was recently proposed~\cite{Cavaglia:2021eqr, Ekhammar:2021pys} and solved numerically for some type of states~\cite{Cavaglia:2022xld}, as well as various deformations, such as $\beta$-deformation~\cite{Marboe:2019wyc,Levkovich-Maslyuk:2020rlp}, $\gamma$-deformation~\cite{Grabner:2017pgm,Kazakov:2018ugh} or $\eta$-deformation~\cite{Klabbers:2017vsm,Klabbers:2017vtw}, or to calculate the Hagedorn temperature non-perturbatively~\cite{Harmark:2018red,Harmark:2021qma,Ekhammar:2023glu}.
We hope our code would be beneficial for the communities studying these systems.

\begin{figure}
    \centering
    \includegraphics[scale=1.4]{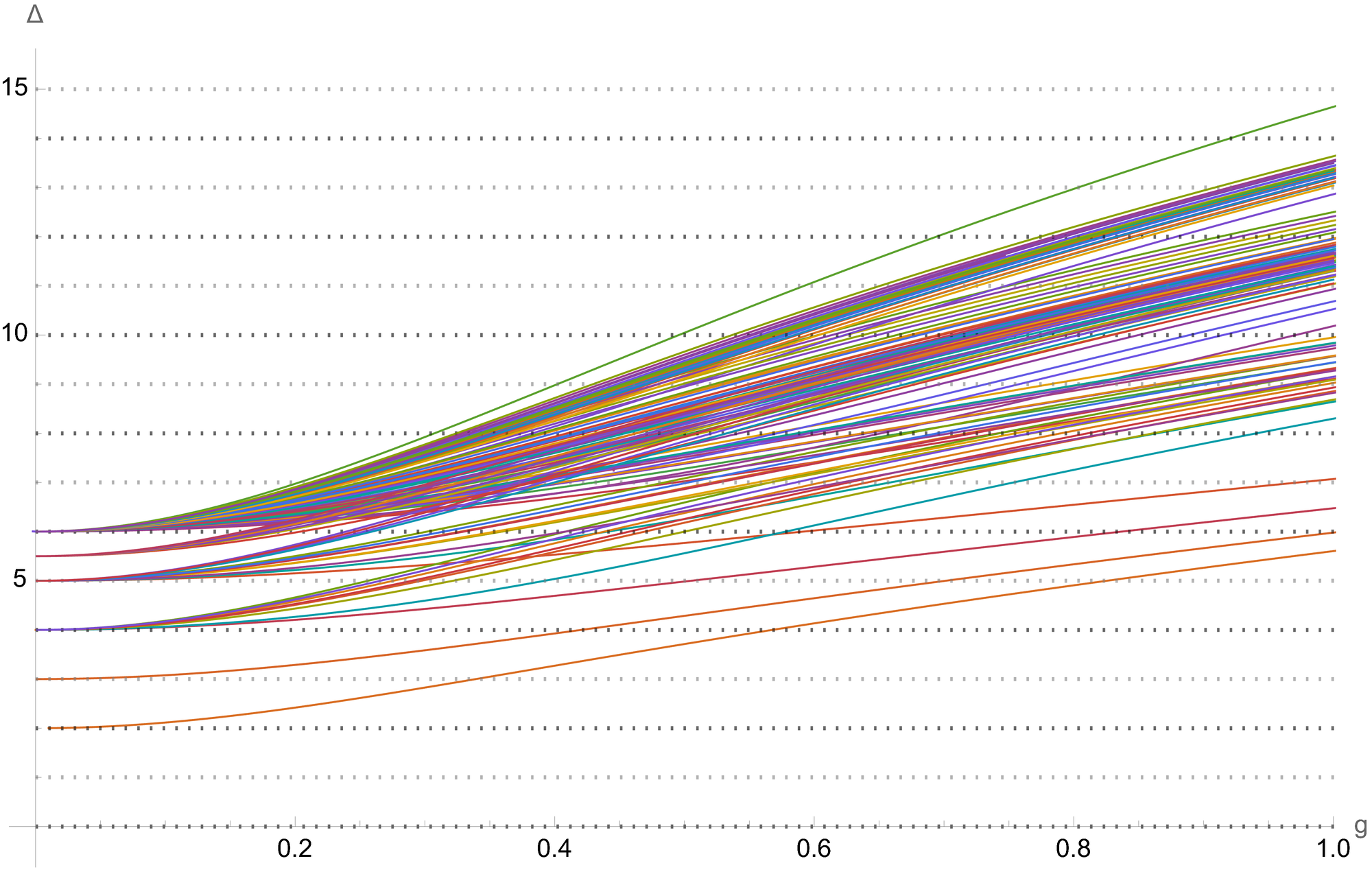}
    \caption{\small\it    The scaling dimension of all 219 states in ${\cal N} = 4$ SYM, with bare dimension $\Delta_0\leq 6$ in the range $g\in [0,1]$ or 't Hooft coupling $\lambda\in [0,16\,\pi^2] \sim [0,158]$.}
    \label{fig:AllStates}
\end{figure}

\begin{figure}
    \centering
    \includegraphics[scale=1.4]{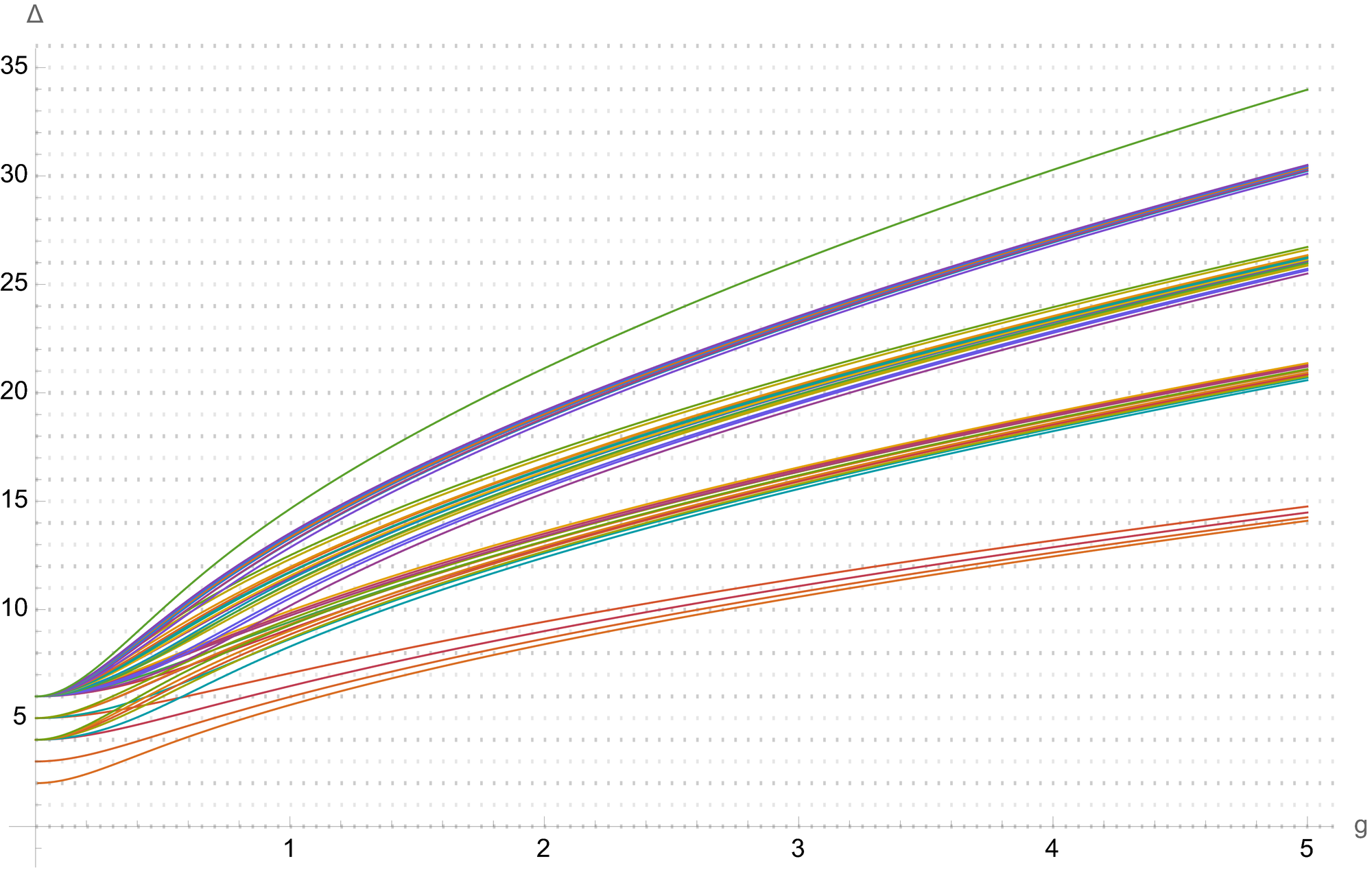}
    \caption{\small\it    The scaling dimension of 45 states in ${\cal N} = 4$ SYM, with bare dimension $\Delta_0\leq 6$, that have additional symmetries ({\it cf.}~type I states, discussed in section~\ref{LReven}) which make them amicable to more efficient numerical computation thereby allowing us to achieve higher range in the 't Hooft coupling. We provide their scaling dimensions in the range $g\in [0,5] = \lambda \in [0,400\,\pi^2]\sim [0,4000]$.}
    \label{fig:LREvenStates}
\end{figure}
\paragraph{Details on the results.}
In addition to making the source code available via \GitHub on {\tt \href{https://GitHub.com/julius-julius/qsc}{https://GitHub.com/julius-julius/qsc}}
alongside the additional tools to test the C++ compilation, to generate the starting points and to control the parameters during the run. We also generated and share the following data: for the 45 states with additional symmetries we computed the spectrum in the range $g=[0,5]$
($\lambda\in[0,\sim 4\,000]$). Their spectral plot is given in figure~\ref{fig:LREvenStates}. For some selected states in a wider range e.g. for Konishi operator we have the data in much wider range $g\in [0,13]$. Then for 92 the operators with less symmetries we considered smaller range $g\in [0,2]$. These are plotted in figure~\ref{fig:LRGenGenEvenStates}. Finally for the most complicated 82 operators we currently have $g\in [0,1]$\footnote{The state ${}_{6}[0\ 2 \ 2\ 2\ 2 \ 2\  2\ 0]_{6,7}$  is computed in the range $g=[0,0.746875]$. This state will be updated.}, which we plan to expand later. They are plotted in figure~\ref{fig:GenGenStates}. The full spectrum of planar ${\cal N} = 4$ SYM, for all states with bare dimension $\Delta_0 \leq 6$, in the range $g\in [0,1]$ is plotted in figure~\ref{fig:AllStates}.

In the paper we also analyse the strong coupling behaviour by fitting of our data. When the accuracy allows, we determine analytic numbers or provide numerical expression for the strong coupling expansion otherwise.

Finally, we use the strong coupling data to find the  structure constants large $g$ coefficients, which were studied recently using conformal bootstrap~\cite{Alday:2022uxp, Alday:2022xwz}, by boosting these results with the spectral data.

\paragraph{Structure of the paper.} In section~\ref{sec:QN} we introduce the general notations for the local operator in ${\cal N}=4$ SYM and introduce notations for their quantum numbers; in section~\ref{sec:qsc} we review the key aspect of the QSC needed for the introduction of the numerical algorithm for the QSC solver in section~\ref{sec:num}. Then in section~\ref{sec:imp} we describe the C++ implementation with the installation and usage instructions as well as the benchmarking of the performance. In section~\ref{sec:data} we present the data for the spectrum and its detailed analysis followed by a conclusion in section~\ref{sec:concl}.

\section{Local operators and quantum numbers}

\la{sec:QN}

In this section we describe the notations we use through this paper for the states and for their quantum numbers.

$\mathcal{N} = 4$ SYM is a superconformal field theory with ${\rm PSU}(2, 2|4)$ symmetry group~\cite{Gliozzi:1976qd, Brink:1976bc, Sohnius:1981sn, Mandelstam:1982cb, Howe:1983sr, Brink:1982pd, Dobrev:1985qv}.
As such, a state in this theory is characterised by six quantum numbers. Namely, the scaling dimension: $\Delta$, the Dynkin labels of the Lorentz group $\rm SO(3,1)$: $(\ell_1\;\ell_2)$ (a.k.a. Lorentz spin labels), and the Dynkin labels of the ${R}$-symmetry group $SO(6)$: ($q_1\; p\; q_2$) (a.k.a. $R$-symmetry labels).

The Lagrangian of ${\cal N} = 4$ SYM depends on two parameters: the Yang-Mills coupling $g_{\tt YM}$, and the rank of the gauge group ${\rm SU}(N)$ \cite{Gliozzi:1976qd, Brink:1976bc}. The integrable structure of the theory manifests itself~\cite{Minahan:2002ve} in the planar, or large-$N$ limit, where we take $g_{\tt YM} \to 0$ and $N \to \infty$ in such a way that the 't Hooft coupling $\lambda \equiv g^2_{\tt YM}\,N$, is held fixed. Another scaling of the 't Hooft coupling is $g \equiv \frac{\sqrt{\lambda}}{4\,\pi}$, which is a standard notation in the integrability literature. We use both $g$ and $\lambda$, preferring $g$ for weak and finite coupling, and $\lambda$ for strong coupling.

In this paper, we work exclusively with planar ${\cal N} = 4$ SYM.
As such, due to large-$N$ factorisation~\cite{tHooft:1973alw, Witten:1979kh} one can focus entirely on linear combinations of the single trace operators
\begin{align}
 {\rm tr}\, {\cal W}_1\,\dots\,{\cal W}_L\;
    \;,
\end{align}
where ${\cal W}_I$ are ``fields'' of ${\cal N} = 4$ SYM, defined as\footnote{Here, the superscript on $\cal D$ is used to denote multiple applications of the covariant derivative, rather than an index associated to the symmetry group.}
\begin{align}\label{eqn:fieldDef}
    {\cal W}_I \in \left\{
    {\cal D}^i\,\Phi\;,
    {\cal D}^j\,\Psi\;,
    {\cal D}^k\,{\bar \Psi}\;,
    {\cal D}^l\,{\cal F}\;
    \right\}
    \;.
\end{align}
Here, ${\cal D}$ is the covariant derivative, and $\Phi$ is a scalar, $\Psi$ and $\bar\Psi$ are gluinos, and $\cal F$ is the gluon field strength.\footnote{Strictly speaking, each of these have Lorentz and/or $R$-symmetry indices, which we have suppressed for convenience. The form of these fields with their associated indices is given in table~\ref{tab:FieldsAndOsc}.} The trace is over the ${\rm SU}(N)$ colour indices, under which all fields transform in the adjoint representation. The number of fields entering under the trace is called the length of the operator and is denoted by $L$, which is only well defined at zero `t Hooft coupling.

In general, the scaling dimension $\Delta$ of a state in this theory is a non-trivial function of the 't Hooft coupling $\lambda$.
As ${\cal N} = 4$ SYM is a conformal field theory (CFT)~\cite{Sohnius:1981sn, Green:1982sw}, the scaling dimension of an operator $\mathcal{O}$ can be extracted from its two-point function\footnote{Assuming the correct normalisation.}${}^{\text{,}}$\footnote{Note that operators dressed with more indices that furnish non-trivial representations of the symmetry group ${\rm PSU}(2,2|4)$ will have a more complicated expression for their two-point function, which are available in the standard references on CFT~\cite{Rychkov:2016iqz,Simmons-Duffin:2016gjk,DiFrancesco:1997nk,Poland:2018epd}.}
\begin{equation}
\langle    \mathcal{O}(x) \mathcal{O}(y) \rangle = \frac{1}{|x-y|^{2\,\Delta}}\;\;,
\end{equation}
where $x$ and $y$ are coordinates in the 4D space.
We can separate the scaling dimension $\Delta$ into the bare/engineering/classical dimension ---  denoted by $\Delta_0$, which is the sum of the bare dimensions of its constituent fields --- and its quantum corrections, called the anomalous dimension.
The bare dimension $\Delta_0$ of an operator may be obtained by setting the 't Hooft coupling $g = \lambda = 0$.

Due to supersymmetry, operators organise themselves into supermultiplets which comprise all operators with the same anomalous dimension. Supermultiplets are denoted by the quantum numbers of the highest weight state (we choose the convention/grading where the state with the lowest dimension $\Delta$~\cite{Marboe:2017dmb})
in the multiplet. All primary members of the multiplet are superdescendants of the superprimary, obtained by acting with a set of $16$ supercharges on the highest weight state. Each time we act with the supercharge we either get zero or an operator with a dimension increased by $1/2$.

Superconformal multiplets can be short, semi-short and long~\cite{Andrianopoli:1998ut,Andrianopoli:1999vr,Lee:1998bxa,Arutyunov:2000ku,Dolan:2002zh}. Here we focus on long multiplets as for them the scaling dimension is not protected. Semi-short multiplets require fine-tuning of the parameters and as a result are only known to exist at $g=0$~\cite{Dolan:2002zh}.

To classify weak coupling solutions of the QSC, it is convenient  to switch to the notation which uses the oscillator numbers following \cite{Beisert:2003jj,Gunaydin:1984fk,Bars:1982ep,Beisert:2004ry,Marboe:2017dmb}. The reason for this is that at $g=0$ there is an additional ``quantum number'' (or rather a label): the length of a single-trace operator $L$.
In addition to the Lorentz spin labels, and the $R$-symmetry labels, these quantum numbers parameterise the length $L$ as well~\cite{Marboe:2017dmb}.
Therefore, we use the following set of oscillator numbers:
\begin{equation}
\label{eqn:oscnumbers}
    \left\{n_{{\bf b}_1}\,,\; n_{{\bf b}_2}\,,\; n_{{\bf f}_1}\,,\; n_{{\bf f}_2}\,,\; n_{{\bf f}_3}\,,\; n_{{\bf f}_4}\,,\;  n_{{\bf a}_1}\,,\; n_{{\bf a}_2} \right\}\;.
\end{equation}
These are directly related to Lorentz spin labels and $R$-symmetry labels as
\begin{gather}
\label{eqn:osctoDynkin}
    \ell_1 = n_{\mathbf{b}_{2}} - n_{\mathbf{b}_{1}},\  \ell_2 = n_{\mathbf{a}_{1}} - n_{\mathbf{a}_{2}},\  q_1 = n_{\mathbf{f}_{1}} - n_{\mathbf{f}_{2}},\  p = n_{\mathbf{f}_{2}} - n_{\mathbf{f}_{3}},\  q_2 = n_{\mathbf{f}_{3}} - n_{\mathbf{f}_{4}}\;,
\end{gather}
and parameterise the bare dimension $\Delta_0$ and length $L$ of single-trace operators as
\begin{align}
\la{Delta0inn}
    \Delta_ 0 &= \sum_{i = 1}^4 \frac{n_{{\bf f}_i}}{2} +\sum_{i = 1}^2 n_{{\bf a}_i}\;, \\ \nonumber
    \la{Linn}
    L &= \sum_{i = 1}^4 \frac{n_{{\bf f}_i}}{2} +
    \sum_{i = 1}^2 \left(\frac{ n_{{\bf a}_i}}{2} - \frac{n_{{\bf b}_i}}{2} \right)\;.
\end{align}
At finite coupling, the symmetry group reduces to $\rm PSU(2,2 |4)$ and a multiplet is described by the usual set of 6 quantum numbers $\left\{\Delta\,,\;\ell_1\,,\;\ell_2\,,\;q_1\,,\;p\,,\;q_2\right\}$~\eqref{eqn:osctoDynkin}, mixing operators with different length $L$~\cite{Beisert:2003ys}.
Even though the integer $L$ is no longer associated with any symmetry at finite coupling, we still can use it to label the states by tracing them back to $g=0$.
Similarly, the bare dimension $\Delta_0$ is not a quantum number, but rather a useful label which can be used to distinguish different states.

Even though the oscillator numbers contain more information than the Dynkin labels, we will still find several operators sharing the same oscillator numbers. In order to distinguish them we introduce an additional label which we call the multiplicity label, it is a positive integer and denoted as {\tt sol}.
Therefore, in order to identify each superprimary state/operator in planar ${\cal N} = 4$ SYM uniquely, we introduce the
$\texttt{State ID}$, which comprises the set of oscillator numbers, the bare dimension and solution number. We have
\begin{gather}
\texttt{State ID}:    {}_{\Delta_0}[n_{{\bf b}_1}\ n_{{\bf b}_2} \  n_{{\bf f}_1}\ n_{{\bf f}_2},n_{{\bf f}_3}\ n_{{\bf f}_4}\  n_{{\bf a}_1}\ n_{{\bf a}_2}]_{\texttt{sol}}\;.
\end{gather}
Another reason why oscillator numbers are particularly convenient is because they enable us to immediately associate a state/operator with its possible field content at zero coupling.
Below, we present the map between oscillator numbers and ``fundamental fields''.\footnote{The difference between fundamental fields and fields used in equation~\eqref{eqn:fieldDef} is that we consider the covariant derivative as a separate fundamental field, rather than it being a constituent of another field~\cite{Marboe:2017dmb}.} We have

\begin{table}[h!]
\centering
\begin{tabular}{ c|C|C|C|C }
 \hline
Type of Field & \text{Field interpretation} & \text{Oscillator Content} & \Delta_0 & L\\
 \hline\hline
Scalar & \Phi_{a\,b} & \mathbf{f}_{a}^{\dagger} \mathbf{f}_{b}^{\dagger}|0 \rangle & 1 & 1 \\
 \hline
Gluino/Fermion & \Psi_{a\, \alpha} & \mathbf{f}_{a}^{\dagger}  \mathbf{a}_{\alpha}^{\dagger} |0 \rangle & \frac{3}{2} & 1 \\
 & \bar{\Psi}_{a\, \dot{\alpha}} & \epsilon_{abcd} \mathbf{f}_{b}^{\dagger} \mathbf{f}_{c}^{\dagger} \mathbf{f}_{d}^{\dagger} \mathbf{b}_{\dot{\alpha}}^{\dagger} |0 \rangle  & \frac{3}{2} & 1 \\
 \hline
Gluon Field Strength & \mathcal{F}_{\alpha\, \beta} & \mathbf{a} _{\alpha}^{\dagger} \mathbf{a}_{\beta}^{\dagger} | 0 \rangle & 2 & 1\\
 & \bar{\mathcal{F}}_{\dot{\alpha}\, \dot{\beta}}  & \mathbf{f}_1^{\dagger} \mathbf{f}_2^{\dagger} \mathbf{f}_3^{\dagger} \mathbf{f}_4^{\dagger} \mathbf{b}_{\dot{\alpha}}^{\dagger} \mathbf{b}_{\dot{\beta}}^{\dagger} | 0 \rangle  & 2 & 1\\
 \hline
Covariant Derivative & \mathcal{D}_{\alpha\, \dot{\alpha}} & \mathbf{a}_{\alpha}^{\dagger} \mathbf{b}_{\dot{\alpha}}^{\dagger} |0 \rangle  & 1 & 0 \\
 \hline
\end{tabular}
\caption{\small\it   The map between fundamental fields of~${\cal N} = 4$ SYM and action of different types of oscillators on a Fock vacuum $|0\rangle$.
The fields have the following indices: the Greek letters $\alpha,\,\beta,\,\dots$ and ${\dot \alpha},\,{\dot \beta},\,\dots$ are ${\rm SU}(2)$ spinor indices, take values $1,\,2$, and together furnish representations of the Lorentz algebra. The Latin letters $a,\,b,\,\dots$ are ${\rm SU}(4) \cong {\rm SO}(6)$ spinor indices, take values $1,\dots,4$, and furnish a representations of the $R$-symmetry algebra.
Content of the table taken from~\cite{Beisert:2004ry,Marboe:2017dmb}.}
\label{tab:FieldsAndOsc}
\end{table}
\noindent
In table~\ref{tab:FieldsAndOsc}, the fermionic creation operators $\mathbf{f}_a^{\dagger}$ increase the corresponding fermionic oscillator number $n_{\mathbf{f}_a}$, $\mathbf{a}_\alpha^{\dagger}$ are counted by $n_{\mathbf{a}_\alpha}$ and $n_{\mathbf{b}_{\dot\alpha}}$ counts the number of $\mathbf{b}_{\dot\alpha}^{\dagger}$, {\it cf.}~\eqref{eqn:oscnumbers}. Below, we present some examples, to illustrate this.

\paragraph{{\tt Konishi} multiplet.}
To denote this multiplet, we need the oscillator content of its highest weight state ${\cal O}_\text{h.w.}$, which is\footnote{The indices used in the r.h.s. of the second equality, are ${\rm SO}(6)$ vector indices, and take the values $I = 1,\dots,6$.}
\begin{align}
    {\cal O}_\text{h.w.} =
    {\rm tr}\,(\mathcal{X} \bar{\mathcal{X}}) +{\rm tr}\,(\mathcal{Y} \bar{\mathcal{Y}}) +{\rm tr}\,(\mathcal{Z} \bar{\mathcal{Z}}) = {\rm tr}\,\Phi_I\,\Phi_I\;.
\end{align}
In table~\ref{tab:ScalarsTab}, we present the 6 scalar fields of ${\cal N} = 4$ SYM with their respective oscillator content.
\begin{table}[h]
\centering
\begin{tabular}{ C|C|C|C }
 \hline
\text{Scalar field} & \text{Oscillator Content} & \Delta_0 & L\\
 \hline\hline
\mathcal{Z} \equiv \Phi_{1\,2} = \Phi_1 + i\,\Phi_2  & [0\ 0\ 1\ 1\ 0\ 0\ 0\ 0] & 1 & 1 \\
 \hline
\mathcal{X} \equiv \Phi_{1\,3} = \Phi_3 + i\,\Phi_4  & [0\ 0\ 1\ 0\ 1\ 0\ 0\ 0] & 1 & 1 \\
 \hline
\mathcal{Y} \equiv\Phi_{1\,4} = \Phi_5 + i\,\Phi_6 & [0\ 0\ 1\ 0\ 0\ 1\ 0\ 0] & 1 & 1 \\
 \hline
\bar{\mathcal{Z}} \equiv\Phi_{3\,4} = \Phi_1 - i\,\Phi_2 & [0\ 0\ 0\ 0\ 1\ 1\ 0\ 0] & 1 & 1 \\
 \hline
\bar{\mathcal{X}} \equiv\Phi_{2\,4} = \Phi_3 - i\,\Phi_4 & [0\ 0\ 0\ 1\ 0\ 1\ 0\ 0] & 1 & 1 \\
 \hline
\bar{\mathcal{Y}} \equiv\Phi_{2\,3} = \Phi_5 - i\,\Phi_6 & [0\ 0\ 0\ 1\ 1\ 0\ 0\ 0] & 1 & 1 \\
 \hline
\end{tabular}
\caption{\small\it    The six polarisations of $\Phi_{a\,b}$ and their respective oscillator content. We also display how they can be re-expressed in terms of the vector representation of ${\rm SO}(6)$, which capture the six real scalar fields of ${\cal N} = 4$ SYM.}
\label{tab:ScalarsTab}
\end{table}
Using it we can construct ${\cal O}_\text{h.w.}$, and therefore write down the {\tt State ID} of the {\tt Konishi} multiplet as
\begin{equation}
    \texttt{Konishi}:\  {}_{2}[0\ 0\ 1\ 1\ 1\ 1\ 0\ 0]_{1}\;,
\end{equation}
The superprimary of this multiplet has the bare dimension $2$ (the left subscript) and as it is the only multiplet with these quantum numbers it has unit multiplicity, and therefore ${\tt sol} = 1$ (the right subscript).

\paragraph{{\rm\bf SL}(2) sector.}
Consider another important example of multiplets in the $\rm SL(2)$ sector. Each such multiplet contains an operator that is constructed out of $L$ scalars ${\cal Z}$, with $S$ covariant derivatives, that can schematically\footnote{
Concretely, such operators are ${\rm tr}\,({\cal D}^{n_1}\,{\cal Z})\dots({\cal D}^{n_1}\,{\cal Z})$, with $\sum_k^L n_k = S$.
} be represented as ${\rm tr}\,\mathcal{D}^{S} {\cal Z}^{L}$.\footnote{Indeed, the {\tt Konishi} multiplet contains the operator ${\rm tr}\,{\cal D}^2\,{\cal Z}^2$ of this form and thus the {\tt Konishi} multiplet is also an $\mathfrak{sl}(2)$ multiplet.} These operators are not the highest weight states in the multiplet by our choice of convention/grading (see \cite{Marboe:2017dmb} for the details). In our notation $\rm SL(2)$ sector multiplets can be identified by the following form of their {\tt State ID}. We have
\begin{equation}
\label{eq:Sl2}
  {\mathfrak{sl}(2)}\texttt{ sector: }  {}_{L+S-2}\left[ 0\ S{-}2 \ L{-}1\ L{-}1\ 1\ 1 \ S{-}2\ 0 \right]_{\texttt{sol}}.
\end{equation}
These states have quantum numbers
\begin{align}\label{eqn:Sl2Dynkin}
     {\mathfrak{sl}(2)}\texttt{ sector: } [\ell\;\ell\;0\;p\;0] = [S{-}2\;S{-}2\;0\;L{-}2\;0]
    \;.
\end{align}
Finally, we assign to each state an integer number \text{{\tt State} {\tt Number}} or
\text{{\tt St.} {\tt No.}} which is just a useful label to enumerate all states in the database.

\section{QSC: A practical user's manual}
\la{sec:qsc}

Here we describe the Quantum Spectral Curve (QSC) \cite{Gromov:2013pga, Gromov:2014caa} a method to solve the spectral problem for planar ${\cal N} = 4$ SYM for both analytical and numerical applications. In this section we cover the basics of the QSC and introduce key notations.

\subsection{Generalities}

Each state in this theory is associated with $2^8$ Q-functions of a complex {\it spectral parameter} $u$, associated with the ${\rm PSU(2,2|4)}$ symmetry of the theory. The Q-functions are not all independent and are related by the QQ-relations described below.
In the well studied cases of the spin-chains the Q-functions (or Baxter polynomials for the compact symmetry group) are the building blocks for the wave-functions of the states in a separation of variables basis (for recent developments see~\cite{Gromov:2016itr, Maillet:2018bim, Ryan:2018fyo, Maillet:2018czd, Ryan:2020rfk, Maillet:2020ykb, Cavaglia:2019pow, Gromov:2020fwh, Cavaglia:2021mft, Gromov:2022waj, Bercini:2022jxo}).
In particular, they contain all the quantum numbers of the state in the large-$u$ asymptotic, which includes, in particular, the exact scaling dimension $\Delta$.

In addition to the large-$u$ asymptotic, injecting the information about the quantum numbers, one has to impose additional analyticity constraints on the Q-functions.
As functions of the complex variable $u$ they may have branch cuts with the branch points at $\pm 2\,g+ i\,n/2$ for an integer $n$
and $g=\frac{\sqrt\lambda}{4\pi}$ related to the 't Hooft coupling $\lambda$.

The monodromy of the Q-functions through branch cuts is linked to a symmetry of the QQ-relations, which can be traced back to the crossing symmetry and which allows to constrain them to a discrete set of solutions, corresponding to the local states of the theory \cite{Gromov:2014caa}. We  describe this property below in more detail.

\subsubsection{Algebraic properties}

A Q-function is labelled by up to 8 indices, with a bar separating up to four indices from the rest.
The Q-functions are absolutely anti-symmetric for each of two groups of indexes.
It is easy to see that this labelling accounts for all $2^8$ Q-functions.

The advantage of this labelling of the Q-function is that the QQ-relations (a.k.a. Wronskian identities) can be written in a compact form (in the conventions of~\cite{Gromov:2014caa}):
\begin{align}\label{eqn:QQrelation1}
    Q_{A \mid I} Q_{A a b \mid I} &= Q_{A a \mid I}^{+} Q_{A b \mid I}^{-}-Q_{A a \mid I}^{-} Q_{A b \mid I}^{+}\;, \\
    \label{eqn:QQrelation2}
    Q_{A \mid I} Q_{A \mid I i j} &= Q_{A \mid I i}^{+} Q_{A \mid I j}^{-}-Q_{A \mid I i}^{-} Q_{A \mid I j}^{+}\;, \\
    \label{eqn:QQrelation3}
    Q_{A a \mid I} Q_{A \mid I i} &= Q_{A a \mid I i}^{+} Q_{A \mid I}^{-}-Q_{A \mid I}^{+} Q_{A a \mid I i}^{-}\;.
\end{align}
Here we use the notation $f^\pm \equiv f(u \pm i/2)$, lower case indices denote single indices, and upper case indices denote sets of indexes.

Additionally, we have the following determinant-type relations~\cite{Gromov:2014caa}:
\begin{align}
    \label{eqn:DetRel1}
    Q_{a_1 \ldots a_{k+n} \mid i_1 \ldots i_k} &= \frac{(n+k) !}{n ! k !} Q_{\left[a_1 \ldots a_n \mid \emptyset\right.} Q_{\left.a_{n+1} \ldots a_{n+k}\right] \mid i_1 \ldots i_k}^{[ \pm n]}
    \;, \\
    Q_{a_1 \ldots a_k \mid i_1 \ldots i_{k+n}} &= (-1)^{n\, k} \frac{(n+k) !}{n ! k !} Q_{a_1 \ldots a_k \mid \left[ i_1 \ldots i_k \right.}^{[ \pm n]} Q_{\left. \emptyset \mid i_{k+1} \ldots i_{k+n} \right]}
    \;,
    \label{eqn:DetRel2}
\end{align}
where $[\dots]$ stands stands for the standard anti-symmetrisation of the indices.

The Q-function with no indices, {\it i.e.} $Q_{\emptyset|\emptyset}$, is set to unity. The eight single index Q-functions ${\bf P}_a \equiv Q_{a|\emptyset}$ and ${\bf Q}_i \equiv  Q_{\emptyset|i}$, are distinguished, since starting with them and using the QQ-relations ~\eqref{eqn:QQrelation1},~\eqref{eqn:QQrelation2} and~\eqref{eqn:QQrelation3}, we can construct all $2^8$ Q-functions.
A special case of the equation~\eqref{eqn:QQrelation3} with $A=\emptyset, I = \emptyset$, allows us to relate the distinguished Q-functions, with two-index Q-functions, and is very useful practically
\begin{align}\label{eqn:QQmain1}
    Q_{a|i}^+ - Q_{a|i}^- = {\bf P}_a {\bf Q}_i\;.
\end{align}
One  defines Hodge-dual Q-functions with upper indices in the following way~\cite{Gromov:2014caa,Gromov:2017blm}:
\begin{align}
    Q^{a_1 a_2 \cdots a_n \mid i_1, i_2 \cdots i_m} \equiv(-1)^{n m} \epsilon^{a_1 a_2 \cdots a_n b_1 b_2 \cdots b_{4-n}} \epsilon^{i_1, i_2 \cdots i_m j_1, j_2 \cdots j_{4-m}} Q_{b_1 b_2 \cdots b_{4-n} \mid j_1 j_2 \cdots j_{4-m}}\;,
\end{align}
with $b_1<\cdots<b_{4-n}$ and $j_1<\cdots<j_{4-m}$.
Then, we see that $Q^{\emptyset|\emptyset} = Q_{1234|1234}$, which we further require to be equal to unity as well.\footnote{This can be linked with the ${\rm P}$ in ${\rm PSU}$~\cite{Gromov:2017blm}.}
We can also define distinguished Q-functions with single upper-indices:
\begin{align}\label{eqn:SingleUpDef1}
    \mathbf{P}^a \equiv Q^{a \mid \emptyset} &= Q^{a \mid i}(u+ i / 2) \mathbf{Q}_i\;, \\
    \mathbf{Q}^i \equiv Q^{\emptyset \mid i} &= Q^{a \mid i}(u+ i / 2) \mathbf{P}_a
    \;.\label{eqn:SingleUpDef2}
\end{align}
Using the determinant relations~\eqref{eqn:DetRel1} and~\eqref{eqn:DetRel2} to write $Q_{1234|1234}$ as a $4\times 4$ determinant of various $Q_{a|i}$, one gets
\begin{align}\label{eqn:Q1234Det}
    Q_{1234 \mid 1234}=\left|\begin{array}{llll}
Q_{1 \mid 1} & Q_{1 \mid 2} & Q_{1 \mid 3} & Q_{1 \mid 4} \\
Q_{2 \mid 1} & Q_{2 \mid 2} & Q_{2 \mid 3} & Q_{2 \mid 4} \\
Q_{3 \mid 1} & Q_{3 \mid 2} & Q_{3 \mid 3} & Q_{3 \mid 4} \\
Q_{4 \mid 1} & Q_{4 \mid 2} & Q_{4 \mid 3} & Q_{4 \mid 4}
\end{array}\right| = 1\;.
\end{align}
Now, since we know that $Q_{1234|1234} = 1$, we see that $Q_{1234|1234}^+-Q_{1234|1234}^- = 0$. At the same time, the r.h.s. of~\eqref{eqn:Q1234Det} gives a non-trivial constraint:
\begin{align}
    \mathbf{Q}_i \mathbf{P}_a Q^{a \mid i}=0
    \;.
\end{align}
Combining with~\eqref{eqn:SingleUpDef1} and~\eqref{eqn:SingleUpDef2}, we get
\begin{align}\la{PPQQ}
    \mathbf{P}_a \mathbf{P}^a=\mathbf{Q}_i \mathbf{Q}^i=0\;.
\end{align}
Expanding the determinant~\eqref{eqn:Q1234Det} on the first row, and repackaging the $3\times 3$ sub-determinants as $Q^{a|i}$, we get
\begin{align}
\label{eqn:QaiInv}
    Q_{i \mid a} Q^{j \mid a}=-\delta_i^j\;.
\end{align}
From equations~\eqref{eqn:SingleUpDef1} and~\eqref{eqn:QaiInv}, we get
\begin{align}
    {\bf P}_{a} = - {\bf Q}^i \, Q_{a|i}^+\;, \\
    {\bf Q}_{i} = - {\bf P}^a \, Q_{a|i}^+\;.
    \label{eqn:QQmain2}
\end{align}
Combining equations~\eqref{eqn:QQmain1} and~\eqref{eqn:QQmain2}, we obtain
\begin{align}
    \label{eqn:QQmain3}
    Q_{a|i}^+ - Q_{a|i}^- = - {\bf P}_a {\bf P}^b Q_{b|i}^+\;.
\end{align}
Equations~\eqref{eqn:QQmain1},~\eqref{eqn:SingleUpDef1},~\eqref{eqn:SingleUpDef2}~\eqref{PPQQ}--\eqref{eqn:QQmain3}
will play the central role in the description of the numerical algorithm, as we describe in section \ref{sec:num}.

The Q-functions are further constrained by the analyticity conditions, which we describe below.

\subsubsection{Symmetries of the Q-system}
The subset of the QQ-relations above \eq{eqn:QQmain1}--\eq{eqn:QQmain3} is invariant under the following $\Lambda$-transformation~\cite{Marboe:2018ugv}
\begin{align}
\label{LambdaTrans}
{\bf P}_a\to x^{+\Lambda} {\bf P}_a\;\;,\;\;
{\bf Q}_i\to x^{-\Lambda} {\bf Q}_i\;\;,\;\;Q_{a|i}\to Q_{a|i}
\end{align}
and
\begin{align}
\label{LambdaTransUp}
{\bf P}^a\to x^{-\Lambda} {\bf P}^a\;\;,\;\;
{\bf Q}^i\to x^{+\Lambda} {\bf Q}^i\;\;,\;\;Q^{a|i}\to Q^{a|i}
\end{align}
where $x$ is some function of the spectral parameter $u$. As we show below, in order for this symmetry to remain the symmetry of the QSC, which also includes the analyticity constraints one can conclude that $x(u)$ should be the Zhukovsky functions, defined in equation~\eqref{eqn:ZhukDef} (or its power).
Note that other Q-functions outside the reduced set transform rather non-trivially under this map.

In addition one has the following linear transformation {\it $H$-symmetry}~\cite{Gromov:2014caa}
\begin{align}
\label{Htrans}
{\bP}_{a}\to \hat H_{a}^{\;\;b}{\bP}_b\;\;,\;\;
{\bQ}_{i}\to \check H_{i}^{\;\;j}{\bQ}_j
\end{align}
for some constant
matrices $\hat H$ and $\check H$.
The transformation of other Q-functions can be deduced from \eq{Htrans} and can be found in \cite{Gromov:2014caa}.
We also impose ${\rm det} \hat H=
{\rm det} \check H =1$ in order to preserve $Q_{\emptyset|\emptyset}=1$.
We also have
\begin{align}
\label{HtransUp}
{\bP}^{a}\to \hat H^{a}_{\;\;b}{\bP}^b\;\;,\;\;
{\bQ}^{i}\to \check H^{i}_{\;\;j}{\bQ}^j
\end{align}
where $\hat H^{a}_{\;\;b}$
and $\check H^{i}_{\;\;j}$ are inverse and transposed matrices of $\hat H_{a}^{\;\;b}$ and $\check H_{i}^{\;\;j}$.

\subsubsection{Asymptotic properties}

Like in the case of the simple spin chains, in the case of the planar ${\cal N}=4$ SYM, the quantum numbers of the states are hidden in the large-$u$ behaviour of the Q-functions.
Since, we can build all the Q-functions starting from the distinguished single-index ones, it is sufficient to specify the asymptotic information only for these functions.
The asymptotics of ${\bf P}_a$ and ${\bf Q}_i$ are related to the quantum numbers of the state that they characterise. Introduce the constants $\mathtt{powP}_a$ and $\mathtt{powQ}_i$, so that as $u\to\infty$, we have
\begin{align}
    \label{eqn:PQasymDef}
    {\bf P}_a \simeq \mathbb{A}_a\, u^{\mathtt{powP}_a} \;,\qquad {\bf Q}_i \simeq \mathbb{B}_i\, u^{\mathtt{powQ}_i}\;,
\end{align}
where $\mathbb{A}_a$ and $\mathbb{B}_i$ are constants, defined below. The large-$u$ asymptotic is given by
\begin{align}
    \mathtt{powP} &= \{
     n_{{\bf f}_1} - 2 - \Lambda, n_{{\bf f}_2} - 1 - \Lambda, n_{{\bf f}_3} - \Lambda, n_{{\bf f}_4} + 1 - \Lambda
    \}\;,\\
    \mathtt{powQ} &= \bigg\{
    L + \frac{\gamma}{2} + n_{{\bf b}_1}  + \Lambda, L + \frac{\gamma}{2}  + n_{{\bf b}_2} + 1 + \Lambda, - \frac{\gamma}{2}  - n_{{\bf a}_1} - 2 + \Lambda, - \frac{\gamma}{2}  - n_{{\bf a}_2} -1  + \Lambda
    \bigg\}\;,
\end{align}
where $\Lambda$ is the ambiguity related to the symmetry \eq{LambdaTrans}, $n$'s are the oscillator numbers of the state
and the anomalous dimension $\gamma$ is defined as
\begin{align}
    \gamma \equiv \Delta - \Delta_0\;,
\end{align}
where $\Delta_0$ and $L$ can be written in terms of the oscillator numbers via \eq{Delta0inn}.
Furthermore, the asymptotics of the single-upper-index Q-functions are
\begin{align}
    {\bf P}^a \simeq \mathbb{A}^a\, u^{-\mathtt{powP}_{a} - 1}\;,\qquad {\bf Q}^i \simeq \mathbb{B}^i\,u^{-\mathtt{powQ}_{i}-1} \;.
\end{align}
The constants $\mathbb{A}_a$ and $\mathbb{B}_i$
contain certain freedom as they can be re-defined using $H$-symmetry \eq{Htrans}, However the combinations $\mathbb{A}_a \mathbb{A}^a$ and $\mathbb{B}_i \mathbb{B}^i$ (for each $a$ and $i$) are invariant under the subset of the $H$-symmetry, which preserves the asymptotic \eq{eqn:PQasymDef}.
Those $8$ combinations can be also written explicitly in terms of the quantum numbers of the states~\cite{Gromov:2014caa}. In this paper we adopt the conventions of~\cite{Marboe:2018ugv} where the individual $\mathbb{A}_a$ and $\mathbb{B}_i$ are fixed in terms of \texttt{powP} and \texttt{powQ}
\begin{equation} \label{Aafix}
\mathbb{A}_a =  \, \frac{(\mathtt{powP}_a+\mathtt{powQ}_1 + 1)(\mathtt{powP}_a+\mathtt{powQ}_2+1)}{ \prod\limits_{b > a} i (\mathtt{powP}_a-\mathtt{powP}_b)}\;,
\end{equation}
\begin{equation} \label{Aupfix}
\mathbb{A}^a = \frac{(\mathtt{powP}_a+\mathtt{powQ}_3 + 1)(\mathtt{powP}_a+\mathtt{powQ}_4+1)}{ \prod\limits_{b < a} i (\mathtt{powP}_a-\mathtt{powP}_b)}\;,
\end{equation}
\begin{equation}
  \mathbb{B}_j =\begin{cases}
    \dfrac{1}{ \prod_{k > j} i  (-\mathtt{powQ}_j + \mathtt{powQ}_k)}\;, & \text{for }j=1,2\;,\\[15pt]
    \dfrac{\prod_{k} (\mathtt{powP}_k + \mathtt{powQ}_j + 1)}{ \prod_{k > j}  i(\mathtt{powQ}_j - \mathtt{powQ}_k)}\;, & \text{for }j=3,4\;,
  \end{cases}
\end{equation}
\begin{equation*}
    \\[2pt]
\end{equation*}
\begin{equation}
  \mathbb{B}^j =\begin{cases}
    \dfrac{\prod_{k} (\mathtt{powP}_k + \mathtt{powQ}_j+1)}{   \prod_{k < j} i(\mathtt{powQ}_j -\mathtt{powQ}_k)}\;, & \text{for }j=1,2\;,\\[15pt]
    \dfrac{1}{   \prod_{k < j} i  (-\mathtt{powQ}_j +\mathtt{powQ}_k)}\;, & \text{for }j=3,4\;.
  \end{cases}
\end{equation}

\subsubsection{Analytic constraints}
We will mainly focus on the analytic properties of ${\bf P}_a$ and ${\bf Q}_i$ and the Hodge dual ${\bf P}^a$ and ${\bf Q}^i$.

Firstly, on their main sheet, the ${\bf P}_a$ and $\bP^a$ have only two branch points, which are located at $-2g$ and $+2g$, and joined by a short branch cut, ${\it i.e.}$ a branch cut passing through zero.
As such, they can be efficiently parameterised by the Zhukovsky variable $x(u)$ defined by
\begin{align}\label{eqn:ZhukDef}
    x(u) \equiv \frac{u + \sqrt{u-2g}\sqrt{u+2g}}{2g}\;.
\end{align}
As a result, we have the following \emph{convergent} expansion for $\mathbf{P}$ which  parameterises it  everywhere on the main sheet and even includes some area around the cut on the next sheet:
\begin{align}\label{eqn:PinZhuk}
    {\bf P}_a = \,(g\,x)^{\mathtt{powP}_a}
    \left(
    \mathbb{A}_{a} +
    \sum_{n = 1}^\infty
    \frac{c_{a,n}}{x^n}
    \right)\;,
     \qquad
     {\bf P}^a =
     \,(g\,x)^{-\mathtt{powP}_a-1}
    \left(
    \mathbb{A}^{a} +
    \sum_{n = 1}^\infty
    \frac{c^{a,n}}{x^n}
    \right)
    \;.
\end{align}

Note that explicit dependence on $\Delta$ of the state enters through $\mathbb{A}_a$ and $\mathbb{A}^a$ to~\eqref{Aafix} and~\eqref{Aupfix}.
The expansion coefficients $c_{a,n}$, $c^{a,n}$ and $\Delta$ are some functions of $g$ for each particular state and contain complete information on all other Q-functions~\cite{Gromov:2015wca} as we also argue below. In particular one can reconstruct $\bQ_i$ and $\bQ^i$.

The expansion \eq{eqn:PinZhuk} is also convergent near the brunch-cut $[-2\,g,2\,g]$ on the second sheet, where the $\bP$-functions contains infinite series of other cuts $[-2\,g+i\,{\mathbb Z},2\,g+i\,{\mathbb Z}]$. Denote the analytic continuation of ${\bf P}_a$ on this sheet by $\tilde{\bf P}_a$. Due to the parametrisation~\eqref{eqn:PinZhuk} in terms of the Zhukovsky variable, we can obtain an expression for $\tilde{\bf P}_a$ by just sending $x\to 1/x$ in~\eqref{eqn:PinZhuk}.

Note that the fact that the branch cut $[-2\,g,2\,g]$, which becomes a unit circle in $x$ variable, is inside the convergence domain of the Laurent series in $x$ guarantees fast convergence with the coefficients $c_{a,n}$ decaying exponentially at sufficiently large $n$. For numerical applications this allows to truncate the series at finite $n$ while keeping the high precision of the parametrisation in the vicinity of the branch-cut and the whole main sheet.

One should note, that due to the $H$-symmetry
\eq{Htrans}, partially fixed by the asymptotic \eq{eqn:PQasymDef}, there is still some freedom in re-defining the coefficients $c_{a,n}$, which we fix below in a particular way, depending on a type of the state. We refer to this symmetry as \textit{gauge symmetry}.

Finally, in order to constrain the infinite set of constraints $c_{a,n},\;c^{a,n}$ and $\Delta$ we have to impose the following analytic conditions on the ${\bf Q}$'s:
\begin{itemize}
\item The ${\bf Q}_i$ should be analytic in the upper half-plane, and have infinitely many branch points on the lower-half plane, located at $\pm 2g - i\, n$, with $n\in\mathbb{Z}$ and $n\geq 0$, joined by short cuts. This property is guaranteed by construction and follows from the representation of $\bP$ in \eq{eqn:PinZhuk}.
\item Upon passing through the branch cut on the real axis, we obtain $\tilde{\bf Q}_i$, which are analytic in the lower half-plane, and have an infinite series of cuts on the upper half plane.

Note that the branch structure of $\tilde{\bf Q}_i$ is similar to that of $\bar{\bf Q}_i$ (i.e. complex conjugate function), the complex conjugate of ${\bf Q}_i$.
In fact, we have to impose the following ``gluing'' conditions~\cite{Gromov:2014caa, Gromov:2015vua, Alfimov:2018cms}\footnote{This condition is essentially a reflection of the unitarity of the ${\cal N}=4$ SYM. However, for some applications, such as complex 't Hooft coupling analytic continuation, complex spin or Fishnet limit this may not be general enough. In general one can always construct two sets of $\bQ_i$ with upper and lower half analyticity and glue those sets on the cut $[-2\,g,2\,g]$ on the real axis. In the present case, when the coupling is real, the complex conjugation is simply a quick way to construct the lower-half-plane analytic set of the $\bQ$-functions for numerical applications.}:
\begin{align}\la{Qgluing}
    \tilde{\bf Q}_i = G_{i\,j}\,\bar{\bf Q}^{j} \;, \qquad \tilde{\bf Q}^i = G^{i\,j}\,\bar{\bf Q}_{j} \;,
\end{align}
where the most general form of them can be written as
\begin{align}
    G_{i\,j} =
    \left(\begin{array}{c c c c}
        0 & \alpha_1 & 0 & 0 \\
        \alpha_2 & 0 & 0 & 0 \\
        0 & 0 & 0 & \beta_1 \\
        0 & 0 & \beta_2 & 0
    \end{array}\right)_{i\,j}
    \;,
    \qquad
    G^{i\,j} =
    \left(\begin{array}{c c c c}
        0 & \alpha_3 & 0 & 0 \\
        \alpha_4 & 0 & 0 & 0 \\
        0 & 0 & 0 & \beta_3 \\
        0 & 0 & \beta_4 & 0
    \end{array}\right)^{i\,j}
    \label{GluingMatrice}
    \;,
\end{align}
which relate $\tilde{\bf Q}_i$ and $\bar{\bf Q}_i$. The coefficients in the gluing matrices are related as we should have $\bar G^{i\,j}$ is inverse matrix of $G_{i\,j}$.
Furthermore, $G_{i\,j} = \bar{G}_{j\,i}$~\cite{Alfimov:2018cms}.
Therefore, only two parameters are independent, {\it i.e.}
\begin{align}~\label{eqn:gluparams}
    \alpha_1 = \bar{\alpha}_2 =  1/\alpha_3 = 1/\bar{\alpha}_4 = \alpha\;, \qquad 1/\beta_1 = 1/\bar{\beta}_2 =  \beta_3 = \bar{\beta}_4 = \beta
    \;.
\end{align}
\end{itemize}

The gluing conditions \eq{Qgluing} impose strong constraints on $\{\Delta,c_{a,n},c^{a,n}\}$, giving a discrete set of solutions (up to the gauge symmetry). The way to impose and solve this constraint numerically was initially developed in~\cite{Gromov:2015wca}, and it is generalised in section \ref{sec:num} for all types of states. Next we review our notation for the $4$ types of states, classified according to some discrete symmetries of the system.

\subsection{Types of States and discrete Symmetries}
\label{sec:TypesStates}
\la{sec:LR}

Based on quantum numbers, we can divide the states of $\mathcal{N}=4$ SYM into four broad groups according to additional symmetries of their $\mathbf{P}_{a}$ functions. For that we will use two types of symmetries: \textit{left-right symmetry} and \textit{parity symmetry}, which we explore below.

\paragraph{Left-right Symmetry.}
There is an additional symmetry of interchanging $Q$ functions with their Hodge dual. This symmetry interchanges two ${\rm SU}(2|2)$ subgroups of the ${\rm PSU}(2,2|4)$ and is usually referred to as left-right Symmetry.

The state is invariant under the left-right symmetry if there exists a combination of $H$- and $\Lambda$-transformations which allows to relate $\mathbf{P}_{a}$ and $\mathbf{P}^{a}$ in the following way
\begin{equation}
\label{lr}
    \mathbf{P}_{a} = \chi_{ab} \mathbf{P}^{b},\
\end{equation}
where

\begin{equation}
  \chi_{ab} = -\chi^{ab}  =  \begin{pmatrix}
0 & 0 & 0 & 1 \\
0 & 0 & -1 & 0\\
0 & 1 & 0 & 0 \\
-1 & 0 & 0 & 0\\
\end{pmatrix}.
\end{equation}
The relation \eq{lr} implies, via the QQ-relations that one can also assume that $\bQ_i=\chi_{ij}\bQ^j$.
Thus, the anti-symmetric tensors $\chi_{ab}$ and $\chi_{ij}$ identifies the two ${\rm PSU}(2|2)$ subgroups reducing the symmetry to ${\rm OSp}(4|4)$ (up to a real form).

In this case we do not need to use both set of $\mathbf{P}_{a}$ and $\mathbf{P}^{a}$, and we have only one set of parameters $\{c_{a, n} \}$ corresponding to the lower index, which gives considerable acceleration of the numeric.

\paragraph{Parity Symmetry.}
Another $Z_2$ symmetry of the local operators of ${\cal N}=4$ SYM is the symmetry of reversing the order of operators under the trace. In the language of the spectral parameter this corresponds to changing $u$ to $-u$.

There is a distinct set of states which is invariant under this symmetry. In this case one can assume that $\mathbf{P}$'s have certain parity under $u \to -u$, so that it is possible to simplify the ansatz for $\mathbf{P}_{a}$ as follows, skipping every second power of $x$
\begin{equation}
    \mathbf{P}_{a} = (g x)^{\texttt{powP}_{a}} \left( \mathbb{A}_a + \sum_{n=1}^{+\infty} \frac{c_{a, 2n}}{x^{2n}} \right)\;.
\end{equation}
Again this amounts to efficiently reducing the number of parameters in the system by a factor of two, considerably improving the numerical performance for these states.

\paragraph{Types of states.}
As we explained above we have two types of $Z_2$ symmetries: \textit{left-right symmetry} and \textit{parity symmetry}. Each operator in SYM can be either in a singlet or doublet representation of each of the two symmetries. Accordingly, we have $4$ types of states, which require slightly different numerical treatment, as the additional reflection symmetries are beneficial for the reduction of the parameters. At the same time if the state is not invariant under a reflection then there should be another state with exactly the same dimension $\Delta$, which may also be used to save the computational time.
We introduce the  notation for the four type of states as shown in the Table~\ref{tab:types}.\footnote{There could be states invariant only w.r.t. to both reflection performed simultaneously. As these states are quire rare we have not introduce a separate code for those and they are treated as general.}
\begin{table}[h]
    \centering
    \begin{tabular}{c|c|c|c}
        Notation & LR-symmetry & $u\leftrightarrow -u$ symmetry & \# of states related by the symmetries\\ \hline\hline
        Type I & x & x & 1 \\
        Type II & x & o & 2\\
        Type III & o & x & 2\\
        Type IV & o & o & 2\text{ or }4\\
    \end{tabular}
    \caption{\small\it    Four types of the states. The type I states require the smallest number of parameters needed to reach a given accuracy and are the simplest to deal with numerically. At the same time the type IV states require roughly $4$ times more parameters and are the ``heaviest'' to deal with numerically. Note that it is not always totally trivial to identify the specific type of a state, as that requires finding the $H$-symmetry transformation which maps, Left to Right.}
    \label{tab:types}
\end{table}

\section{Setting up the Numerical Problem}
\label{sec:num}

In this section we describe the main steps of the numerical algorithm.
We setup our numerical problem essentially following~\cite{Gromov:2015wca}. The main novelty in our approach is elucidated in the next section, where we describe a method which considerably improves the performance at weak coupling, which is essential for highly excited states, that get more an more degenerate with increased bare dimension. As we describe below, this allows us to initialise the numerical procedure from weak coupling, using the perturbative QSC solver of~\cite{Marboe:2017dmb, Marboe:2018ugv}.

\subsection{Basics}
\label{sec:numBasic}
We describe the main steps of the algorithm for states with no additional symmetry (type IV or general states). There are some further simplifications to the procedure if we consider states with $Z_2$ symmetries (left-right symmetry or parity symmetry, {\it cf.}~section~\ref{sec:LR}), and some additional subtleties appear when we relax symmetries. Here for the most general states we describe the main steps of the algorithm and will comment on deviations, depending on the type of the states, in the section \ref{SectionTypesOfStates}.

\paragraph{Step 1: Parameters of the problem.}

We begin with the ansatz~\eqref{eqn:PinZhuk}. We truncate the infinite sum at some fixed value of $n$, which we denote as $\mathtt{cutP}$:
\begin{align}\la{Prepcut}
    {\bf P}_a = \,(g\,x)^{\mathtt{powP}_a}
    \left(
    \mathbb{A}_{a} +
    \sum_{n = 1}^\mathtt{cutP}
    \frac{c_{a,n}}{x^{n}}
    \right)\;,
\end{align}
\begin{align}\la{Prepcut1}
    {\bf P}^a = \,(g\,x)^{-\mathtt{powP}_a + 1}
    \left(
     \mathbb{A}^{a} +
    \sum_{n = 1}^\mathtt{cutP}
    \frac{c^{a,n}}{x^{n}}
    \right)\;.
\end{align}
Thus, the set of parameters of our problem is $\{\Delta, c_{a,n}, c^{a,n} \}$, with $a = 1,\dots,4,\; n = 1,\,\dots\,\mathtt{cutP}$. However, there is a nontrivial condition, following from the QQ-relations \eq{PPQQ}, which we use to constrain a set of  $\mathtt{cutP}$ parameters by setting

\begin{equation}\label{eqn:PaPa}
    \mathbf{P}_{a} \mathbf{P}^{a} = 0+{\cal O}(1/x^{\mathtt{cutP}})\;.
\end{equation}
Therefore,
the total number of parameters to fix from the gluing conditions \eq{Qgluing} is therefore $7\,\mathtt{cutP} + 1$ (which also includes $\Delta$).

The gauge symmetry ({\it i.e.}~left over $H$-symmetry) allows us to fix six coefficients
\begin{align}\label{eqn:gengauge}
\nonumber
    c_{2, \mathtt{powP}_2-\mathtt{powP}_1}=0\;,  \quad c_{3, \mathtt{powP}_3-\mathtt{powP}_1}=0\;, \quad c_{3, \mathtt{powP}_3-\mathtt{powP}_2}=0\;, \\  \quad  c_{4, \mathtt{powP}_4-\mathtt{powP}_1}=0\;, \quad  c_{4, \mathtt{powP}_4-\mathtt{powP}_2}=0\;, \quad  c_{4, \mathtt{powP}_4-\mathtt{powP}_3}=0\;,
\end{align}
which reduces the total number of free parameters to $7\,\mathtt{cutP} - 5$. In the case of the most general operators there are two additional constraints on the parameters as described below in subsection \ref{sec:gengen}. In cases of less symmetric states we have bigger gauge symmetry and thus more gauge conditions to impose. In table~\ref{table:numberparameters}, we summarise the number of independent parameters for the different types of states.

\begin{table}[h!]
\centering
 \begin{tabular}{c|c|c}
 \hline
 State type & $\#$ gauge conditions & $\#$ independent parameters \\
 \hline\hline
 I & 1 & 2 \texttt{cutP}\\
 II & 4 & 4 \texttt{cutP} - 4\\
 III & 2 & 7 \texttt{cutP}/2 - 1 \\
 IV & 6 & 7 \texttt{cutP} - 7 \\
 \hline
 \end{tabular}
 \caption{\small\it    Number of independent parameters for each state type.}
 \label{table:numberparameters}
\end{table}

\paragraph{Step 2: Finding $Q_{a|i}$ at large $u$.}
 Having the ansatz for $\bP$'s at hand, we  firstly go to the large $u$ limit. In this limit we can find the whole set of functions $\mathbf{P}_{a}$, $\mathbf{Q}_{i}$ and  $Q_{a|i}$ asymptotically.

For instance, given the asymptotics of $\mathbf{P}_{a}$ and $\mathbf{Q}_{i}$, one can read the asymptotic for $Q_{a | i}$ from the QQ-relation (\ref{eqn:QQmain3}). Therefore, one can approximate $Q_{a|i}$ at large $u$ as the following asymptotic expansion
\begin{equation}
\label{eqn:QaiLarge}
    Q_{a|i} \simeq u^{\texttt{powP}_{a} + \texttt{powQ}_{i} + 1} \sum_{n = 0}^{\texttt{cutQai}} \frac{B_{a|i,n}}{u^{n}}\;,
\end{equation}
introducing coefficients $B_{a|i,n}$ and the  cut-off for the number of terms in the asymptotic expansion $\texttt{cutQai}$. Our goal is to find the set of coefficients $B_{a|i,n}$ in terms of ${c_{a, n}}$ and ${c^{a, n}}$. For that we use the QQ-relation (\ref{eqn:QQmain3}) substituting the asymptotic expansions of all functions.\footnote{Note that due to the expected infinite series of branch-cuts in the lower half plane accumulating at $-i\,\infty$ the series above can only be asymptotic and has zero convergence radius.
In practice this means that one should increase ${\rm Im}\;u$ together with $\texttt{cutQai}$ to get a better approximation for $Q_{a|i}$.}

At the leading order, coefficients $B_{a|i,0}$ can be fixed in terms of $\mathbb{A}_a$, $\mathbb{B}_i$, $\mathtt{powP}_{a}$ and $\mathtt{powQ}_{i}$:
\begin{align}
    B_{a|i,0} = -i\, \frac{\mathbb{A}_a\, \mathbb{B}_i}{\mathtt{powP}_a + \mathtt{powQ}_i + 1}\;.
\end{align}
Expanding the QQ-relation ($\ref{eqn:QQmain3}$) in large $u$, we obtain a linear system of $16\times\texttt{cutQai}$ equations to determine $B_{a|i,n}$ where $a = 1,\dots, 4$, $i = 1, \dots, 4$ and $n = 1\dots\ \texttt{cutQai}$.
We have
\begin{align}\label{eqn:LinSys}
    T^{b}_{a|i,n}\,B_{b|i,n} = F_{a|i,n}
    \;,\qquad n = 1,\dots,\,{\tt cutQai}
    \;,
\end{align}
where
\begin{align}
\label{eqn:Tmatrix}
    T^{b}_{a|j,n} = i\,\delta^{b}_a\, ({\tt powP}_a + {\tt powQ}_j + 1 - n) +
    \mathbb{A}_{a} \mathbb{A}^{b}\;,
\end{align}
and $F_{a|j,n}$ is a function of all previously computed $\{B_{a|j,k}:\; k = 1,\dots,\,n-1\}$ and $\{
c_{a,m}\;,c^{a,m}:\;m = 1,\dots,\,n
\}$.

We can solve the linear system~\eqref{eqn:LinSys} to obtain the $B_{a|i,n}$ coefficients in terms $c_{a,k}$ and $c^{a, k}$, for $k = 1,\dots,\,n$, and the quantum numbers of the state.
As we explain in section~\ref{sec:gengen} this step should be taken with extra care, as the linear system may not have a solution.
In some cases, as we will see in section~\ref{SectionTypesOfStates}, the linear system~\eqref{eqn:LinSys} may have zero-modes, which need to be handled with care. We explain how to do this in detail in section~\ref{SectionTypesOfStates}.

As a result of solving the linear system~\eqref{eqn:LinSys}, we obtain $Q_{a|i}$ at large $u$ up to $\texttt{cutQai}$ orders in $1/u$.
In the next steps, we need to explore $Q_{a|i}$ near the branch cut.

\paragraph{Step 3: Approaching the cut.} Equipped with $Q_{a|i}$ at large $u$, we can proceed to approximating   $Q_{a|i}$ numerically close to the branch cut.
We start by evaluating~\eqref{eqn:QaiLarge} at some particular value of $u = u_0 + i\,\mathtt{QaiShift} - i/2$, where $\mathtt{QaiShift}$ is a large positive integer, so that the expansion~\eqref{eqn:QaiLarge} is reliable at this point.
Then we rewrite~\eqref{eqn:QQmain3} as
\begin{equation}
\label{ShiftingToCut}
    Q_{a|k}\left(u - \frac{i}{2}\right) = \left[ \delta_{a}^{b} + \mathbf{P}_{a} (u) \mathbf{P}^{b} (u) \right] Q_{b|k}\left(u + \frac{i}{2}\right) \equiv U_{a}^{b}(u)\  Q_{b|k}\left(u + \frac{i}{2}\right) \;.
\end{equation}
This allows us to recursively reduce the magnitude of the imaginary part of the argument of $Q_{a|i}$, until we reach just above the real axis. We have (with a matrix multiplication under the square brackets)
\begin{align}
    Q_{a|k}\left(u_0 + \frac{i}{2}\right) = \Bigg[ U(u_0 + i) \dots U(u_0 + (\texttt{QaiShift}-1)\ i) \Bigg]_{a}^{b}\,Q_{b|k}\left(u_0 + \mathtt{QaiShift}\  i - \frac{i}{2}\right)\;.
\end{align}
Notice that it is possible to do this because the ansatz for ${\bf P}_a$ in~\eqref{eqn:PinZhuk} is convergent, and therefore we can evaluate the product ${\bf P}_a\,{\bf P}^b$ at any value of $u$ in the upper-half plane, whereas the expansion~\eqref{eqn:QaiLarge} for ${Q_{a|i}}$ is asymptotic, and only gives a good approximation at large enough ${\rm Im}(u)$.

We repeat the procedure for
$\mathtt{nPoints}$ probe points on the cut. More precisely we use Chebyshev nodes in the interval $[-2g,2g]$ i.e. $u_{0,k}=2g \cos\left(\frac{2k-1}{2\;\mathtt{nPoints}}\pi\right)$ for $k=1,\dots,\mathtt{nPoints}$.
The Chebyshev nodes are particularly convenient because: firstly, they are more dense near the branch points, where the $\bQ$-functions are the most unpredictable; secondly, they are mapped points with equal separation on the unit circle under the Zhukovsky map~\eqref{eqn:ZhukDef}.

Having computed $Q_{a|i}^+$ we can easily find both ${\bf Q}_i$ and its analytic continuation under the cut on the real axis $\tilde{\bf Q}_i$ from~\eqref{eqn:QQmain2}:
\begin{equation}
    {\mathbf{Q}}_{i} = -Q_{a|i}^{+} {\mathbf{P}}^{a}\;\;,\;\;
    \tilde{\mathbf{Q}}_{i} = -Q_{a|i}^{+} \tilde{\mathbf{P}}^{a}\;,
\end{equation}
where we remember that $\tilde{\bf P}_a$ can be obtained from \eq{Prepcut} ${\bf P}_a$, by sending $x\to \tilde x= 1/x$.

Here let us notice that one can find $Q^{a|i}$ directly from $Q_{a|i}$ by using \eq{eqn:QaiInv}, which states that $Q^{a|i}$ is minus inverse of the transpose of $Q_{a|i}$. Then one constructs $\mathbf{Q}^{i}$ and $\mathbf{Q}_{i}$ in the same way, using
\begin{equation}
    {\bQ}^{i} = Q^{a|i, +} {\bP}_{a}\;\;,\;\;
        \tilde{\bQ}^{i} = Q^{a|i, +} \tilde{\bP}_{a}\;.
\end{equation}

In the end, we obtain lists of $\mathbf{Q}$-functions: $\mathbf{Q}_{i}$, $\mathbf{Q}^{i}$,  $\tilde{\mathbf{Q}}_{i}$ and $\tilde{\mathbf{Q}}^{i}$  evaluated at  $\mathtt{nPoints}$  Chebyshev nodes on the cut. We denote the lists as $\texttt{Qidownlist}$, $\texttt{Qiuplist}$, $\texttt{Qidowntlist}$ and $\texttt{Qiuptlist}$ respectively.
After we obtain the $\mathbf{Q}$-functions just above the branch cut, we can impose the gluing conditions \eq{Qgluing}.

In the next section, we will see how the gluing conditions are used to obtain a system of equations for the free parameters of our problem which is then solved using Newton's method.

\subsection{Closing the system: Newton method for harmonics on the cut}\label{sec:Newt}
Here we describe the main novelty of the current implementation.

First let us briefly summarise the previous method by~\cite{Gromov:2015wca}. We started from the $\bP$-functions, solved the QQ-relations to construct $Q_{a|i}$ which by itself enabled us to compute the $\bQ$-functions both above and below the branch cut. Basically for any set of parameters $\{c\} \equiv \{c_{a,n},c^{a,n}\}$ and $\Delta$, we would be able to conduct this procedure. In order to close the equations we have to impose the gluing condition, which relates $\bQ_i(u\pm i0)$ to $\bar\bQ^i(u\mp i0)$ for $u\in[-2g,2g]$. Schematically
\begin{align}
\bP_a,\;\bP^a\;\to\;Q_{a|i}\;\to\;\bQ_i(\{c\},\Delta|u),\;\tilde\bQ^i(\{c\},\Delta|u)
\end{align}
where we added the parameters into arguments of the $\bQ$-functions to emphasise that they can be constructed for given values of parameters.

The method of \cite{Gromov:2015wca} consisted in using an optimisation algorithm to adjust parameters in order to minimise the mismatch in the gluing condition evaluated at the Chebyshev nodes $u_m$:
\begin{align}
\min\sum_m\left|\bQ^k(\{c\},\Delta|u_{m}\pm i0)-
G^{kj}\;\overline{\bQ_j(\{c\},\Delta|u_{m}\pm i0)}\right|^2\;.
\end{align}
This works well at finite $g$, but loses precision quickly at very small couplings. The reason for this is that the convergence radius of the expansion \eq{eqn:PinZhuk} of the $\bP$-functions into inverse powers of $x$ at weak coupling has the convergence radius $1/|x\,(\pm 2\,g\pm i)|\simeq g$, meaning that the expansion coefficients $c_{a,n}\sim g^n$ all vanish quickly but at a different rate. The fact that they vanish quickly of course allows for very efficient truncation, which becomes exact at a given order in small coupling, however, the difference in the scale makes the optimisation problem ill-conditioned.

In order to resolve this problem in this paper we propose an alternative method. Denoting
\begin{align}
\bP^{\tt Updated}_a\equiv
- Q_{a|k}^+
G^{kj}\;\overline{\bQ_j(\{c\},\Delta|u\pm i0)}\;\;,\;\;u\in[-2g,2g]
\end{align}
such that on the solution of the QSC we should have $\bP^{\tt Updated}_a=\bP_a$, but ``off-shell'' it is a function with rather different analytic properties -- it is almost guaranteed to have infinitely many cuts even on the defining sheet. Nevertheless, we can use the Zukovsky map to resolve the cut on the real axis, which produces an analyticity annulus, containing the unit circle where we can define the Laurent expansion
\begin{align}
\bP^{\tt Updated}_a =
\,(g\,x)^{\mathtt{powP}_a}
    \sum_{n=-\infty}^{+\infty}
    \frac{c^{\tt Updated}_{a,n}}{x^n}
\end{align}
the prefactor we introduced for convenience, without loss of generality. We notice that on-shell ({\it i.e.}~when the gluing condition is satisfied) we should have all positive powers coefficients to vanish $c^{\tt Updated}_{a,n}=0,\;n<0$, zero power coefficient should become $1$ and the rest should coincide with the initial parameters $c_{a,n}$.

Technically, finding coefficients $c_{a,n}^{\tt Updated}$ is identical to the Fourier transformation in the angle $\phi$ going around the unit circle $x=e^{i\phi}$. Furthermore, since we have already values of the $\bP$-functions computed at the Chebyshev nodes we can conveniently use the fast-Fourier transformation method.
In a similar way we construct $\mathbf{P}^{a\ \texttt{Updated}}$ and its Fourier modes $c^{a,n\;{\tt Updated}}$.

Finally, we have to impose
\begin{equation}
\label{eqn:Newton}
\vec F(\{c\},\Delta)=\left\{c_{a,n}^{{\tt Updated}}(\{c\},\Delta)-c_{a,n},\;
c^{a,n\;{\tt Updated}}(\{c\},\Delta)-c^{a,n}
\right\}=0\;,
\end{equation}
which we do with the Newton method.
Notice, however, that the above equation is over-defined
as $c_{a,n}$ are only non-zero for positive $n$ and, furthermore, some of them are fixed to be zero for the gauge-fixing, as discussed  around equation~\eq{eqn:gengauge}. At the same time we have an extra parameter $\Delta$ in addition to $\{c\}$. For the Newton method we have to equalise the number of equations and the number of parameters. To do that in practice we drop all negative $n$'s, we use some of $n=0$ equations to solve for the parameters in the gluing matrix (as they are simply linear equations for $\alpha$ and $\beta$) while keeping one to match with $\Delta$. We also remove the equations corresponding to the gauge fixing conditions.
The exact details depends on the type of the state and can vary for a concrete implementation. Finally, one can use the discarded equations in order to estimate the accuracy of the gluing. Note that the Newton method, if convergent, finds zero in the multidimensional space of parameters with the precision which only depending on the number of iterations and is not suitable for precision control of the numerical approximation made at all other steps.

Let us note that for the Newton method we have to compute the derivative of the vector valued function $\vec F$, which we do by the simplest first order finite difference formula $\partial \vec F= \frac{1}{\epsilon}\big[\vec F(\dots+\epsilon)-\vec F\big]$. This is of course the slowest part of the algorithm as it requires approximately $8 \times \texttt{cutP}$ evaluations of the faction $\vec F$.
Furthermore, one should take into account the following subtlety in the technical implementation.
The coefficients $c_{1,n}$ and $c_{3,n}$ are purely real, whereas $c_{2,n}$ and $c_{4,n}$ are purely
imaginary.
When computing numerical derivatives, these parameters need to be shifted by a small constant $\epsilon$, as described above.
To ensure the correct direction of the shifting, for real parameters $c_{1, n}$ and $c_{3, n}$ the real part of equations \eqref{eqn:Newton} is taken, and for imaginary parameters $c_{2, n}$ and $c_{4, n}$ the imaginary part. Furthermore, this restricts the gluing coefficients $\alpha$ and $\beta$ in~\eqref{eqn:gluparams} to be real.

\begin{table}[h]
\centering
\begin{tabular}{  m{4cm} | m{10cm}  }
  \hline\hline
  \textbf{Step 1: Parameters of the problem} & Start with initial parameters $\Delta$, $c_{a,n}$ $c^{a,n}$, $a=1\dots4$, $n=1\dots \texttt{cutP}$, setting some $c_{a, n}$ to zero by the gauge choice. \\
  \hline
  \textbf{Step 2: Finding $Q_{a|i}$ at large $u$} & Go to the large $\text{Im}\; u$ limit. Solve the linear system of coefficients of $Q_{a|i}$ $B_{a|i,n}$, $a=1\dots4$, $i=1\dots4$, $n=1 \dots \texttt{cutQai}$.  \\
  \hline
  \textbf{Step 3: Approaching the cut} & Starting from large $\text{Im}\,u$ limit, shift $Q_{a|i}$ down to the branch cut. Get $Q_{a | i}$ just above the branch cut and calculate $\mathbf{Q}_{i}$, $\mathbf{Q}^{i}$, $\tilde{\mathbf{Q}}_{i}$ and $\tilde{\mathbf{Q}}^{i}$. \\
  \hline
  \textbf{Newton-Fourier} & \begin{itemize}
      \item Use Steps 1--3 to find $c_{a, n}^{\texttt{Updated}}$ $c^{a, n\; \texttt{Updated}}$ as functions of the initial parameters by performing a Fourier transform on the unit circle in $x$.
      \item Use Newton method to find the values of parameters such that
$c_{a,n}^{\texttt{Updated}}=c_{a,n}$ and $c^{a,n\;\texttt{Updated}}=c^{a,n}$.
\item
Read-off the value of $\Delta$ among other parameters at the solution of the equation $c_{a,n}^{\texttt{Updated}}=c_{a,n}$ and $c^{a,n\;\texttt{Updated}}=c^{a,n}$.
  \end{itemize} \\
  \hline
\end{tabular}
\caption{\small\it    The main steps of the numerical algorithm.}
\label{tab:algorithm}
\end{table}

The main steps of the algorithm are summarised in table~\ref{tab:algorithm}.

\subsection{Details of the numerical procedure for different types of states}
\label{SectionTypesOfStates}

Based on the behaviour of the states w.r.t. the two discrete symmetries (left-right symmetry and parity symmetry) we split all operators into $4$ types as explained in section~\ref{sec:LR}. Accordingly we have $4$ slightly different implementations of the QSC solver for each of these types and in this section we describe the specific differences.

\subsubsection{Left-right and parity symmetric case (Type I)}
\label{LReven}

This case is the most symmetric one and therefore the simplest. Here we will outline the simplifications in the algorithm which arise because of symmetries.

First of all, we have the least amount  of independent parameters among all cases. Due to the left-right symmetry, at $\textbf{Step 1}$ we introduce coefficients only for $\mathbf{P}_{a}$:
\begin{equation}
\label{PLReven}
    \mathbf{P}_{a} =  (g\,x)^{\texttt{powP}_{a}} \left(\mathbb{A}_{a} + \sum_{n=1}^{\texttt{cutP}/2} \frac{c_{a,2\,n}}{x^{2\,n}} \right)\;,
\end{equation}
as $\mathbf{P}^{a}$ can be obtained by raising the index $\mathbf{P}^{a} = \chi^{ab} \mathbf{P}_{b}$. Here let us note that we convert $\mathbf{P}_{a}$ and $\mathbf{P}^{a}$ available in the perturbative solver \cite{Marboe:2017dmb, Marboe:2018ugv} into the left-right symmetric form~\eqref{lr}. The details of this conversion are provided in the appendix \ref{app:LRconvert}. The expansion (\ref{PLReven}) includes only even powers because of the parity symmetry. In addition, we impose the following gauge conditions:
\begin{align}\label{eqn:LRgauge}
    c_{2, \mathtt{powP}_2-\mathtt{powP}_1}=0\;,  \quad c_{3, \mathtt{powP}_3-\mathtt{powP}_1}=0\;, \quad c_{3, \mathtt{powP}_3-\mathtt{powP}_2}=0\;, \quad  c_{4, \mathtt{powP}_4-\mathtt{powP}_1}=0\;.
\end{align}
In practice, three of the gauge conditions are already satisfied trivially, due to parity symmetry requiring that particular coefficients are zero.
Because of this, we end up with only $2\ \texttt{cutP}$ parameters ({\it cf.}~table~\ref{table:numberparameters}).

To start the algorithm, we need to find $Q_{a|i}$ at the large-$u$ limit at \textbf{Step 2}. The linear system~\eqref{eqn:LinSys} for the coefficients $B_{a|i,2\,n}$ where $a = 1\dots4, i=1\dots4, n=1\dots \texttt{cutQai/2}$ can be solved as a function of $c_{a, n}$ without complications, ${\it i.e.}$~there are no zero-modes in this case.

The step $\textbf{Step 3}$, where we compute $Q_{a|i}$ does not require a modification.

The gluing conditions are simplified as they connect only $\mathbf{Q}_{i}$ and $\tilde{\mathbf{Q}}_{i}$. The lists of $\mathbf{Q}_{i}$ and $\tilde{\mathbf{Q}}_{i}$ can be obtained as
\begin{equation}
    {\mathbf{Q}}_{i} = - Q_{a|i}^{+} \chi^{ab}{\mathbf{P}}_{b} \;\;,\;\;
    \tilde{\mathbf{Q}}_{i} = - Q_{a|i}^{+} \chi^{ab} \tilde{\mathbf{P}}_{b}\;.
\end{equation}
The gluing matrix \eqref{GluingMatrice} is simplified and involves the only one independent gluing constant
\begin{align}
    \tilde{\mathbf{Q}}_{i} = G_{\quad\; i}^{(\texttt{LR})\ j}\, \bar{\mathbf{Q}}_{j}\;,\qquad
\label{eqn:LRgluing}
    G_{\quad\; i}^{(\texttt{LR})\ j} =
    \left(\begin{array}{c c c c}
        0 & 0 & \alpha & 0 \\
        0 & 0 & 0 & -\bar{\alpha} \\
        \frac{1}{\alpha} & 0 & 0 & 0 \\
        0 & -\frac{1}{\bar{\alpha}} & 0 & 0
    \end{array}\right)_{i\,j}
    \;,
\end{align}
where $G_{\quad\; i}^{(\texttt{LR})\ j} = G_{ij} \chi^{jk} = \chi_{ik} G^{kj} $ is obtained by lowering the index $\mathbf{Q}^{i} = \chi^{i\, j}\, \mathbf{Q}_{j}$ of the general gluing conditions~\eqref{Qgluing} and noticing that $\beta = \bar{\alpha}$.
To close the system and set up a numerical search we follow the procedure of section \ref{sec:numBasic}.  Regarding the Newton search at the last step, the
main procedure is unchanged, and is the same as that described in section~\ref{sec:Newt}, albeit with only $\Delta$ and $c_{a,n}$ as parameters.

\subsubsection{Left-right symmetric states with general parity (Type II)}
\label{LRgen}

For this case, the parity symmetry is relaxed but the left-right symmetry is preserved. Therefore, in \textbf{Step 1} the ansatz for $\textbf{P}_{a}$ should contain both even and odd powers in the expansion:
\begin{equation}
    \textbf{P}_{a} = (g x)^{\texttt{powP}_{a}} \left( \mathbb{A}_{a} +  \sum_{n = 1}^{\texttt{cutP}} \frac{c_{a, n}}{x^{n}} \right)\;,
\end{equation}
which immediately gives twice more coefficients to define $\mathbf{P}_{a}$ up to the same $\texttt{cutP}$. However, we still can use the left-right symmetry to define $\mathbf{P}^{a} = \chi^{ab}\mathbf{P}_{b}$. In addition, all 4 gauge conditions for the left-right symmetric case (\ref{eqn:LRgauge}) are non-trivial. This gives us $4\,\texttt{cutP}-3$ parameters ({\it cf.}~table~\ref{table:numberparameters}) to the numerics.

The main important difference here is due to the additional obstruction in the linear system for the asymptotic expansion of $Q_{a|i}$. In the absence of the parity symmetry we notice that $\bQ_1$ and $\bQ_2$ can mix, {\it i.e.}~the large-$u$ expansion of $\bQ_2$ will be invariant under
\begin{align}\label{eqn:QiGaugeLRGen}
    \bQ_2 \rightarrow \bQ_2 + \eta\, \bQ_1\;,
\end{align}
for some constant $\eta$, as well as the large-$u$ expansion of $\mathbf{Q}_4$ will be invariant under
\begin{align}\label{eqn:QiGaugeLRGen2}
    \bQ_4 \rightarrow \bQ_4 + \mu\, \bQ_1\;,
\end{align}
for some constant $\mu$.
In the parity symmetric case one can avoid this mixing by requiring  $\mathbf{Q}_i$ to have definite parity, which then prevent the mixing as, for instance, $\bQ_1$ and $\bQ_2$ must have opposite parity due to the asymptotic. We notice that in the left-right symmetric case, due to the gluing condition \eqref{eqn:LRgluing}, we must have $\mu = - \eta\,\frac{\alpha}{\bar\alpha}$,
where $\alpha$ is the gluing constant.

This freedom affects the \textbf{Step 2}, where we build a large-$u$ expansion for $\Qai$.
The ambiguity in the definition of $\bQ_i$ translates into existence of a zero mode in the linear system~\eqref{eqn:LinSys}, responsible for the redefinition of $Q_{a|2}\to Q_{a|2}+\eta Q_{a|4}$ and
$Q_{a|4}\to Q_{a|4}+\mu Q_{a|1}$ in accordance with \eq{eqn:QiGaugeLRGen} and \eq{eqn:QiGaugeLRGen2}.
Thus the zero mode occurs at $n=n_\eta$ in~\eqref{eqn:LinSys}, which, as follows from the asymptotics, should be equal to
\begin{align}
    n_\eta \equiv \mathtt{powQ}_2 - \mathtt{powQ}_1 =
    n_{\mathbf{a}_{1}} - n_{\mathbf{a}_{2}} + 1 = n_{\mathbf{b}_{2}} - n_{\mathbf{b}_{1}} + 1
    \;,
\end{align}
which is the first level in the linear system~\eqref{eqn:LinSys}, where $Q_{a|1}$ and $Q_{a|2}$ can mix.
Technically this implies that $T_{n_\eta}^{a\,b|2}$ in \eqref{eqn:Tmatrix} has a null vector.
To be more specific, the four components of the right null vector   $B^{\texttt{zero\;mode}}_{a|2,n_\eta} $ are
\begin{align}
\begin{split}
    B^{\texttt{zero\;mode}}_{1|2,n_\eta}  &=
    -\frac{i \left(2 p-2 \Delta -2 \ell_1+3 q_1+q_2\right)}{\left(q_1+1\right) \left(p+q_1+2\right) \left(p+q_1+q_2+3\right) \left(2 p+2 \Delta +2 \ell_1+q_1+3 q_2+12\right)}\;,\\
    B^{\texttt{zero\;mode}}_{2|2,n_\eta}  &=\frac{2 p-2 \Delta -2 \ell_1-q_1+q_2-4}{(p+1) \left(p+q_2+2\right) \left(2 p+2 \Delta +2 \ell_1+q_1+3 q_2+12\right)}\;,\\
    B^{\texttt{zero\;mode}}_{3|2,n_\eta}  &=\frac{i \left(-2 p-2 \Delta -2 \ell_1-q_1+q_2-8\right)}{\left(q_2+1\right) \left(2 p+2 \Delta +2 \ell_1+q_1+3 q_2+12\right)}\;,\\
    B^{\texttt{zero\;mode}}_{4|2,n_\eta}  &= 1\;.
\end{split}
\end{align}
Note that we can always choose the $4$th component to be $1$, because the combination $2 p+2 \Delta +2 \ell_1+q_1+3 q_2+12>0$.
Since the $B^{\texttt{zero\;mode}}_{4|2,n_\eta}$ is always nonzero we can impose an additional constraint $B_{4|2,n_\eta} = 0$ and solve
\begin{align}
    T^{b}_{a|2,n_\eta}\, B_{b|2,n_\eta} = F_{a|2,n_\eta}\;,
\end{align}
for remaining $a = 1,\dots,3$. However, in order for this reduction to be consistent, one should also impose that the r.h.s. of the linear system \eqref{eqn:LinSys} shares the left null-vector with $T$. This is not guaranteed for arbitrary values of the parameters and imposes and additional constraint on the expansion coefficients $c_{a,n}$ of the $\bP_a$, with $n = n_\eta$.

To give the simplest example for such a constraint, for the states ${}_{4}[0\ 0\ 4\ 3\ 2\ 1\ 0\ 0]_{\tt sol}$, with ${\tt sol} = 1,2$, the value $n_\eta=1$ and the corresponding constraint is\footnote{There is a
freedom to choose which of the coefficients are considered to be the
dependent ones. In our implementation we always treat $c_{1,n_{\nu}}$ as a dependent
parameter for all type II states.}
\begin{equation}
\label{eqn:constraintlr}
    c_{1, 1} = \frac{6(\Delta - 5)(\Delta + 7)(\Delta - 1)^2 c_{3, 1} - 12(\Delta - 5)(\Delta + 3)^2 c_{2, 1} + 5(\Delta - 5)(\Delta - 1)(\Delta + 3) c_{4, 1}}{336 (\Delta - 1)(\Delta + 3) (
    \Delta + 7)^2}\;.
\end{equation}
Conceptually, this constraint means that the power-like asymptotic for $\bQ_i$ is consistent with the ansatz for $\bP_a$, the same way as for the general type of states the coefficients ${\mathbb A}_a$ are partially fixed by the quantum numbers.

Once $Q_{a|i}$ at large $u$ is defined, we can proceed with {\bf Step 3}, {\it i.e.}~shifting to the cut. Just above the cut we obtain $\mathbf{Q}_{i}$ and $\tilde{\mathbf{Q}}_{i}$ the same way as it is done for the left-right and parity symmetric case in section~\ref{LReven}.

During the Newton search, we need to treat the coefficient fixed by the constraint as a dependent parameter. Once the independent parameters are shifted by $\epsilon$, we need to calculate the shift in the dependent variable using the constraint ({\it cf}.~in the above example, where~\eqref{eqn:constraintlr} which can be used to fix $c_{1,1}$).

\subsubsection{General and parity symmetric states (Type III)}
\label{sec:geneven}

For this case we do not consider the left-right symmetry any more, though we keep parity symmetry. The ansatz for $\mathbf{P}_{a}$ and $\mathbf{P}^{a}$ in {\bf Step 1} is as follows:

\begin{equation}
    {\bf P}_a = \,(g\,x)^{\mathtt{powP}_a}
    \left(
    \mathbb{A}_{a} +
    \sum_{n = 1}^{\mathtt{cutP}/2}
    \frac{c_{a,2n}}{x^{2n}}
    \right)\;,
     \qquad
     {\bf P}^a =
     \,(g\,x)^{-\mathtt{powP}_a-1}
    \left(
    \mathbb{A}^{a} +
    \sum_{n = 1}^{\mathtt{cutP}/2}
    \frac{c^{a,2n}}{x^{2n}}
    \right)
    \;,
\end{equation}
which satisfy the identity~\eqref{eqn:PaPa}.
Due to this restriction, there are $7\   \texttt{cutP} / 2 $ parameters as one of the sets $\{ c_{a} \}$ or $\{ c^{a} \}$ is not independent.\footnote{
There is a freedom to choose which coefficients are kept made dependent ones, in our implementation we always treat $c^{1, n}$ as dependent parameters.}
The gauge conditions are the most general (\ref{eqn:gengauge})
though some one them are trivially satisfied because of parity symmetry. This results in  $7\  \texttt{cutP}/2 - 1$ parameters to start with ({\it cf.}~table~\ref{table:numberparameters}).

The rest of the algorithm is the same as outlined in the section \ref{sec:numBasic}. However, we will mention a technical subtlety in implementing the Newton's search: some coefficients $\{ c \}$ are not independent due to the identity~\eqref{eqn:PaPa}. Thus, we do not vary them in the Newton search and need to recalculate them after shifting the independent parameters.

\subsubsection{General states (Type IV)}\label{sec:gengen}

As the main steps of the algorithm have been described on the basis of this case, we just highlight two points.
Complications for this case, with least symmetry are essentially a combination of subtleties which we have seen separately for left-right symmetric and general ({\it i.e.}~for type II states), and general and parity symmetric ({\it i.e.}~for type III states) cases.

As mentioned in section~\ref{LRgen}, now there are two independent emergent ambiguity in the definition of $\bQ_i$
\begin{align}
    \bQ_2 \rightarrow \bQ_2 + \eta\, \bQ_1\;, \qquad \bQ_3 \rightarrow \bQ_3 + \mu\, \bQ_4\;,
\end{align}
for two independent constants $\eta$ and $\mu$.
This implies an analogous ambiguity in $\Qai$:
\begin{align}
    Q_{a \mid i} \rightarrow Q_{a \mid i} + \eta\, \delta_{i\, 2}\, Q_{a \mid 1} + \mu\, \delta_{i\, 4}\, Q_{a \mid 3}
    \;.
\end{align}
Therefore, we have zero-modes at two levels now:
\begin{align}
    n_\eta \equiv {\tt powQ}_2 - {\tt powQ}_1
    = n_{\mathbf{b}_{2}}-n_{\mathbf{b}_{1}} + 1
    \;,\qquad
    n_\mu \equiv {\tt powQ}_4 - {\tt powQ}_3
    = n_{\mathbf{a}_{1}}-n_{\mathbf{a}_{2}} + 1
    \;.
\end{align}
There are two cases, $n_\eta = n_\mu$ and $n_\eta \neq n_\mu$.
Both of the right null vectors  $B^{\texttt{zero\;mode}}_{a|2,n_\eta}$ and $B^{\texttt{zero\;mode}}_{a|4,n_\mu}$ can be chosen to have a fourth component to be one, and this can be used to resolve the zero-modes at $n = n_\eta$ and $n = n_\mu$, analogously to how it was done in section~\ref{LRgen}. The left null vectors
induce constraints on the coefficients $c_{a,n_\eta}$ and $c_{a,n_\mu}$.

When $n_{\eta} = n_{\mu}$ we need to impose two constraints at the same order. Let us provide the simplest example of the constraint for the oscillator numbers ${}_{6}[0\ 0\ 4\ 3\ 3\ 2\ 0\ 0]_{{\tt sol}}$, with ${\tt sol} = 4,\dots,\,9$:\footnote{Our choice of independent parameters in the implementation is $c^{2, n_{\eta}}$ and $c^{3, n_\mu}$ when $n_{\eta}=n_{\mu}$,  $c^{2, n_{\eta}}$ and  $c^{2, n_\mu}$ when $n_{\eta}\neq n_{\mu}$.}, which has $n_\eta=n_\mu=1$. In this case we get
\begin{gather}
\la{constr1}
    c^{2, 1} = \frac{(\Delta+2)\Delta(\Delta+6)^2c_{1,1}+60(\Delta+2)(\Delta^2+4\Delta-20)c_{2,1} + 8(\Delta+2)(\Delta-4)^2c_{4,1}}{40(\Delta-4)\Delta^2(\Delta+6)}- \\ \nonumber \frac{5\Delta(\Delta+2)(\Delta-4)(\Delta+6)c_{3,1}-600(\Delta+2)(\Delta+6)^2c^{4,1}-2880(\Delta+2)c_{2,1}}{40(\Delta-4)\Delta^2(\Delta+6)},\\
    c^{3,1}=\frac{180\Delta(\Delta+6)^2c_{1,1}-30\Delta(\Delta^2 +4\Delta-20)c_{2,1}+5\Delta^2(\Delta-4)(\Delta+2)(\Delta+6)c_{3,1}}{90(\Delta-4)(\Delta+2)(\Delta+6)}-\\ \nonumber \frac{16\Delta(\Delta-4)c_{4,1}+900(\Delta+6)^2\Delta c^{4,1} + 1440 c_{2,1}}{90(\Delta-4)(\Delta+2)(\Delta+6)}.
\end{gather}

When $n_\eta \neq n_\mu$, we need to impose constraints on different orders. Here one of the simplest examples is the constraints for  ${}_{11/2}[0\ 0\ 4\ 3\ 1\ 1\ 1\ 0]_{\tt sol}$ multiplets, with ${\tt sol} = 1,2$. In this case $n_{\eta}=1$ and $n_{\mu}=2$. The first constraint which looks the following:
\begin{gather}\la{constr2}
    c^{2,1} = -\frac{6(2\Delta+13)(2\Delta+21)c_{1,1}}{4(\Delta-7)\Delta+33}+\frac{3(2\Delta+5)(2\Delta+13)c_{2,1}}{(3-2\Delta)^2}-\frac{8(2\Delta-7)(2\Delta+1)c_{3,1}}{5(2\Delta-3)(2\Delta+9)}\\ \nonumber
    \frac{11-2\Delta}{16}c_{4,1}-\frac{30(2\Delta+9)c^{3,1}}{2\Delta-3}+\frac{36(2\Delta+13)c^{4,1}}{2\Delta-3},
\end{gather}
and the second one is similar, but a more cumbersome expression for $c^{2,2}$. In general, the constraint for ${n_\mu}>1$ becomes nonlinear in $c_{a,n}$ and $c^{a,n}$, however, it is linear in $c^{a, n_\mu}$ terms. Finally, let us note that there are poles in $\Delta$ in the constraints~\eqref{constr1},\eqref{constr2}. However, for states we have studied, the poles satisfy $\Delta \leq \Delta_0$ and as $\Delta$ is expected to grow monotonically with the coupling, they should not cause a problem at finite real $g$.

\section{C++ Implementation}\la{sec:imp}
This section is devoted to give the most necessary technical details for using the C++ implementation of the QSC solver algorithm described in the previous section.
We give step by step instructions on the installation and usage.

The source-code is available at \GitHub alongside with the {\tt Mathematica} prototype and a database for the initial points and perturbative data.

\subsection{Fast QSC Solver: User Manual}
\subsubsection{Installation}

The numerical solution of the QSC equations requires higher than machine precision arithmetic. This implies, that the usual double arithmetic
of C or C++ is not enough for our purposes. This is why, we used the Class Library of Numbers (CLN), which is a
 special library for efficient computations with all kinds of numbers in arbitrary precision. All information about this free library can be found
 in the library's homepage: \url{https://www.ginac.de/CLN}. Nevertheless, for practical use of our C++ codes, knowing all subtle details is not necessary.
 The CLN package is available at Unix-like operating systems\footnote{E.g. Linux and MAC.}, but can also be used under Windows operating system, provided a  Windows Subsystem for Linux (WSL) is installed\footnote{\url{https://learn.microsoft.com/en-us/windows/wsl/install}}.
 Then all linux programs can be executed from the Windows PowerShell. The only difference with respect to usual linux is that each command should be
 anticipated by typing \texttt{wsl} and a space to indicate that the WSL system is used.

 The prerequisites to run the C++ codes is that, a C++ compiler and the CLN package should be installed on the computer.
 This requires the installation of three packages either using the package manager or from the terminal window. Namely,
these are the \texttt{g++} compiler\footnote{Certainly, depending on taste other popular compilers (e.g. \texttt{clang}) can be used, as well.},
 and the \texttt{libcln} and \texttt{libcln-dev} packages for the usage of arbitrary precision
 arithmetic\footnote{The actual package names may change depending on the Linux distribution. For example in  OpenSuse the \texttt{cln} and \texttt{cln-devel} packages should be installed.}.
 In Debian linux systems, they can be installed easily from terminal window by the following commands:
 \newline
 \texttt{> sudo apt install g++ }
 \newline
 \texttt{> sudo apt install libcln6 }
 \newline
 \texttt{> sudo apt install libcln-dev}\footnote{Just to show an example for windows users, the equivalent command from PowerShell using the WSL, is \texttt{> wsl sudo apt install libcln-dev}. }
 \newline
Then if the C++ source file can be found in the current directory, the compilation can be done easily from terminal by typing:
\newline
\texttt{> g++ source.cpp -lm -lcln -o executable.out}
\newline
This command creates an executable output file named \emph{executable.out} in the current directory out of the \emph{source.cpp} C++ source file.

There are four C++ codes and four notebook files can be found in the \GitHub repository under the names:
\newline
\texttt{TypeI\_core.cpp}, and  \texttt{TypeI\_example.nb},
\newline
\texttt{TypeII\_core.cpp},   and  \texttt{TypeII\_example.nb},
\newline
\texttt{TypeIII\_core.cpp}, and \texttt{TypeIII\_example.nb},
\newline
\texttt{TypeIV\_core.cpp}, and \texttt{TypeIV\_example.nb}.
\newline
They correspond to the four special cases, namely left-right and parity symmetric, left-right symmetric and general parity,
general and parity symmetric, and general states respectively, following the classification given in section~\ref{sec:LR}.
The \texttt{.cpp} codes are the C++ source codes and the \texttt{.nb} files serve as examples of how to run the C++ codes correctly with examples of an input, details of which we describe in the next section.
The \texttt{.nb} files will work only if the names of executables in the four special cases are as follows:
\newline
\texttt{TypeI\_exec.out}, \texttt{TypeII\_exec.out}, \texttt{TypeIII\_exec.out} and
\newline
 \texttt{TypeIV\_exec.out}.
Thus compilation should be done such that the output files should have these names. For example:
\newline
 \texttt{> g++ TypeI\_core.cpp -lm -lcln -o TypeI\_exec.out}
etc.

The source codes should be compiled only once.

\subsubsection{Parameters and outputs}
The parameters of an individual run are passed as arguments to the executable code.
Here under the word parameters we mean all the data, which is necessary for a proper execution. For example: cutoff values, precision of internal computations,
number of sampling points, quantum numbers of the states, initial values of $\Delta$ and $c_{a,n}$ etc.
The simplified syntax for an execution from a terminal window is as follows:
\newline
\texttt{> executable.out } followed by numbers separated by space characters, such that each number corresponds to an argument of the
executable code. The meaning of each argument can be read off from the example notebook files, where the string corresponding to the  appropriate command for the execution
is created and the C++ code is run from \texttt{Mathematica} as an external program.

\paragraph{Parameters.}
Here, we give the detailed list of arguments below, such that the numbers in front of the description of a given argument, indicates its position in the list of arguments.

\begin{enumerate}

\item {\tt WorkingPrecision}: is an integer for specifying the precision of internal computations.

\item Integer cutoff: $\texttt{cutP}$ or $\texttt{cutP}/2$ in the general parity and parity symmetric cases respectively.

\item Integer cutoff: $\texttt{cutQai}$ or $\texttt{cutQai}/2$ in the general parity and parity symetric cases respectively.

\item Integer cutoff: $\texttt{QaiShift}$.

\item An integer denoting the number of Chebyshev sampling points on the cut: \texttt{nPoints}.

\item {\tt DH}: is an integer for giving the small shift $\epsilon=10^{-\texttt{DH}}$ for
computing the numerical derivatives by a first order formula.

\item {\tt precSSF}: is an integer for specifying one of the quit conditions, as described below.

\item {\tt precDelta}: is another integer for giving the 2nd condition for quitting the iteration cycle, as described below.

\item\la{in:maxiter} {\tt maxiter}: is an integer giving the maximal number of iterations.

In terms of the previous three integers, one can give the condition at the fulfilment of which the
program quits the iteration process and returns. This is as follows:

The program returns after the $n^\text{th}$ iteration,
if either ${n \geq \tt maxiter}$ or if the following two conditions are simultaneously satisfied:
\begin{equation} \label{}
\begin{split}
\sum\limits_{a,n} \bigg\{ | c_{a,n}-c_{a,n}^{\texttt{Updated}}|^2
+| c^{a,n}-c^{a,n, \texttt{Updated}}|^2 \bigg\}&<10^{-{\tt precSSF}}, \nonumber \\
|\Delta^{(n)}-\Delta^{(n-1)}|&<10^{-{\tt precDelta}}, \nonumber
\end{split}
\end{equation}
where $\Delta^{(n)}$ denotes the value of $\Delta$ after the $n$th iteration.

\item--\; 17. Integer oscillator quantum numbers: $n_{{\bf b}_1}, n_{{\bf b}_2}, n_{{\bf f}_1}, n_{{\bf f}_2}, n_{{\bf f}_3}, n_{{\bf f}_4}, n_{{\bf a}_1}, n_{{\bf a}_2},$ \setcounter{enumi}{17}

\item An integer for giving the multiplicity label {\tt sol} of the state,

\item A real number for the 't Hooft coupling $g.$

\item A real number for the initial value of the \emph{anomalous} dimension $\gamma=\Delta-\Delta_0.$

Next come the initial values for $c_{a,n}$ and $c^{a,n}$ as real numbers:

\item $\text{Im}(c_{1,0}) \quad \text{Im}(c_{1,1})\quad  \dots \quad \text{Im}(c_{1,N^{(in)}})$

 $\text{Re}(c_{2,0}) \quad \text{Re}(c_{2,1}) \quad \dots \quad \text{Re}(c_{2,N^{(in)}})$

 $\text{Im}(c_{3,0}) \quad \text{Im}(c_{3,1})\quad  \dots \quad \text{Im}(c_{3,N^{(in)}})$

 $\text{Re}(c_{4,0}) \quad \text{Re}(c_{4,1}) \quad \dots \quad \text{Re}(c_{4,N^{(in)}})$

 $\text{Re}(c^{1,0}) \quad \text{Re}(c^{1,1})\quad  \dots \quad \text{Re}(c^{1,N^{(in)}})$

 $\text{Im}(c^{2,0}) \quad \text{Im}(c^{2,1}) \quad \dots \quad \text{Im}(c^{2,N^{(in)}})$

 $\text{Re}(c^{3,0}) \quad \text{Re}(c^{3,1})\quad  \dots \quad \text{Re}(c^{3,N^{(in)}})$

 $\text{Im}(c^{4,0}) \quad \text{Im}(c^{4,1}) \quad \dots \quad \text{Im}(c^{4,N^{(in)}}).$

\end{enumerate}
A few important comments are in order. In the parity symmetric cases, only the $c_{a,2n}$ and $c^{a,2n}$ coefficients must be given in the
list of arguments. In the left-right symmetric case, the initial values of $c^{a,n}$ must not be put in the list of arguments.
All arguments must be positive integers or real floating point numbers. In the list of arguments an integer valued argument must
be given as an exact integer. E.g. 5 and not 5.0.
On the other hand giving the real floating point numbers might be also cumbersome.
The code cannot interpret exact rational numbers. Thus e.g. $1/10$ as a specification for $g$ is uninterpretable for the code.
Instead the value $1/10$ should be given as $0.1000000....0000000000000000$ with {\tt WorkingPrecision} $- 1$
zeroes followed by the digit $1.$ The initial value $0.1$ for $g$ would be interpreted as a machine precision number, but
since in the code {\tt WorkingPrecision} digit arithmetic is set, the code generates a {\tt WorkingPrecision} precision number for
$g$ with a value deviating from $1/10$ in the order of $10^{-16}$. At that random generated value of $g,$ the code will serve
with precise {\tt WorkingPrecision} precision computations and results. Though, this number representation seems to be a bit cumbersome
it can be simply and safely treated e.g. in \texttt{Mathematica}. Because of such issues, the output of the C++ code also writes out the
parameters of the actual calculation.

It is also very important to note, that the C++ implementation uses
a bit different from \eqref{Prepcut} and \eqref{Prepcut1}
conventions for the ${\bf P}$-functions.
Both in the input and  the output,
the coefficients $c_{a,n}$ and $c^{a,n}$ should be meant as coefficients
of the series representation as follows:

\begin{align}\label{eqn:PinZhukC++}
    {\bf P}_a = \, x^{\mathtt{powP}_a}
    \sum_{n = 0}^\infty
    \frac{c_{a,n}}{x^n}
    \;,
     \qquad
     {\bf P}^a =
     \, x^{-\mathtt{powP}_a-1}
    \sum_{n = 0}^\infty
    \frac{c^{a,n}}{x^n}
     \;.
\end{align}
Thus, the C++ code requires in the list of arguments, and in the output gives, the coefficients in the convention defined by \eqref{eqn:PinZhukC++}.
The relation between the coefficients of the two conventions can be given by  simple formulas:
\begin{equation}
c_{a,n}^{(\text{C++})}=  g^{\,\mathtt{powP}_a}    \, c_{a,n}, \qquad c^{a,n,(\text{C++})}=  g^{-\mathtt{powP}_a-1}   \, c^{a,n},
\end{equation}
where
$c$ and $c^{(\text{C++})}$ stands for the coefficients of \eqref{Prepcut}-\eqref{Prepcut1}
and (\ref{eqn:PinZhukC++}), respectively.
When listing the arguments, and the output of the C++ codes, for short we neglected to put the C++ label on the coefficients.
\paragraph{Output.}
The output contains numbers in a form immediately interpretable by \texttt{Mathematica}. If somebody would like to get the numbers in a
different form, than it can be reached after appropriate string manipulations on the output string.
The structure and content of the output is explained in the example notebook files, but for completeness we summarise them here, too.

The output is a huge \texttt{Mathematica} compatible table of numbers and vectors. Now, we describe the meaning of each  element of this table
one after the other.{\footnote{In the sequel, the number in front of a table element corresponds to its position
 within the table.}}

\begin{enumerate}

\item The $1^\text{st}$ element is a constant, which can take values $0$ or $1.$ If it is equal to $0,$ then the program reached the
desired convergence within less then ${\tt maxiter}$ iterations and the final result is within the desired precision given by the
{\tt precSSF} and {\tt precDelta} input parameters. Otherwise, it takes the value $1,$ meaning that the program failed to converge in {\tt maxiter} number of iterations to the desired precision.

\item The value of $g$.

\item Oscillator numbers and multiplicity label of the state in a vector of the form: \newline
$\{\{ n_{{\bf b}_1},n_{{\bf b}_2}, \{ n_{{\bf f}_1},n_{{\bf f}_2},n_{{\bf f}_3},n_{{\bf f}_4}  \},n_{{\bf a}_1},n_{{\bf a}_2} \}, \texttt{sol} \}.$

\item The anomalous dimension of the state: $\gamma=\Delta-\Delta_0.$

\item A vector for $\{ c_{1,n} \}\big|_{n=0,...,\texttt{cutP}}$ in the general parity case or
$\{ c_{1,2n} \}\big|_{n=0,...,\texttt{cutP}/2}$ in the parity symmetric case.

\item A vector for $\{ c_{2,n} \}\big|_{n=0,...,\texttt{cutP}}$ in the general parity case or
$\{ c_{2,2n} \}\big|_{n=0,...,\texttt{cutP}/2}$ in the parity symmetric case.

\item A vector for $\{ c_{3,n} \}\big|_{n=0,...,\texttt{cutP}}$ in the general parity case or
$\{ c_{3,2n} \}\big|_{n=0,...,\texttt{cutP}/2}$ in the parity symmetric case.

\item A vector for $\{ c_{4,n} \}\big|_{n=0,...,\texttt{cutP}}$ in the general parity case or
$\{ c_{4,2n} \}\big|_{n=0,...,\texttt{cutP}/2}$ in the parity symmetric case.

The next four elements are present only in the non-LR symmetric cases:

\item A vector for $\{ c^{1,n} \}\big|_{n=0,...,\texttt{cutP}}$ in the general parity case or
$\{ c^{1,2n} \}\big|_{n=0,...,\texttt{cutP}/2}$ in the parity symmetric case.

\item A vector for $\{ c^{2,n} \}\big|_{n=0,...,\texttt{cutP}}$ in the general parity case or
$\{ c^{2,2n} \}\big|_{n=0,...,\texttt{cutP}/2}$ in the parity symmetric case.

\item A vector for $\{ c^{3,n} \}\big|_{n=0,...,\texttt{cutP}}$ in the general parity case or
$\{ c^{3,2n} \}\big|_{n=0,...,\texttt{cutP}/2}$ in the parity symmetric case.

\item A vector for $\{ c^{4,n} \}\big|_{n=0,...,\texttt{cutP}}$ in the general parity case or
$\{ c^{4,2n} \}\big|_{n=0,...,\texttt{cutP}/2}$ in the parity symmetric case.

\item {(LR 9.)}\footnote{Just to avoid confusion, with this notation we would like to indicate that this
vector is the $9^\text{th}$ element of the output table in the left-right symmetric case, since in that case the previous
upper case $c$-vectors are not listed in the output.} The next element is a vector
containing the parameters and some error estimates of the
actual calculation. The meanings of its elements in order are as follows:
\begin{enumerate}
\item Number of iterations done

\item The value of the sum:
\begin{equation} \label{}
\begin{split}
\sum\limits_{a,n} \bigg\{ | c_{a,n}-c_{a,n}^{\texttt{Updated}}|^2
+| c^{a,n}-c^{a,n, \texttt{Updated}}|^2 \bigg\} \nonumber
\end{split}
\end{equation}
at the end of the last iteration.

\item The absolute value of the difference of the values of $\Delta$ obtained after the last and the
penultimate iterations:
\begin{equation} \label{}
\begin{split}
 |\Delta^{(\text{last})}-\Delta^{(\text{last}-1)}| \nonumber
\end{split}
\end{equation}

\item The value used for $\texttt{cutP}$ or $\texttt{cutP}/2$ in the general parity and parity symmetric cases respectively.

\item The value used for $\texttt{cutQai}$ or $\texttt{cutQai}/2$ in the general parity and parity symmetric cases respectively.

\item The value used for $\texttt{QaiShift}$.

\item The number of Chebyshev sampling points \texttt{nPoints} used.

\item The value used for {\tt DH} in the actual calculation. I.e. $\epsilon=10^{-{\tt DH}}$ was the small shift parameter for computing the numerical derivatives by a first order formula.

\item The actual value of {\tt WorkingPrecision} used.

\item The actual value of {\tt precDelta} used.

\item The actual value of {\tt precSSF} set.

\item The value of {\tt maxiter} for specifying the  maximal number of iterations.

\item Another number measuring the ``goodness'' of the actual calculation:
$\sum\limits_a \bigg\{ \big| c^{{\tt Updated}}_{a,-1}\big|^2 +\big| c^{a\;{\tt Updated}}_{-1}\big|^2 \bigg\}$ in the general parity or
$\sum\limits_a \bigg\{ \big| c^{\tt Updated}_{a,-2}\big|^2 +\big| c^{a\;{\tt Updated}}_{-2}\big|^2 \bigg\}$ in the even
cases{\footnote{In the left-right symmetric cases the superscripted terms are not involved.}}. This should be as small as possible on the actual solution of QSC and is a good indicator of the overall precision.

\end{enumerate}

\item\label{out:GC} {(LR 10.)} A number measuring the ``goodness'' of the cutoffs:

\begin{equation} \label{}
\begin{split}
\sqrt{\sum\limits_{i,n} \bigg\{ \bigg|\mathbf{Q}_i(u_n)\big|_{{}_\texttt{QaiShift}}\!\!\!\!-\!\mathbf{Q}_i(u_n)\big|_{{}_{\texttt{QaiShift}-1}}  \bigg|^2
\! \! + \!
\bigg|\mathbf{Q}_i(u_n)\big|_{{}_\texttt{QaiShift}}\!\!\!\!-\!\mathbf{Q}_i(u_n)\big|_{{}_{\texttt{QaiShift}-1}}  \bigg|^2  \bigg\}}, \nonumber
\end{split}
\end{equation}
where $\mathbf{Q}_i(u_n)\big|_{{}_{\texttt{QaiShift}}}$ denotes the numerical value of $\mathbf{Q}_i(u_n)$ computed with cutoff value \texttt{QaiShift}.
This indicates how precise the solution of the QQ-system is within the given approximation.

\item\la{out:gluing} {(LR 11.)} A quantity, which is proportional to the $L^2$-norm of the deviations of the $\mathbf{Q}/\bar{\tilde{\mathbf{Q}}}$ ratios from the gluing matrix elements:

$\frac{1}{4\ \texttt{nPoints}}\sqrt{  \sum\limits_{n} \bigg \{
 \bigg|\frac{\mathbf{Q}_1(u_n)}{\alpha_{1} \, \bar{\tilde{\mathbf{Q}}}^2(u_n) }-1\bigg|^2 \! \! + \!
 \bigg|\frac{\mathbf{Q}_2(u_n)}{\alpha_{2} \, \bar{\tilde{\mathbf{Q}}}^1(u_n) }-1 \bigg|^2 \! \! + \!
 \bigg|\frac{\mathbf{Q}_3(u_n)}{\beta_{1} \, \bar{\tilde{\mathbf{Q}}}^4(u_n) }-1\bigg|^2  \! \! + \!
 \bigg|\frac{\mathbf{Q}_4(u_n)}{\beta_{2} \, \bar{\tilde{\mathbf{Q}}}^3(u_n) }-1\bigg|^2
\bigg\}  }$

This number is the indicator of the overall precision of the numerical solution, which is sensitive to both precision of the QQ-system and also measures the precision of the gluing condition fulfilment.
\end{enumerate}

\subsection{Management of the parameters, precision control and extrapolation}

In this section we discuss our strategies for adjusting various parameters controlling precision of the solver and our adaptive strategy for selecting the step in the coupling $g$.

For managing and adjusting these parameters in the real time we provide a simple Python script, which is also available on \GitHub.

\subsubsection{Adjusting precision controlling parameters}
The are quite a few parameters we have to adjust to ensure optimal performance within certain precision. Let us remind the main parameters:
\begin{itemize}
    \item \texttt{cutP} - controls the number of parameters. Bigger \texttt{cutP} gives better approximation of $\bP$-functions but also slows the performance as $\sim \texttt{cutP}^3$.
    \item \texttt{QaiShift} - controls the number of jumps from the real axis to the domain where we use the asymptotic expansion. Increasing this parameter increases the precision of the $\bQ$-functions computed for given $\bP$-functions.
    \item \texttt{cutQai} - controls the number of terms in the asymptotic expansion of $Q_{a|i}$ to use. It also increases the precision of the $\bQ$-functions for given $\bP$-functions but due to the asymptotic nature of the large $u$ expansion increase in this parameter has to be balanced with an increase in  \texttt{QaiShift}.
    \item $\texttt{nPoints}$ - number of the probe points to use. As roughly each probe points gives $8$ gluing equations we can simply fix it to $\texttt{cutP}+2$ for the general Type IV states,
    for Type I and III: $\texttt{cutP}+4$ and $\texttt{cutP}+8$ for the state of the Type II.
\end{itemize}

\paragraph{Step A. Optimizing \texttt{QaiShift} and \texttt{cutQai}.}
One has to ensure that at the given starting points for the parameters $\{c\}$ and $\Delta$ the QQ-relations are solved to a sufficiently good precision. To test this we rely on the output \ref{out:GC} with the goal to keep it below our targeted precision. If the output \ref{out:GC} is not small enough we have to increase either
\texttt{QaiShift} or \texttt{cutQai}. In practice, to decide which of the parameters needs to be increased we run the C++ code with the input parameter \ref{in:maxiter} ${\tt maxiter}=0$, which then make the solver to stop before performing any Newton method iterations, and returns the
output \ref{out:GC} very quickly. Then we run it multiple times, changing \texttt{QaiShift} and \texttt{cutQai} separately to check which parameter has the biggest impact on the precision and then adjusting them to reach the desired precision of the output~\ref{out:GC}.

\paragraph{Step B. Optimizing \texttt{cutP}.} This parameter needs to be changed when we are confident that the Q-system is solved to a sufficiently high precision (Step A). Then the indicator that \texttt{cutP} needs an increment is 1) Newton method converges with a good enough precision, but at the same time 2) the output \ref{out:gluing}, measuring the ``goodness'' of the gluing condition is too large. In this situation we increase \texttt{cutP} by $1$ or by $2$, depending on the type of the state, and run again.

\subsubsection{Extrapolation and  step in $g$}
The convergence of the QSC Solver relies a lot on good set of initial points for $c$'s and $\Delta$. At small $g$ we use the perturbative QSC Solver of \cite{Marboe:2017dmb, Marboe:2018ugv}.  The main advantage of the current algorithm is that it is very stable at very small values of $g$ such as $10^{-5}$, where the precision of the perturbative solution (which we provide for up to $g^{10}$ order) is excellent and provides good starting points for our numerical QSC Solver. This way we managed to initialise all $219$ states.
Then having sufficiently many numerical points (say $n \times 10^{-5}$, $n=1\dots 10$) one can use polynomial extrapolation to create starting points for the next value of $g$. Here we explain the strategy we used to manage the step in $g$.

We start from the step $dg=1/10/2^n$ where typically $n=7$. Then if in $3$ consecutive runs for different $g$'s the convergence was achieved with less than $4$ iterations of the Newton method we increase $dg$ by factor of $2$ (making sure $dg$ is below $1/10$).
If the convergence failed within the limit of the iterations specified, this likely means that the starting point are too far from the true values, in this case we decrease the step by to  $dg\to dg/2$ and step back, close to the previously computed point.
Finally, when increasing $g$ we make sure that the solver runs through all values of $g$ for which $10 g$ is an integer.

\subsection{Examples of Parameters and Benchmarking}\la{sec:benchmarking}
In this section we give the details on how fast our QSC solver is for various regimes in $g$ and different types of states. We also compare the timings for the C++ code with our {\it Mathematica} implementation for a Type I state. We were using Mac Book Pro with 2.3 GHz Quad-Core Intel Core i7  and {\it Mathematica} version 12.3.1.0 for the benchmarking.
We also give typical parameters and the corresponding precision estimate for the result for $\Delta$ for various types of states.
\begin{table}[h]
\centering
\begin{tabular}{c|c|c|c|c|c|c|c|c|c}
     $g$ & $\Delta$ & \tt cutP & \tt cutQai & \tt QaiShift & Iterations & Error & $t_{C++}$ & $t_{M}^{\text{4 cores}}$& $t_{M}^{\text{1 core}}$ \\
    \hline\hline
     1/10 & 2.11551 & 22  & 24 & 60 & 1 & $10^{-37}$ & 00:37 & 00:55 & 02:37\\
     1/5 & 2.41886 & 22  & 36 & 60 & 1 & $10^{-25}$ & 00:52 & 02:15 & 06:43 \\
     1/2 & 3.71272 & 36  & 36 & 60 & 2 & $10^{-23}$ & 02:56 & 08:28 & 23:14\\
     1 & 5.60407 &  50 & 36 & 60 & 2 & $10^{-21}$ & 04:50 & 13:47 & 37:46\\
     2 & 8.40482 & 72 & 44 & 60 & 2 & $10^{-21}$ & 11:58 & 37:31\\
     5 & 14.0998  & 112 & 64 & 70 & 2 & $10^{-21} $ & 54:36\\
\end{tabular}
\caption{\small\it    Parameters we used for various states, and couplings, and comparison between C++ and {\it Mathematica} implementations for {\tt St.} {\tt No.} {\tt 1} (Type {I}).}
\label{tab:benchmarkingStNo1}
\end{table}
\begin{table}[h]
\centering
\begin{tabular}{c|c|c|c|c|c|c|c}
     $g$ & $\Delta$ & \tt cutP & \tt cutQai & \tt QaiShift & Iterations & Error & $t_{C++}$  \\
    \hline\hline
     1/10 & 6.04812 & 24  & 40 & 50 & 2 & $10^{-17}$ & 01:59 \\
     1/5 & 6.19283 & 26  & 40 & 50 & 2 & $10^{-15}$ & 02:05 \\
     1/2 & 7.19999 & 38  & 40 & 50 & 2 & $10^{-18}$ & 03:23 \\
     1 & 10.1817 & 50  & 40 & 50 & 2 & $10^{-17}$ & 04:39 \\
     2 & 15.37 & 78  & 48 & 60 & 2 & $10^{-21}$ & 14:55 \\
    5 & 25.5 & 106  & 80 & 70 & 2 & $10^{-14}$ & 01:17:18 \\
\end{tabular}
\caption{\small\it    Parameters we used for various states, and couplings, and timings of the C++ implementation for {\tt St.} {\tt No.} {\tt 78} (Type I).}
\label{tab:benchmarkingStNo78}
\end{table}
\begin{table}[h]
\centering
\begin{tabular}{c|c|c|c|c|c|c|c}
     $g$ & $\Delta$ & \tt cutP & \tt cutQai & \tt QaiShift & Iterations & Error & $t_{C++}$  \\
    \hline\hline
     1/10 & 4.14474  & 17 & 26 & 100 & 1 & $10^{-25}$ & 01:40 \\
     1/5 & 4.52873 & 23 & 26 & 100 & 1 & $10^{-27}$ & 02:47 \\
     1/2 & 6.2573 & 33 & 26 & 100 & 1 & $10^{-24}$ & 04:05  \\
     1 & 8.92989 & 45 & 30 & 100 & 2 & $10^{-22}$ & 20:12 \\
\end{tabular}
\caption{\small\it    Parameters we used for various states, and couplings, and timings of the C++ implementation for {\tt St.} {\tt No.} {\tt 9} (Type II).}
\label{tab:benchmarkingStNo9}
\end{table}

\begin{table}[h]
\centering
\begin{tabular}{c|c|c|c|c|c|c|c}
     $g$ & $\Delta$ & \tt cutP & \tt cutQai & \tt QaiShift & Iterations & Error & $t_{C++}$  \\
    \hline\hline
     1/10 & 5.06741 & 22 & 30 & 100 & 1 & $10^{-33}$ & 00:56 \\
     2/10 & 5.26376 & 24 & 54 & 190 & 1 & $10^{-25}$ & 03:40 \\
     1/5 & 6.41518 & 34 & 54 & 190 & 2 & $10^{-21}$ & 12:29 \\
     1 & 8.82232 & 46 & 54 & 190 & 2 & $10^{-18}$ & 24:22 \\
     2 & 12.7445 & 76 & 54 & 190 & 2 & $10^{-21}$ & 01:03:42 \\
\end{tabular}
\caption{\small\it    Parameters we used for various states, and couplings, and timings of the C++ implementation for {\tt St.} {\tt No.} {\tt 13} (Type III).}
\label{tab:benchmarkingStTypeIII}
\end{table}

For {\tt St.} {\tt No.} {\tt 1} (Type I), which is in the Konishi multiplet we made a comparison between C++ and {\it Mathematica} performance in table~\ref{tab:benchmarkingStNo1}. We see that when compared on single core C++ is about $8$-times faster. Introducing $4$ cores into {\it Mathematica} computation only gives the factor of $\sim 2.5$ in efficiency of the {\it Mathematica} code, due to the overheads in the parallelisation procedure for this range of parameters. At the same time having the code running on one core allows computing several states simultaneously which gives proper multiplicative efficiency gain. Our computational code is open-source and readily available for usage, eliminating the necessity for costly licensing agreements associated with proprietary software such as {\it Mathematica}. We respectfully request that users of this code attribute its original source by citing the present paper.

States with bigger $\Delta_0$ usually require higher values of parameters. For comparison we give the parameters and timings for another, more challenging, {\tt St.} {\tt No.} 78 (Type I) in table~\ref{tab:benchmarkingStNo78}.

Finally in table~\ref{tab:benchmarkingStNo9} and table~\ref{tab:benchmarkingStTypeIII} the data for less symmetric states {\tt St.} {\tt No.} 9 (Type II) and
 {\tt St.} {\tt No.} 13 (Type III) is given. Let us emphasise once again that the C++ code require only one CPU core, so in practice we were running $\sim 20$ states simultaneously on an HPC.

\section{Data Analysis at Strong Coupling}\la{sec:data}

In this section we present analysis of the numerical data obtained by the Fast QSC Solver for the first 219 states/operators -- all the single trace operators with the bare dimension $\Delta_0\leq6$.
Whereas at weak coupling, the perturbative expansion can be obtained analytico-numerically by using the Perturbative QSC solver~\cite{Marboe:2018ugv}, at the moment, the strong coupling behaviour of these states can only be obtained by fitting the high precision numerical data. The main content of this section thus dedicated to strong coupling analysis.

In section~\ref{sec:strongexplain}, we explain the fitting procedure which we use to obtain strong coupling predictions, and compare the predictions to those in the literature.
In section~\ref{sec:KKtowers}, we analyse our database of states, to see the match to counting formulas available in the literature, and furthermore propose some predictions for strong coupling behaviour of a subset of the states we studied numerically, but also giving prediction for infinitely many states with large bare dimension.
Finally, in section~\ref{sec:bootstrability}, we see how our spectral data can be used to initiate the bootstrability program for local operators of ${\cal N} = 4$ SYM at strong coupling.

\begin{figure}
     \centering
     \begin{subfigure}[t]{0.49\textwidth}
         \centering
         \includegraphics[width=\textwidth]{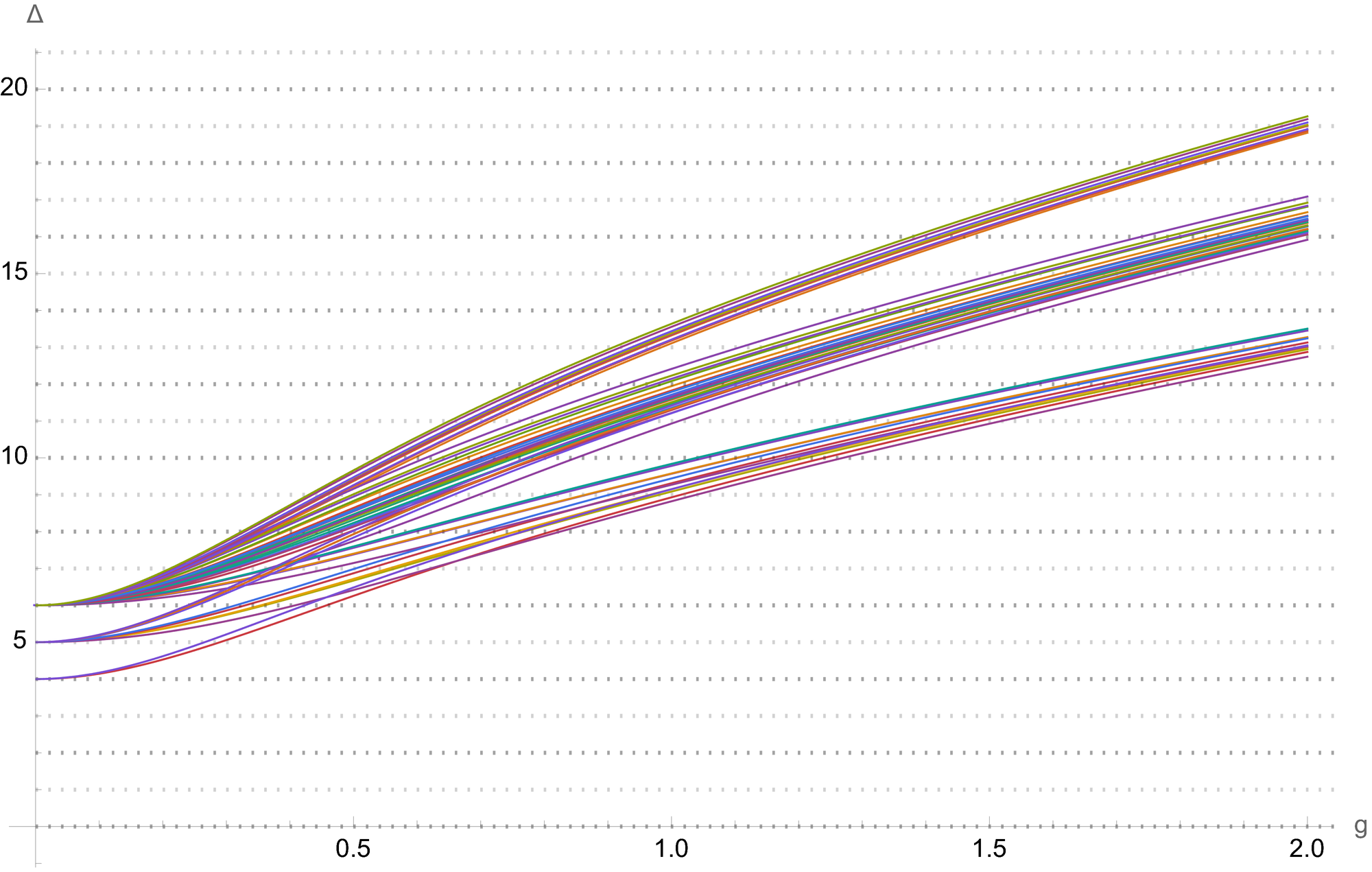}
         \caption{\small\it    The scaling dimension of 92 states in ${\cal N} = 4$ SYM, with bare dimension $\Delta_0\leq 6$, that
         are of type II~({\it cf.} section~\ref{LRgen}) and type III~(
         {\it cf.} section~\ref{sec:geneven}
         ).
         We provide their scaling dimensions in the range $g\in [0,2] = \lambda \in [0,64\,\pi^2]\sim [0,630]$.
         }
    \label{fig:LRGenGenEvenStates}
     \end{subfigure}
     \hfill
     \begin{subfigure}[t]{0.49\textwidth}
         \centering
         \includegraphics[width=\textwidth]{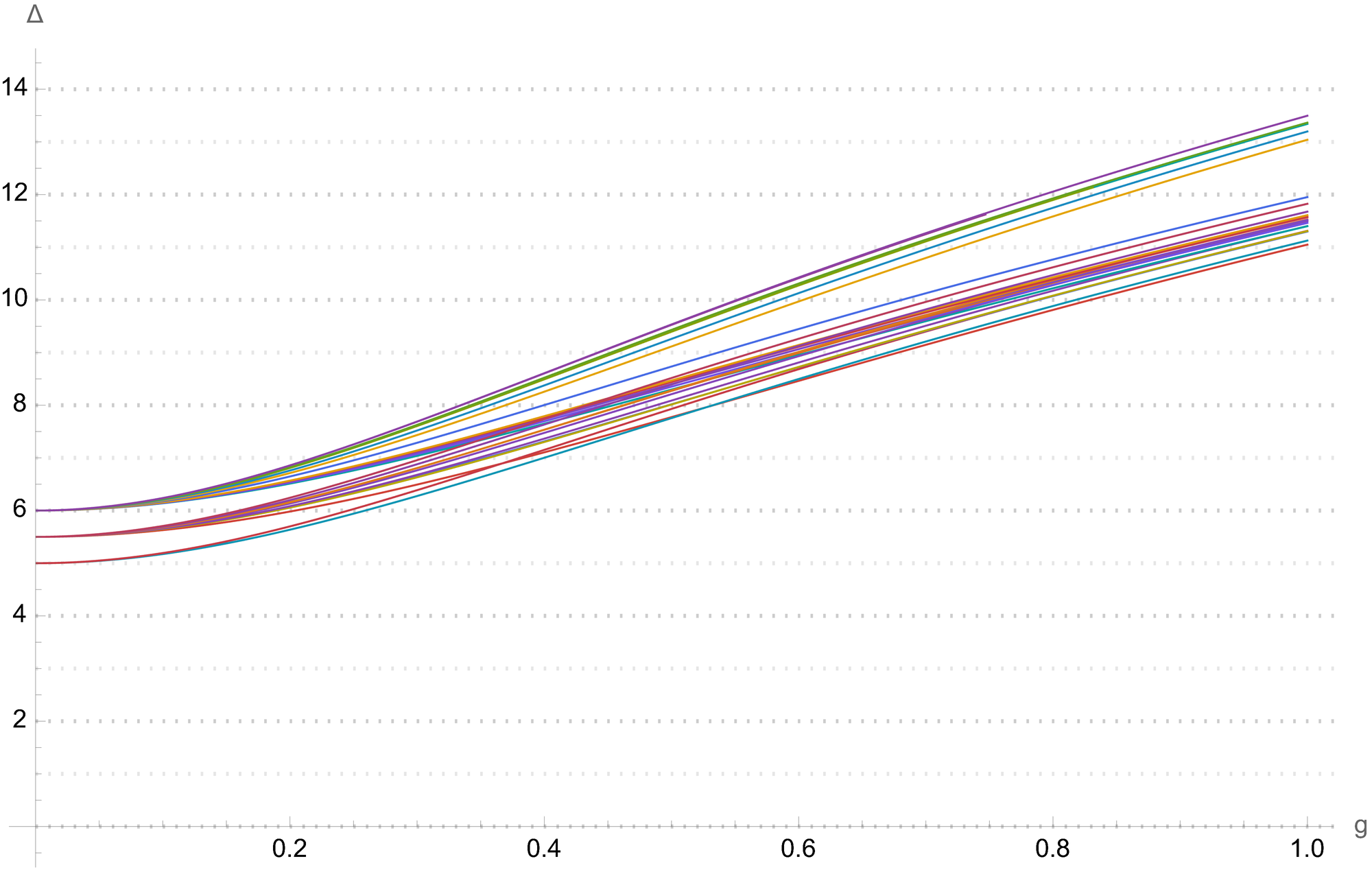}
         \caption{\small\it    Here we have the of 82 states in ${\cal N} = 4$ SYM, with bare dimension $\Delta_0\leq 6$, that have no additional symmetry ({\it cf.}~type IV states, discussed in section~\ref{sec:gengen}). Their scaling dimensions are given in the range $g\in [0,1] = \lambda \in [0,16\,\pi^2]\sim [0,160]$.}
    \label{fig:GenGenStates}
     \end{subfigure}
\end{figure}

\subsection{Strong coupling analysis of the numerical data}
\label{sec:strongexplain}

A convenient way to arrange the states at strong coupling was proposed in~\cite{Gubser:1998bc}.
It was argued there that in the regime where the quantum numbers $[\ell_1\;\ell_2\;q_1\;p\;q_2] \sim 1$,
the states in ${\cal N} = 4$ SYM, should map to massive string states on AdS${}_5\times S^5$.
In particular, in this regime, the spectrum should
organise itself into ``string mass levels'', parameterised by a positive integer $\delta$, as
\begin{align}\label{eqn:StrongLeading}
    \Delta \simeq 2\,\sqrt{\delta}\, \lambda^{1/4} \;.
\end{align}
Physically, in this regime the strings carry a lot of energy so that their wave-function is localised and can only explore a small part of the curved target space. Even though the 't Hooft coupling is large, this is a quantum regime, as opposed to classical strings, for which the quantum numbers scale as $ [\ell_1\;\ell_2\;q_1\;p\;q_2] \sim \sqrt\lambda$.

From \eq{eqn:StrongLeading} we see that the natural expansion parameter at strong coupling is the combination $\delta^2\,\lambda$, which indeed simplifies the structure of the expansion coefficients as we show below.
Even though there is no first principle derivation based on the string theory which would allow for a systematic large $\lambda$ expansion, it is expected that the structure of the asymptotic strong coupling expansion should be of the following form~\cite{Gromov:2009zb,Roiban:2009aa,Tseytlin:2009fw,Frolov:2010wt,Passerini:2010xc,Gromov:2011de, Roiban:2011fe, Vallilo:2011fj}:
\begin{align}
    \label{eqn:reg+const}
        \Delta &= \Delta_\mathtt{reg} + \Delta_\mathtt{const}\;,\\
        \Delta_\mathtt{reg} &=  ({\delta\,\sqrt\lambda})^{1/2}\left(2
        + \sum_{n = 1}^\infty
        \frac{d_n}{(\delta\,\sqrt{\lambda})^{n}}
        \right)\;,
        \label{eqn:regdef}
\end{align}
where $\Delta_\mathtt{const}$ and $ d_n$ are undetermined coefficients.
The subscript {\tt reg} is used to remind that the expansion of ${\Delta_{\tt reg}}/{\lambda^{1/4}}$ has an (asymptotic) expansion in integer powers of $1/\sqrt{\lambda}$.

\subsubsection{Fixing $\Delta_\mathtt{const}$}
In the absence of the general first principle derivation, in order to fix the constant term in the strong coupling expansion \eq{eqn:regdef} we present a heuristic argument.
Consider a state in ${\cal N} = 4$ SYM, with dimension $\Delta$, and Dynkin labels $\left[ \ell_1\;\ell_2\;q_1\;p\;q_2 \right] $. The quadratic Casimir of this representation is given by~\cite{Beisert:2004ry}
\begin{multline}\label{eqn:Casimir}
    J^2 =
    \frac{1}{2} (\Delta + 2)^2 - 2
    + \frac{1}{4} \ell_1\left(\ell_1+2\right)+\frac{1}{4} \ell_2\left(\ell_2+2\right)
    \\
    -\frac{1}{4} q_1\left(q_1+2\right)-\frac{1}{4} q_2\left(q_2+2\right)-\frac{1}{8}\left(2 p+q_1+q_2\right)^2-\left(2 p+q_1+q_2\right)\;.
\end{multline}
It is reasonable to assume that the strong coupling expansion of this operator is regular in powers of $1/\sqrt{\lambda}\;$. If we substitute the form of the strong coupling expansion of $\Delta$ from equation~\eqref{eqn:reg+const} into~\eqref{eqn:Casimir}, we see that the choice $\Delta_\mathtt{const} = -2$ renders the strong coupling expansion of $J^2$ to be only in terms of $\Delta_\mathtt{reg}^2$, thereby fulfilling this assumption.
Based on this we conclude that
\begin{align}\label{eqn:deltaconst}
    \Delta_\mathtt{const} = -2\;,
\end{align}
for \emph{all} states in planar ${\cal N} = 4$ SYM. In the next section, we give credence to this conjecture by performing some numerical fits. Of course for some states, such as Konishi~\cite{Gromov:2009zb,Roiban:2009aa,Tseytlin:2009fw,Frolov:2010wt,Passerini:2010xc,Gromov:2011de, Roiban:2011fe, Vallilo:2011fj}, and those with quantum numbers $[\ell\;\ell\;0\;0\;0]$ with even $\ell$~\cite{Alday:2022uxp,Alday:2022xwz}, this is already well established, but here we argue that should be true in general.

\subsubsection{Fitting procedure}

For each of the 219 states in our database, our goal is to make numerical fits of our data, and obtain predictions for $d_n$ in~\eqref{eqn:regdef}.
To this end,
we fit our numerical data against the truncated  expansion \eq{eqn:regdef}.
We also assume that $\Delta_\mathtt{const}=-2$. Then for each state we proceed as follows:

\paragraph{Step 1: Determine $\delta$.}  This is relatively easy to do, as for the range of 't Hooft coupling that we consider, we can clearly see how the states divide into bands with $2\,\sqrt{\delta}\,\lambda^{1/4}$ asymptotics, where $\delta = 1,2,\dots$ ({\it cf.}~figures~\ref{fig:LREvenStates},~\ref{fig:LRGenGenEvenStates}, and~\ref{fig:GenGenStates})
Thus, we can assign a positive integral value of $\delta$, to each state in our database. Let $\delta_{\tt m}$ be the value of $\delta$ associated with the state whose {\tt St.} {\tt No.} is ${\tt m}$.

\paragraph{Step 2: Find the best hyperparameters.}
Let $\lambda_{\mathtt{max};\,{\tt m}}$ be the highest value of the 't Hooft coupling for which we have a data point for the state ${\tt m}$.
Consider the set of 't Hooft couplings for which we have data points for this state, in between some value of $\lambda$, denoted as $\lambda_{\mathtt{min}}$ and $\lambda_{\mathtt{max};\,{\tt m}}$.
For a particular choice of $\lambda_{\mathtt{min}}$, we choose a range of powers of $1/\sqrt{\lambda}$ up to $1/\lambda^{n_{\mathtt{max}}/2}$.
Thus, a choice of the tuple $\{\lambda_\mathtt{min},\,n_\mathtt{max}\}$, defines the hyperparameters of our fit.
We would like to fit our numerical data against the ``model''~\eqref{eqn:regdef}. Let ${d}_{n;\,{\tt m}}$ be the value of the coefficient ${d}_n$, associated with the state ${\tt m}$.
Thus, we need to determine the tuple $\{\lambda_\mathtt{min},n_{\mathtt{max}}\}$, for which the we get the best fit.
For each choice of this tuple, we subtract the first two terms of~\eqref{eqn:regdef}, evaluated at each value of $\lambda$ in between $\lambda_{\mathtt{min}}$ and $\lambda_{\mathtt{max};\,{\tt m}}$, where we have a data point.
Then we fit this ``refined'' data with the following powers of $\lambda$: $\{\lambda^{1/4},\dots,\lambda^{1/4 - n_{\mathtt{max}}/2}\}$, and measure the coefficient of $\lambda^{1/4}$, which should be zero, according to~\eqref{eqn:regdef}. We choose that tuple $\{\lambda_\mathtt{best},n_{\mathtt{best}}\}$, for which this coefficient is has the least magnitude. As an independent verification of the precision we also usually find the sub-leading coefficient $d_1$ to be some simple rational numbers.

\paragraph{Step 3: Make a prediction for ${d}_{1}$.} We fit the same refined data for $\lambda$ between $\lambda_\mathtt{best}$ and $\lambda_{\mathtt{max};\,{\tt m}}$, with the following powers of $\lambda$: $\{\lambda^{-1/4},\dots,\lambda^{1/4 - n_{\mathtt{best}}/2}\}$. The coefficient of $\delta_{\tt m}^{-1/2}\,\lambda^{-1/4}$ gives us a prediction for ${d}_{1;\,{\tt m}}$, the value of ${d}_1$, associated with state ${\tt m}$.

\paragraph{Step 4: Measure the precision of ${d}_{1}$.} We repeat {\bf Step 3}, but use the powers: $\lambda$: $\{\lambda^{-1/4},\dots,\lambda^{1/4 - n_{\mathtt{best}}/2-1/2}\}$. Then we
check at which digit
the coefficient of $\delta_{\tt m}^{-1/2}\,\lambda^{-1/4}$,
differs from our prediction for ${d}_{1;\,{\tt m}}$. This gives us a measure of the precision of ${d}_{1;\,{\tt m}}$.

In table~\ref{tab:allStates}, we provide the fitted values of ${d}_1$, for every state in our database, up to the best precision, that we could obtain using this procedure.

\subsubsection{Predictions from the literature.
}
\label{sec:dkLit}
Recall, from section~\ref{sec:QN}, that $\mathfrak{sl}(2)$ operators are characterised by the number of derivatives $S$ and length $L$.
The {\tt State ID} of such operators is of the form~\eqref{eq:Sl2}. Their quantum numbers are given by~\eqref{eqn:Sl2Dynkin}.
As described in detail in appendix~\ref{app:sl2}, the scaling dimension of such operators, at one-loop at weak coupling, may be obtained by solving Bethe ansatz equations.
As explained there, this procedure associates a set of integral mode numbers, which can be used as an alternative to the multiplicity label to distinguish between the states with the same quantum numbers.
There are 27 $\mathfrak{sl}(2)$ states in our database.

For $\mathfrak{sl}(2)$ states, which are parity symmetric, and only have two distinct mode numbers $n$ and $-n$, there are predictions in the literature~\cite{Gromov:2009zb,Roiban:2009aa,Tseytlin:2009fw,Frolov:2010wt,Passerini:2010xc,Basso:2011rs, Gromov:2011de, Roiban:2011fe, Vallilo:2011fj,Gromov:2011bz,Frolov:2012zv,Gromov:2014bva}, for some of the coefficients ${d}_k$. Below, we will test them.

\paragraph{States with $n = 1$.}
This is the most studied case.
In this case, there is a prediction for ${d}_1$ in~\cite{Gromov:2009zb,Roiban:2009aa,Tseytlin:2009fw,Frolov:2010wt,Passerini:2010xc,Basso:2011rs, Gromov:2011de, Roiban:2011fe, Vallilo:2011fj}, a prediction for ${d}_2$ in~\cite{Gromov:2011bz}, and finally a prediction for ${d}_3$ in~\cite{Gromov:2014bva}, giving
\begin{multline}\label{eqn:sl2n1}
    \Delta_{S,\,L,\,n = 1}
    \simeq
    \sqrt{2\,S}\,\lambda^{1/4}
    -2
    +
    \frac{2\, L^2+(3\, S-2)\,S}{4\, \sqrt{2\, S}}\,\frac{1}{\lambda^{1/4}}
    \\
    +
    \frac{1}{32\, (2\, S)^{3 / 2}}\,
    \bigg[{-21\, S^4+\left(24-96\, \zeta_3\right) S^3+4\,\left(5\, L^2-3\right) S^2+8\, L^2\, S-4\, L^4}\bigg]\,\frac{1}{\lambda^{3/4}}
    \\[2mm]
    +
    \frac{1}{128\,(2\,S)^{5/2}}
    \,\bigg[
    187 S^6+6\left(208 \zeta_3+160 \zeta_5-43\right) S^5+\left(-146 L^2-4\left(336 \zeta_3-41\right)\right) S^4
    \\[2mm]
    +
    \left(32\left(6 \zeta_3+7\right) L^2-88\right) S^3+\left(-28 L^4+40 L^2\right) S^2-24 L^4 S+8 L^6
    \bigg]\,\frac{1}{\lambda^{5/4}}
    \;.
\end{multline}
In our database, we found 9 states with $n = 1$. In table~\ref{tab:sl2n1test} below, we test the predictions for the strong coupling expansion of the scaling dimension of these states and find perfect agreement within the numerical precision of our fit. The results are given in the table~\ref{tab:sl2n1test}.

{
\scriptsize
\setlength\LTleft{-24mm}\setlength\LTright{-0mm}
\begin{longtable}{@{\extracolsep{\fill}}|C|C|C|C|C|C|C|C|C|C|C|@{}}
\hline
\text{{\tt St.} {\tt No.}} & \texttt{State ID} & [\ell_1\;\ell_2\;q_1\;p\;q_2] & [S\;L\;n] & \Delta _ 0 & \Delta _ 0\text{ \
+ $\#$ }g^2 & \delta & \text{Discrepancy}
 & \text{Discrepancy} & \text{Discrepancy} & \text{Discrepancy} \\

 &  &  &  &  &  &  & 2\,\sqrt{\delta}\,\lambda^{1/4} -2 & d_1 & d_2 & d_3 \\[3pt]
\hline\hline
\tt 1 & \text{}_ 2 \text{[0 0 1 1 1 1 0 0]}_ 1 & \text{[0 0 0 0 0]} & \text{[2 2 1]} & 2 & 12 & 1 & 4.1\times 10^{-16} & 2.4\times 10^{-16} & 7.8\times 10^{-13} & 1.4\times 10^{-10} \\
\tt 2 & \text{}_ 3 \text{[0 0 2 2 1 1 0 0]}_ 1 & \text{[0 0 0 1 0]} & \text{[2 3 1]} & 3 & 8 & 1 & 8.5\times 10^{-12} & -1.9\times 10^{-10} & 6.5\times 10^{-9} & -6.5\times 10^{-8} \\
\tt 7 & \text{}_ 4 \text{[0 0 3 3 1 1 0 0]}_ 2 & \text{[0 0 0 2 0]} & \text{[2 4 1]} & 4 & 10-2 \sqrt{5} & 1 & 1.6\times 10^{-12} & -2.3\times 10^{-10} & 7.8\times 10^{-10} & -2.2\times 10^{-7} \\
\tt 23 & \text{}_ 5 \text{[0 0 4 4 1 1 0 0]}_ 1 & \text{[0 0 0 3 0]} & \text{[2 5 1]} & 5 & 4 & 1 & 1.8\times 10^{-12} & 8.3\times 10^{-11} & -3.0\times 10^{-10} & 5.0\times 10^{-11} \\
\tt 118 & \text{}_ 6 \text{[0 0 5 5 1 1 0 0]}_ 1 & \text{[0 0 0 4 0]} & \text{[2 6 1]} & 6 & 3.012081585 & 1 & -1.1\times 10^{-14} & 1.4\times 10^{-13} & 4.1\times 10^{-11} & -2.0\times 10^{-9} \\
\tt 12 & \text{}_ 4 \text{[0 2 1 1 1 1 2 0]}_ 1 & \text{[2 2 0 0 0]} & \text{[4 2 1]} & 4 & \frac{50}{3} & 2 & -1.6\times 10^{-9} & 2.4\times 10^{-8} & 5.2\times 10^{-6} & -5.5\times 10^{-6} \\
\tt 39 & \text{}_ 5 \text{[0 2 2 2 1 1 2 0]}_ 1 & \text{[2 2 0 1 0]} & \text{[4 3 1]} & 5 & 12 & 2 & -9.3\times 10^{-10} & 1.6\times 10^{-7} & 1.0\times 10^{-6} & -3.8\times 10^{-4} \\
\tt 206 & \text{}_ 6 \text{[0 2 3 3 1 1 2 0]}_ 3 & \text{[2 2 0 2 0]} & \text{[4 4 1]} & 6 & 8.765533583 & 2 & 3.0\times 10^{-10} & 2.4\times 10^{-8} & 2.9\times 10^{-6} & -3.7\times 10^{-5} \\
\tt 219 & \text{}_ 6 \text{[0 4 1 1 1 1 4 0]}_ 1 & \text{[4 4 0 0 0]} & \text{[6 2 1]} & 6 & \frac{98}{5} & 3 & 1.2\times 10^{-8} & -1.1\times 10^{-6} & 8.9\times 10^{-5} & -3.4\times 10^{-3} \\
\hline
\caption{\small\it    Test of strong coupling predictions for ${\mathfrak{sl}(2)}$ states in our database with $n = 1$.
The difference between the prediction for the various $d_k$ from~\eqref{eqn:sl2n1}, and our fitted value for the same $d_k$ is given in the last three columns.
In all cases, the discrepancy occurs within the error of our fit.
}
\label{tab:sl2n1test}
\end{longtable}
}
\noindent
We can also make a fit ${ d}_4$ for some of these states, in table~\ref{tab:sl2n1predict} below we display those, for which we have reliable precision.

{
\scriptsize
\setlength\LTleft{-0mm}\setlength\LTright{-0mm}
\begin{longtable}{@{\extracolsep{\fill}}|C|C|C|C|C|C|C|C|@{}}
\hline
\text{{\tt St.} {\tt No.}} & \texttt{State ID} & [\ell_1\;\ell_2\;q_1\;p\;q_2] & [S\;L\;n] & \Delta _ 0 & \Delta _ 0\text{ \
+ $\#$ }g^2 & \delta & \text{Best fit}\\

 &  &  &  &  &  &  & \text{for }d_4 \\[3pt]
\hline\hline
\tt 1 & \text{}_ 2 \text{[0 0 1 1 1 1 0 0]}_ 1 & \text{[0 0 0 0 0]} & \text{[2 2 1]} & 2 & 12 & 1 & -91.97602351 \\
\tt 2 & \text{}_ 3 \text{[0 0 2 2 1 1 0 0]}_ 1 & \text{[0 0 0 1 0]} & \text{[2 3 1]} & 3 & 8 & 1 & -68.2938 \\
\tt 7 & \text{}_ 4 \text{[0 0 3 3 1 1 0 0]}_ 2 & \text{[0 0 0 2 0]} & \text{[2 4 1]} & 4 & 10-2 \sqrt{5} & 1 & -52.21062 \\
\tt 23 & \text{}_ 5 \text{[0 0 4 4 1 1 0 0]}_ 1 & \text{[0 0 0 3 0]} & \text{[2 5 1]} & 5 & 4 & 1 & -89.959722 \\
\tt 118 & \text{}_ 6 \text{[0 0 5 5 1 1 0 0]}_ 1 & \text{[0 0 0 4 0]} & \text{[2 6 1]} & 6 & 3.012081585 & 1 & -340.652525 \\
\hline
\caption{\small\it    Best fit of $d_4$ for ${\mathfrak{sl}(2)}$ states in our database with $n = 1$.}
\label{tab:sl2n1predict}
\end{longtable}
}
\noindent
\paragraph{Conjectures for states with $n>1$.}
There are a number of conjectures about the strong coupling dimension of states with mode numbers $n>1$ in the literature~\cite{Basso:2011rs, Gromov:2011bz, Frolov:2012zv, Gromov:2014bva}. In~\cite{Gromov:2011bz}, the following expression is proposed for the dimension of a state with $n = 2$ or $3$, though it must be noted that the authors of that article pointed out that this prediction is based on an assumption about the behaviour in $S$, which definitely breaks down for $n>1$ state. So that is curious to see if this affects the prediction already at the low orders in $1/\sqrt\lambda$, which reads
\begin{multline}\label{eqn:GVConj}
    \Delta_{S,\, L,\, n} \simeq \sqrt{2\, S\,n} \,\lambda^{1 / 4} + \frac{2\, L^2+3\, S^2-2\, S}{4\,(2\, S\,n)^{1 / 2}\, \lambda^{1 / 4}}
    \\
    +\frac{1}{32\,(2\, S\,n)^{3 / 2}}\,\bigg[-21\, S^4+\left(32\, B_2+12\right) S^3+\left(20\, L^2-12\right) S^2+8\, L^2\, S-4\, L^4\bigg]\,\frac{1}{\lambda^{3 / 4}}
    \;.
\end{multline}
Here $B_2$ depends on $n$, and we have
\begin{align}
    B_2 =
    \begin{cases}
    -3\, \zeta_3+\frac{3}{8}\;&, \qquad n = 1 \\
    -24\, \zeta_3-\frac{13}{8}\;&, \qquad n = 2 \\
    -81\, \zeta_3-\frac{24}{8}\;&, \qquad n = 3
    \end{cases}
    \;.
\end{align}
In~\cite{Frolov:2012zv}, the conjuncture~\eqref{eqn:GVConj} was tested for $S = 2$, and various values of $L$ and $n$. It was found that this conjecture holds water for $n = 2$, and deviations were observed at $n = 3$.

For the states with $n = 3$ studied in~\cite{Frolov:2012zv}, apart from the leading contribution, the formula~\eqref{eqn:GVConj} doesn't predict either $d_1$ or $d_2$ correctly in this case. In~\cite{Frolov:2012zv}, a prediction for $d_1$ is given for the case that $n = 3$ and $S = 2$. We have
\begin{align}\label{eqn:FrolovPred}
    \Delta_{2,\,L,\,3}
    \simeq 2\,\sqrt{3}\,\lambda^{1/4} + \frac{L^2}{4\,\sqrt{3}}\,\frac{1}{\lambda^{1/4}}
    \;.
\end{align}
We will now compare these results with our numerics.
\paragraph{States with $n = 2$.}
There are 4 states in our database with $n = 2$.
First, let us consider states with $S = 2$.
We display the 3 of them in table~\ref{tab:sl2n2S2test}. For all these states our fit appear to agree within the numerical error with \eq{eqn:GVConj}.

{
\scriptsize
\setlength\LTleft{-14mm}\setlength\LTright{-0mm}
\begin{longtable}{@{\extracolsep{\fill}}|C|C|C|C|C|C|C|C|C|C|@{}}
\hline
\text{{\tt St.} {\tt No.}} & \texttt{State ID} & [\ell_1\;\ell_2\;q_1\;p\;q_2] & [S\;L\;n] & \Delta _ 0 & \Delta _ 0\text{ \
+ $\#$ }g^2 & \delta & \text{Discrepancy}
 & \text{Discrepancy} & \text{Discrepancy} \\

 &  &  &  &  &  &  & 2\,\sqrt{\delta}\,\lambda^{1/4} -2 & d_1 & d_2  \\[3pt]
\hline\hline
\tt 6 & \text{}_ 4 \text{[0 0 3 3 1 1 0 0]}_ 1 & \text{[0 0 0 2 0]} & \text{[2 4 2]} & 4 & 10+2 \sqrt{5} & 2 & -4.6\times 10^{-10} & 2.4\times 10^{-8} & 2.0\times 10^{-2} \\
\tt 24 & \text{}_ 5 \text{[0 0 4 4 1 1 0 0]}_ 2 & \text{[0 0 0 3 0]} & \text{[2 5 2]} & 5 & 12 & 2 & 1.8\times 10^{-11} & 8.1\times 10^{-9} & -7.6\times 10^{-3} \\
\tt 119 & \text{}_ 6 \text{[0 0 5 5 1 1 0 0]}_ 2 & \text{[0 0 0 4 0]} & \text{[2 6 2]} & 6 & 9.780167472 & 2 & 2.7\times 10^{-13} & -3.0\times 10^{-11} & -2.1\times 10^{-3} \\
\hline
\caption{\small\it    Test of strong coupling predictions for ${\mathfrak{sl}(2)}$ states in our database with $S = 2$ and $n = 2$.
The difference between the prediction for the various $d_k$ from~\eqref{eqn:GVConj}, and our fitted value for the same $d_k$ is given in the last two columns.
In all cases, the discrepancy occurs within the error of our fit.
}
\label{tab:sl2n2S2test}
\end{longtable}
}
\noindent
The {\tt State} {\tt Number} ${\tt 6}$, with $[S\;L\;n] = [2\;4\;2]$ was studied in~\cite{Gromov:2011bz}, and the states number ${\tt 24}$ and ${\tt 119}$, with $[2\;5\;2]$ and $[2\;6\;2]$ respectively, were studied in~\cite{Frolov:2012zv} and we are able to corroborate the result, that~\eqref{eqn:GVConj} correctly predicts the scaling dimension at strong coupling for these states.

Next, consider the 1 state in our database with $S>2$ as shown in table~\ref{tab:sl2n2S>2test}.

{
\scriptsize
\setlength\LTleft{-14mm}\setlength\LTright{-0mm}
\begin{longtable}{@{\extracolsep{\fill}}|C|C|C|C|C|C|C|C|C|C|@{}}
\hline
\text{{\tt St.} {\tt No.}} & \texttt{State ID} & [\ell_1\;\ell_2\;q_1\;p\;q_2] & [S\;L\;n] & \Delta _ 0 & \Delta _ 0\text{ \
+ $\#$ }g^2 & \delta & \text{Discrepancy}
 & \text{Best fit} & \text{Prediction} \\

 &  &  &  &  &  &  & 2\,\sqrt{\delta}\,\lambda^{1/4} -2 & \text{for }d_1 & \text{of~\cite{Gromov:2011bz}}  \\[3pt]
\hline\hline
\tt 204 & \text{}_ 6 \text{[0 2 3 3 1 1 2 0]}_ 1 & \text{[2 2 0 2 0]} & \
\text{[4 4 2]} & 6 & 23.18267020 & 4 & 7.5\times 10^{-9} & \
12.00000  & 9 \\
\hline
\caption{\small\it    Test of strong coupling predictions for the ${\mathfrak{sl}(2)}$ state in our database with $S > 2$ and $n = 2$. We find a disagreement with the prediction~\eqref{eqn:GVConj}.}
\label{tab:sl2n2S>2test}
\end{longtable}
}
\noindent
Thus, for the {\tt State} {\tt Number} ${\tt 204}$, with $S = 4$, we see that the formula~\eqref{eqn:GVConj} does not seem to work, even at sub-leading order at strong coupling. It would be interesting to understand the reason for this disagreement and find a general formula which works for all $S$.

\paragraph{States with $n = 3$.}

There is 1 state in our database with $n = 3$, which is displayed  in table~\ref{tab:sl2n3test}.

{
\scriptsize
\setlength\LTleft{-2mm}\setlength\LTright{-0mm}
\begin{longtable}{@{\extracolsep{\fill}}|C|C|C|C|C|C|C|C|C|@{}}
\hline
\text{{\tt St.} {\tt No.}} & \texttt{State ID} & [\ell_1\;\ell_2\;q_1\;p\;q_2] & [S\;L\;n] & \Delta _ 0 & \Delta _ 0\text{ \
+ $\#$ }g^2 & \delta & \text{Discrepancy}
 & \text{Discrepancy} \\

 &  &  &  &  &  &  & 2\,\sqrt{\delta}\,\lambda^{1/4} -2 & d_1  \\[3pt]
\hline\hline
\tt 120 & \text{}_ 6 \text{[0 0 5 5 1 1 0 0]}_ 3 & \text{[0 0 0 4 0]} & \text{[2 6 3]} & 6 & 15.20775094 & 3 & 4.1\times 10^{-12} & -2.0\times 10^{-10} \\
\hline
\caption{\small\it    Test of strong coupling predictions for the ${\mathfrak{sl}(2)}$ state in our database with $S > 2$ and $n = 2$.
The difference between the prediction for $d_1$ from~\eqref{eqn:FrolovPred}, and our fitted value is given in the last column.
The discrepancy occurs within the error of our fit.
}
\label{tab:sl2n3test}
\end{longtable}
}
\noindent
This state, number ${\tt 120}$, is the same state as was studied in~\cite{Frolov:2012zv}. Therefore we are only able to obtain a consistency check, rather than a confirmation of the prediction~\eqref{eqn:FrolovPred}.

In general we conclude that even in the simplest ${\mathfrak{sl}(2)}$ sector we still do not know an analytic way of computing even the leading non-trivial strong coupling coefficient $d_1$, which would work for all states.

\subsection{Counting of states and Kaluza-Klein towers}\label{sec:KKtowers}

The authors of~\cite{Alday:2023flc} perform a counting of states of ${\cal N} = 4$ SYM, that scale as $\sim\lambda^{1/4}$ at strong coupling.
In order to do this,
they take the flat-space limit of AdS${}_5\times S^5$,
and perform a counting of representations of  ${\rm SO}(9)$, the massive little group of $\mathbb{R}^{1,9}$.
In particular, they count representations of ${\rm SO}(4)\times{\rm SO}(5)$, corresponding to the split into AdS$_5$ and $S^5$. Compactifying five directions to $S^5$, implies that that each representation $\left[m\;n\right]$ of ${\rm SO}(5)$ gets replaced by a Kaluza-Klein (KK-) tower~\cite{Bianchi:2003wx} of ${\rm SO}(6)$ representations.
Introduce the following notation for the KK-tower of states with Lorentz spin labels $\left(\ell_1\;\ell_2\right)$, and ${\rm SO}(5)$ labels $[m\;n]$:~\cite{Bianchi:2003wx}
\begin{multline}\label{eqn:KKtows}
    \left[\ell_1\;\ell_2;\;m\;n\right]
    =
    \sum_{r=0}^m \sum_{s=0}^n \sum_{p=m-r}^{\infty}\left[\ell_1\,,\;\ell_2\,,\;r+n-s\,,\; p\,,\; r+s\right]
    \\
    +\sum_{r=0}^{m-1} \sum_{s=0}^{n-1} \sum_{p=m-r-1}^{\infty}\left[\ell_1\,,\;\ell_2\,,\;r+n-s\,,\; p\,,\; r+s+1\right]\;.
\end{multline}
This counting is useful at infinite coupling, where states are split according to the string level $\delta$. States within a KK-tower will share the same $\delta$ but their $\Delta$ will in general split already at $1/\lambda^{1/4}$ order.
There are infinitely many states in each KK-tower, which then implies that there should be infinitely many states for a given string level $\delta$. However, the number of KK-towers at each $\delta$ is finite and we will give some examples below.

Thus, for each $\delta$, one can introduce a counting function ${\tt count}_\delta$, following~\cite{Alday:2023flc}, which contains the KK-towers with that value of $\delta$. For example
\begin{align}
    \label{eqn:count1}
    \mathtt{count}_1 &=
    [0\;0 ;\; 0\;0]
    \;,\\
    \label{eqn:count2}
    \mathtt{count}_2 &=
    2\,[0\;0 ;\; 0\;0]
    +[0\;0 ;\; 2\;0]
    +[0\;0 ;\; 0\;2]
    +2\,[1\;1 ;\; 1\;0]
    +[2\;2 ;\; 0\;0]
    +[2\;0 ;\; 0\;0]
    +[0\;2 ;\; 0\;0]
    \;.
\end{align}
The  ${\tt count}_3$ is also known explicitly~\cite{AHSprivate}, and is given in appendix~\ref{app:delta3states}.
Below we identify states belonging to different KK-towers and check the counting~\eq{eqn:count1} and
~\eq{eqn:count2}.

The process of assigning a specific state to a given Kaluza-Klein tower solely based on quantum numbers can generally be ambiguous. This is due to the degeneracies in the spectrum and the absence of additional input from the wave functions.
As we mentioned already, the states within the same tower could have different sub-leading coefficients $d_1$. However, we noticed that when computing the expansion of the quadratic Casimir instead of $\Delta$, the states within the same tower have the same sub-leading coefficient, which can in turn be used to assign states to different KK-towers. Furthermore, if this observation is correct, it implies a non-trivial set relations on the $d_1$ coefficients in the expansion of $\Delta$ among infinite set of states within one KK-tower. More precisely one gets the following identity:
\begin{multline}
\label{eqn:d1pred}
    d_1 =
    \frac{p^2}{4}+\frac{p}{4} \left(q_1+q_2+4\right)
    +\frac{1}{16} \bigg[
    16
    -2\, \ell_1\, \left(\ell_1+2\right)-2\, \ell_2\, \left(\ell_2+2\right)
    \\
    +3\, q_1 \left(q_1+4\right)
    +3\, q_2 \left(q_2+4\right)
    + 2\,q_1\,q_2
    \bigg]+\frac{j_1}{2}
    \;,
\end{multline}
where $j_1$ is the sub-leading coefficient in the strong coupling expansion for $J^2\simeq 2\,\delta\sqrt\lambda+j_1$, which we conjecture to be the same for all states in one tower, based on the data we have.
We also checked that the next coefficient $j_2$ is already different for states in a KK-tower.

Below we analyse the $\delta=1,2,3$ cases, confirming our conjecture and also confirming the counting of~\cite{Alday:2023flc,AHSprivate}.

\paragraph{States with $\delta = 1$.}
There are 5 such states in our database.
All of them have Lorentz spin labels $\left(0\;0\right)$.
We present them in table~\ref{tab:delta1states}.

{
\scriptsize
\setlength\LTleft{-0mm}\setlength\LTright{-0mm}
\begin{longtable}{@{\extracolsep{\fill}}|C|C|C|C|C|C|C|C|C|C|@{}}
\hline
\text{{\tt St.} {\tt No.}} & \texttt{State ID} & [\ell_1\;\ell_2\;q_1\;p\;q_2] & \Delta _ 0 & \Delta _ 0\text{ \
+ $\#$ }g^2 & \delta & d_1 & j_1 & [S\;L\;n] & \text{Type}  \\

 &  &  &  &  &  &  &  &  & \\[3pt]
\hline\hline
1 & \text{}_ 2 \text{[0 0 1 1 1 1 0 0]}_ 1 & \text{[0 0 0 0 0]} & 2 & \
12 & 1  & 2 & 2 & \text{[2 2 1]} & \text{I} \\
 2 & \text{}_ 3 \text{[0 0 2 2 1 1 0 0]}_ 1 & \text{[0 0 0 1 0]} & 3 \
& 8 & 1  & \frac{13}{4} & 2 & \text{[2 3 1]} & \text{I}
\\
 7 & \text{}_ 4 \text{[0 0 3 3 1 1 0 0]}_ 2 & \text{[0 0 0 2 0]} & 4 \
& 10-2 \sqrt{5} & 1  & 5 & 2 & \text{[2 4 1]} & \text{I} \
\\
 23 & \text{}_ 5 \text{[0 0 4 4 1 1 0 0]}_ 1 & \text{[0 0 0 3 0]} & 5 \
& 4 & 1  & \frac{29}{4} & 2 & \text{[2 5 1]} & \text{I} \\
 118 & \text{}_ 6 \text{[0 0 5 5 1 1 0 0]}_ 1 & \text{[0 0 0 4 0]} & \
6 & 3.012081585 & 1  & 10 & 2 & \text{[2 6 1]} & \text{I} \\
\hline
\caption{\small\it    All states in our database with $\delta = 1$.}
\label{tab:delta1states}
\end{longtable}
}
\noindent
Clearly, all these states are members of the KK-tower $\left[0\;0;\;0\;0\right]$, in exact agreement with~\eqref{eqn:count1}, and thus the results of~\cite{Alday:2023flc}.
Furthermore we see that for all these states $j_1=2$.
All the states are $\mathfrak{sl}(2)$ states of the form $[S, L, n] = [2, p+2, 1]$, as can be seen by comparing with section~\ref{sec:dkLit}. The strong coupling dimensions of these states, at first three sub-leading orders is given by equation~\eqref{eqn:sl2n1}.

All the other states in the tower should have bigger bare dimension due to the unitarity bounds (see e.g.~\cite{Cordova:2016emh}):
\begin{align}\label{eqn:unibound}
    \Delta >
    2 + {\tt max}\left\{
    \ell_1 + \frac{1}{2}\left(
    3\,q_1 + 2\,p + q_2
    \right)\,,\;
    \ell_2 + \frac{1}{2}\left(
    q_1 + 2\,p + 3\,q_2
    \right)
    \right\}
    \;,
\end{align}
which for the current set of states would imply
\begin{align}\label{eqn:uniboundK0000}
    \Delta > p + 2\;.
\end{align}
Thus, we see that a state with quantum numbers $[0\;0\;0\;5\;0]$, must have $\Delta > 7$, and thus are not present in the current data set.

\paragraph{States with higher $\delta$.}
At $\delta>1$ we can face an ambiguity when assigning the states from our database to various KK-towers (given in~\eqref{eqn:count2} for $\delta = 2$ and in~\eqref{eqn:delta3count} for $\delta = 3$). We see that there are many states, whose quantum numbers are consistent with being a member of more than one KK-tower. But this ambiguity is lifted if we look at the $j_1$ coefficient. Or in other words we verified the conjecture \eq{eqn:d1pred}
which relates the sub-leading coefficients in $\Delta$ between various members of the KK-tower.

We have a total of 38 states in our database, with $\delta = 2$. We have classified them into the $9$ KK-towers given in~\eqref{eqn:count2}. The assignment is described in appendix~\ref{app:delta2states}, specifically in tables~\ref{tab:delta2l00} --~\ref{tab:delta2l22}.
We have a total of 128 states in our database, with $\delta = 3$. We have classified them into KK-towers given in~\eqref{eqn:delta3count}. The assignment is described in appendix~\ref{app:delta3states}, specifically in tables~\ref{tab:delta3l00} --~\ref{tab:delta3l44}.
It is also shown there that $j_1$ is indeed a good classifier of states, into KK-towers.

In other words, where we have sufficient precision, our fits confirm the formula~\eqref{eqn:d1pred}. It is important to notice that, by using the formula~\eqref{eqn:d1pred}, it is possible to obtain the $d_1$ coefficients for every state in the KK tower, even ones which are not practically accessible by numerical methods.\footnote{Such as states with very high bare dimension $\Delta_0$, where current methods are not able to obtain perturbative data.}

In the next section, we will see how this new spectral information, can be injected into the conformal bootstrap constraints obtained in~\cite{Alday:2022uxp,Alday:2022xwz}.

\subsection{Bootstrability}
\label{sec:bootstrability}
Following the philosophy of~\cite{Cavaglia:2021bnz,Cavaglia:2022qpg,Caron-Huot:2022sdy}, the results for the spectrum, joined with the constraints from the conformal crossing relations for some correlators could give narrow bounds, or even analytic results for some structure constants.
In the current set-up of the local operators our predictions at strong coupling can be fed into the conformal constraints
obtained in~\cite{Alday:2022uxp,Alday:2022xwz}.

One of the complications of the consideration based purely on conformal constraints is typically related to the mixing problem --- when there are several states with the same quantum numbers and same $\delta$.
As we know how the degeneracy is lifted from our data, we should be able to inject this information and improve the results based on  conformal bootstrap (with some input from localisation).

Below, we describe the setup of~\cite{Alday:2022uxp,Alday:2022xwz}.
They consider a four point function of four $\mathbf{20^\prime}$ $1/2$-BPS operators in ${\cal N} = 4$ SYM. These are operators which transform under the rank-$2$ symmetric traceless representation of the ${\rm SO}(6)$ $R$-symmetry group. They are defined as
\begin{align}
    {\cal O}_2(\vec{x},\,Y) \equiv {\rm Tr}\, {\Phi}_{I}(\vec{x})\,{\Phi}_{J}(\vec{x})\,Y^I\,Y^J\;,
\end{align}
where $\Phi_I$ is a fundamental real scalar field of ${\cal N} = 4$ SYM, and $Y^I$ is a polarisation null vector. After extracting the kinematic factors, the four-point function of four such operators can be written in terms of a function of the conformal and $R$-symmetry cross ratios. We have
\begin{align}
    \left\langle\mathcal{O}_2\left(\vec{x}_1, Y_1\right) \mathcal{O}_2\left(\vec{x}_2, Y_2\right) \mathcal{O}_2\left(\vec{x}_3, Y_3\right) \mathcal{O}_2\left(\vec{x}_4, Y_4\right)\right\rangle=\frac{Y_{1\,2}^2 Y_{3\,4}^2}{x_{1\,2}^4 x_{3\,4}^4} \mathcal{S}(U, V ; \sigma, \tau)\;,
\end{align}
where $Y_{I\,J} \equiv Y_I \cdot Y_J$, $x_{i\,j} \equiv x_i - x_j$, $U$ and $V$ are the conformal cross-ratios, and $\sigma$ and $\tau$ are $R$-symmetry cross ratios. These are defined as
\begin{align}
    U \equiv \frac{x_{1\,2}^2 x_{3\,4}^2}{x_{1\,3}^2 x_{2\,4}^2} \equiv z\,{\bar z}\;, \quad V \equiv \frac{x_{1\,4}^2 x_{2\,3}^2}{x_{1\,3}^2 x_{2\,4}^2}  \equiv (1-z)\,(1-{\bar z})\;,
    \quad
    \sigma \equiv
    \frac{Y_{1\,3}.\,Y_{2\,4}}{Y_{1\,2}.\,Y_{3\,4}}
    \;, \quad
    \tau \equiv
    \frac{Y_{1\,4}.\,Y_{2\,3}}{Y_{1\,2}.\,Y_{3\,4}}
    \;.
\end{align}
Using superconformal symmetry, ${\cal S}$ can be written as~\cite{Dolan:2001tt}
\begin{align}
    \mathcal{S}(U, V ; \sigma, \tau) &=\mathcal{S}_{\mathtt{free }}(U,V ;  \sigma, \tau)+\Theta(U, V ; \sigma, \tau)\,\mathcal{T}(U, V)\;, \\
    \Theta(U, V ; \sigma, \tau) &\equiv \tau+[1-\sigma-\tau]\, V+\tau\,[\tau-1-\sigma]\, U+\sigma\,[\sigma-1-\tau]\, U\, V+\sigma\, V^2+\sigma\, \tau\, U^2\;, \\
    \mathcal{S}_{\mathtt{free }}(U, V ; \sigma, \tau) &= 1+U^2\, \sigma^2+\frac{U^2}{V^2}\, \tau^2+\frac{1}{c}\,\left(U\, \sigma+\frac{U}{V}\, \tau+\frac{U^2}{V}\, \sigma\, \tau\right)
    \;.
\end{align}
Here ${\cal S}_\mathtt{free}$ denotes the free theory correlator, ${\cal T}(U,V)$ is called the reduced correlator, and $c$ is the central charge, which is given by $c = \frac{N^2 - 1}{4}$. The reduced correlator satisfies a crossing equation ${\cal T}(U,V) = {\cal T}(1/U,V/U) = 1/V^2\,{\cal T}(U/V,1/V)$, and can be expanded using the operator product expansion (OPE). We have
\begin{align}\label{eqn:20barOPE}
    \mathcal{T}(U, V)=U^{-2} \sum_{T, \ell} C_{T, \ell}^2\, G_{T+4, \ell}(U, V)
    \;.
\end{align}
Here, the four-dimensional conformal block $G$ is given by~\cite{Dolan:2003hv}
\begin{align}
    G_{T, \ell}(U, V)=\frac{z\, \bar{z}}{z-\bar{z}}\left(k_{T+2\, \ell}(z) k_{T-2}(\bar{z})-k_{T+2\, \ell}(\bar{z}) k_{T-2}(z)\right)
    \;,
\end{align}
where $k_h$ is
\begin{align}
    k_h(z) \equiv z^{\frac{h}{2}}{ }_2 F_1(h / 2, h / 2, h, z)\;.
\end{align}
In the planar limit, the superprimaries exchanged in the OPE~\eqref{eqn:20barOPE} include single- and double-trace operators, and belong to both short and long multiplets.
In~\cite{Alday:2022uxp,Alday:2022xwz} it was shown that it is possible to rewrite the crossing equation in a way that focuses on the long multiplets corresponding to the  ``stringy'' operators exchanged in the OPE~\eqref{eqn:20barOPE} i.e. those operators that we studied in this paper with $\Delta\sim\lambda^{1/4}$ scaling.
The stringy operators exchanged in the OPE~\eqref{eqn:20barOPE} have quantum numbers $[\ell\;\ell\;0\;0\;0]$ with $\ell = 0, 2, 4 \dots$.

In order to make a comparison we introduce further notations from~\cite{Alday:2022uxp,Alday:2022xwz}. For twists $T\equiv\Delta-\ell$ we have
\begin{align}
    T(\lambda)
    \simeq 2\,\sqrt{\delta}\,\lambda^{\frac{1}{4}} - 2 -\ell +
    \frac{T_1}{\lambda^{1/4}}\;\;,\;\;T_1\equiv \frac{d_1}{\sqrt\delta}
    \;.
\end{align}
Then, for the OPE coefficients, we have~\cite{Alday:2022uxp,Alday:2022xwz}
\begin{align}\label{eqn:C^2pre}
    C^2(\lambda) &= \frac{\pi^3}{2^{12}} \frac{2^{-2\, T(\lambda)} T(\lambda)^6}{\sin ^2\left(\frac{\pi\, T(\lambda)}{2}\right)} \frac{1}{2^{2\, \ell}(\ell+1)} f(\lambda)\;.
\end{align}
The prefactor of $C^2$ in~\eqref{eqn:C^2pre} is highly oscillating when $\lambda\to\infty$ and becomes infinite each time the twist $T$ crosses an even integer. At the same time the function $f(\lambda)$ has a regular asymptotic expansion:
\begin{align}
f(\lambda) &\simeq f_0 + \frac{f_1}{\lambda^{1/4}} +
    \frac{f_2}{\lambda^{1/2}}
    \;.
\end{align}
The result of~\cite{Alday:2022uxp,Alday:2022xwz} includes formulas for the CFT-data of the exchanged operators at strong coupling.
In order to compare we introduce an integer ``Regge trajectory number"~\cite{Chew:1962eu} $t\equiv \delta-\ell/2$.

In table~\ref{tab:AHSStates}, we present all the 23 states in our database, that can be exchanged in the OPE~\eqref{eqn:20barOPE} for given $\delta$ and $\ell$.
\begin{table}[h]
    \centering
    \begin{tabular}{c|C|C|C|C|C}
    \diagbox{$\delta$}{$\ell$} & 0 & 2 & 4 & 6 & 8\\
    \hline\hline
     1  & \cellcolor{Apricot} {\tt 1} &  &  & &  \\
     2  & \cellcolor{Apricot} {\tt 3,\;4} & \cellcolor{Apricot} {\tt 12} &  & & \\
     3  & \cellcolor{Apricot} {\tt 76,\;78,\;81,\;83,\;84,\;85} & \cellcolor{Apricot} {\tt 196,\;197,\;198,\;199} & \cellcolor{Apricot} {\tt 219} & & \\
     4 &  {\tt 77,\;79,\;80,\;86,}\;?\;(22) & {\tt 195,\;200,\;201,}\;?\;(24) & ?\;(6) & ?\;(1) & \\
     5 & {\tt80,}\;?\;(99) & ?\;(157) & ?\;(40) & ?\;(6) & ?\;(1) \\
    \hline
    \end{tabular}
    \caption{\small\it
    States in our database with quantum numbers $[\ell\;\ell\;0\;0\;0]$ with even $\ell$, {\it i.e.}~those which are exchanged in the OPE~\eqref{eqn:20barOPE}.
    Each entry contains the {\tt St.} {\tt No.} of the state(s) with the corresponding $\delta$ and $\ell$.
    The cells where we have all the exchanged states in our database are shaded.
    In the cases we have some or none of the exchanged states in our database, we write the {\tt St.} {\tt Nos.} of the states that are there, followed by a ``$\ ?$'' with the total number of states with that $\delta$ and $\ell$ in parenthesis. The counting of states in this table is from~\cite{Alday:2023flc}.
    }
    \label{tab:AHSStates}
\end{table}

Thus, for each value of $t$, we have a set of states, with particular values of $\delta$ and $\ell$, associated with it. These states are said to be ``on the Regge trajectory $t$.''
In the papers~\cite{Alday:2022uxp,Alday:2022xwz} the mixing problem appeared to be an obstacle to disentangle various exchanged operators with the same values of $\delta$ and $\ell$, only  ``average'' formulas were obtained that compute a sum over such operators, which they denote by $\langle\dots\rangle$.
Namely, they compute $\langle f_0\rangle$~\cite{Alday:2022uxp}, $\langle f_1 \rangle$, $\langle T_1\,f_0\rangle$ and $\langle f_2\rangle$~\cite{Alday:2022xwz}, for the states on various Regge trajectories.

\paragraph{Regge trajectory $1$.}
On the leading Regge trajectory, {\it i.e.}~for the states with $t = 1$, the degeneracy of states is trivial, up to $\delta = 7$~\cite{Alday:2023flc}, and therefore, the average formulas give immediately results for the coefficients in the strong coupling expansion of the CFT-data.
Correspondingly, the first $3$ coefficients $f_0,\;f_1$ and $f_2$ are already known analytically for these states (see~\cite{Alday:2022uxp,Alday:2022xwz} for explicit results).

\paragraph{Predictions on Regge trajectory $2$.}
States on this Regge trajectory have $\ell = 2\,(\delta -2)$.
The average of the OPE coefficients is given by~\cite{Alday:2022uxp}
\begin{align}\label{eqn:f0av}
    \left\langle f_0\right\rangle=\frac{r_1(\delta)}{3}\left(2 \delta^2+3 \delta-8\right)
    \;,
\end{align}
where
\begin{align}\label{eqn:rdef}
    r_n(\delta)=\frac{4^{2-2 \delta} \delta^{2 \delta-2 n-1}(2 \delta-2 n-1)}{\Gamma(\delta) \Gamma\left(\delta-\left\lfloor\frac{n}{2}\right\rfloor\right)}\;.
\end{align}
The average of the leading OPE coefficients, weighted by the twist is~\cite{Alday:2022xwz}
\begin{align}\label{eqn:f0T1av}
    \left\langle f_0\, T_1\right\rangle=\frac{r_1(\delta)}{18 \sqrt{\delta}}\left(18 \delta^4+25 \delta^3-57 \delta^2+50 \delta-72\right)
    \;.
\end{align}
Note that the number of states with a given spin contributing to~\eq{averages} depends on the spin.
For this trajectory we only have two cells in $\ell,\delta$ table~\ref{tab:AHSStates} with complete number of states: $\ell=0,\delta=2$ (two states) and $\ell=2,\delta=3$ (four states) according to ~\cite{Alday:2023flc} and the counting in section~\ref{sec:KKtowers}.

Consider first $\ell=0$ case.
The two states are {\tt St.} {\tt No.} ${\tt 3}$ and ${\tt 4}$ (as defined in  table~\ref{tab:allStates}).
For those state we have
\begin{align}\label{eqn:T1ours}
    T_{1;\;{\tt 3}} = \sqrt{2}\;,\qquad T_{1;\;{\tt 4}} = 4\,\sqrt{2}\;,
\end{align}
where the second index corresponds to the {\tt State} {\tt Number}. Then from~\eqref{eqn:f0av} and~\eqref{eqn:f0T1av} we get
\beqa
    \left\langle f_0 \right\rangle &=&
    f_{0;\;{\tt 3}} + f_{0;\;{\tt 4}}
    = \frac{1}{4}
    \;,\\
\la{averages}
    \left\langle f_0\,T_1 \right\rangle &=&
    f_{0;\;{\tt 3}}\,T_{1;\;{\tt 3}} + f_{0;\;{\tt 4}}\,T_{1;\;{\tt 4}} = \sqrt{2}\;.
\eeqa
resulting in
\begin{align}
    \boxed{
    f_{0;\;{\tt 3}} = 0\;, \qquad f_{0;\;{\tt 4}} = \frac{1}{4}}\;.
\end{align}
Curiously only one of the two OPE coefficients is nonzero, which may indicate some additional simplification at strong coupling.

For $\ell=2$ there are four states with {\tt St.} {\tt Nos.}  {\tt 196}, {\tt 197}, {\tt 198} and {\tt 199}.
Furthermore, {\tt 198}, and {\tt 199}, which are Type II states, form a parity dublet, and thus their dimensions are exactly degenerate, as discussed in section~\ref{sec:TypesStates}.
Since the external BPS operators are parity invariant, their OPE coefficients should be equal as well,
which reduces the system to $3$ unknowns.
Using that
\begin{align}\label{eqn:T1oursd3l2}
    T_{1;\;{\tt 196}} = \frac{8}{\sqrt{3}}\;, \qquad T_{1;\;{\tt 197}} = \frac{17}{\sqrt{3}}\;, \qquad T_{1;\;{\tt 198}} =  \frac{13}{\sqrt{3}}\;,
\end{align}
from~\eqref{eqn:f0av} and~\eqref{eqn:f0T1av} we get
\beqa\label{eqn:f0oursd3l2}
   \left\langle f_0 \right\rangle &=&
    f_{0;\;{\tt 196}} + f_{0;\;{\tt 197}} + 2\,f_{0;\;{\tt 198}}
    = \frac{513}{1024}
    \;,\\
\label{eqn:f0T1oursd3l2}
    \left\langle f_0\,T_1 \right\rangle &=&
    f_{0;\;{\tt 196}}\,T_{1;\;{\tt 196}} + f_{0;\;{\tt 197}}\,T_{1;\;{\tt 197}} + 2\,f_{0;\;{\tt 198}}\,T_{1;\;{\tt 198}} = \frac{2547\,\sqrt{3}}{1024}\;,
\eeqa
which allows to express everything in terms of $f_{0; {\tt 196}}$:
\begin{align}\label{eqn:delta2ell2sols}
    f_{0;\;{\tt 197}} =
    \frac{5}{4} f_{0;\;{\tt 196}} +\frac{243}{1024}\;,\qquad
    f_{0;\;{\tt 198}} &=
    -\frac{9}{8} f_{0;\;{\tt 196}} +
    \frac{135}{1024}\;.
\end{align}
Furthermore, as the OPE coefficients have to be non-negative, we have three further inequalities which impose:
\begin{align}\label{eqn:ineqd3l2}
    f_{0;\;{\tt 196}} \geq 0\;,\qquad
    f_{0;\;{\tt 197}} \geq 0\;,\qquad
    f_{0;\;{\tt 198}} \geq 0\;.
\end{align}
Subjecting the solution~\eqref{eqn:delta2ell2sols} to the inequalities~\eqref{eqn:ineqd3l2}, we get the allowed regions
\begin{align}\label{eqn:delta2ell2bounds}
    f_{0;\;{\tt 196}} \in \left[0,\frac{15}{128}\right]\;,
    \quad
    f_{0;\;{\tt 197}} \in \left[0,\frac{135}{1024}\right]\;,
    \quad
    f_{0;\;{\tt 198}} \in \left[\frac{243}{1024},\frac{393}{1024}\right]\simeq \[0.237305,0.383789\]\;.
\end{align}
Optimistically one could hoped that only one of the OPE coefficients is non-zero, however, clearly $f_{0;\;{\tt 196}}$ and $f_{0;\;{\tt 197}}$ cannot vanish simultaneously due to~\eq{eqn:delta2ell2sols}.

\paragraph{Next order on Regge trajectory 2.}
At the next order, we are able to obtain some predictions for the $d_2$ coefficient of states ${\tt 3}$, ${\tt 4}$,  ${\tt 196}$ and ${\tt 197}$. They are presented below.
\begin{table}[h]
    \centering
    \begin{tabular}{c|c|c|c|c|c}
    {\tt St.} {\tt No.} & $d_1$ numeric & $d_1$ prediction & $d_2$ numeric & $d_2$ guess & discrepancy\\ \hline\hline
    {\tt 3} & $1.9999999$ & $2$ & $-24.34936$ & $\frac{9}{2} - 24 \zeta_3$ & $-5.4 \times 10^{-7}$  \\
    {\tt 4} & $8.00000$ & $8$ & $-66.34945$ & $-\frac{75}{2} - 24 \zeta_3$ & $8.7 \times 10^{-6}$ \\
    {\tt 196} & $8.00000$ & $8$ & $-140.366$ &  &  \\
    {\tt 197} & $16.99999$ & $17$ & $-293.365$ &  &  \\
    {\tt 198}, {\tt 199} & $12.999$ & $13$ & $-179.3$ &  &  \\
    \end{tabular}
    \caption{sub-leading and sub-sub-leading dimension at strong coupling for states on Regge trajectory $2$.}
    \label{tab:my_label}
\end{table}

{\it Note added:} while this paper was at the final stage of preparation for publication we received a communication confirming that in an upcoming paper~\cite{Alday:2023mvu}, analytical expressions for $T_1$, $T_2$ and $f_0$ for the state numbers ${\tt 3}$ and ${\tt 4}$, are obtained using the analytical conformal bootstrap, and they confirm our predictions for these quantities.

\paragraph{Predictions on Regge trajectory 3.} For states with $\ell = 2\,(\delta - 3)$
the average of the OPE coefficients for this trajectory is given by the following expression~\cite{Alday:2022uxp}
\beqa\label{eqn:f0avd3}
&&    \left\langle f_0\right\rangle=
    \frac{r_2(\delta)}{45}\left(10\, \delta^4+43\, \delta^3+8\, \delta^2-352\, \delta-192\right)
    \;,
\\
\nonumber&&    \left\langle f_0\, T_1\right\rangle
    =
    \frac{r_2(\delta)}{1350\, \sqrt{\delta}}\left(450\, \delta^6+1985\, \delta^5+1043\, \delta^4-12782\, \delta^3-2552\, \delta^2-35712\, \delta-11520\right)
    \;.
\eeqa
where $r_2$ is defined in~\eqref{eqn:rdef}.

For this trajectory only for  $\ell = 0$ and $\delta = 3$ we have complete set of $6$ states in the database: {\tt St.} {\tt Nos.}  {\tt 76}, {\tt 78}, {\tt 81}, {\tt 83}, {\tt 84} and {\tt 85}.
Again states {\tt 76} and {\tt 85} are exactly degenerate due to the LR-symmetry, which is also the symmetry of the external operators, implying that $f_{0;\;{\tt 76}}=f_{0;\;{\tt 85}}$.
From table~\ref{tab:allStates} we know the sub-leading coefficients in their dimensions to be (within the numerical accuracy of our data)
\begin{align}\label{eqn:T1d3our}
    T_{1;\;{\tt 76}} = \frac{31}{4\,\sqrt{3}}\;,\quad
    T_{1;\;{\tt 78}} = \frac{1}{\sqrt{3}}\;,\quad
    T_{1;\;{\tt 81}} = \frac{10}{\sqrt{3}}\;,\quad
    T_{1;\;{\tt 83}} = \frac{11}{\sqrt{3}}\;,\quad
    T_{1;\;{\tt 85}} = \frac{19}{\sqrt{3}}\;.
\end{align}
From~\eqref{eqn:f0avd3}, we get
\begin{align}
\begin{split}\label{eqn:delta2ell0sols}
    f_{0;\;{\tt 83}} &=  -\frac{45}{16}\, f_{0;\;{\tt 76}} -\frac{9}{4}\, f_{0;\;{\tt 78}} -\frac{9}{8}\, f_{0;\;{\tt 81}} +\frac{25}{1024}\;, \\
    f_{0;\;{\tt 85}} &=  \frac{13}{16}\, f_{0;\;{\tt 76}} +\frac{5}{4}\, f_{0;\;{\tt 78}} +\frac{1}{8}\, f_{0;\;{\tt 81}} +\frac{81}{1024}
    \;.
\end{split}
\end{align}
Furthermore, we have the inequalities
\begin{align}\label{eqn:ineqd3}
    f_{0;\;{\tt 76}} \geq 0\;,\quad
    f_{0;\;{\tt 78}} \geq 0\;,\quad
    f_{0;\;{\tt 81}} \geq 0\;,\quad
    f_{0;\;{\tt 83}} \geq 0\;,\quad
    f_{0;\;{\tt 85}} \geq 0\;.
\end{align}
We can subject the solution~\eqref{eqn:delta2ell0sols} to the inequalities~\eqref{eqn:ineqd3}, to get the following allowed regions:
\begin{align}\label{eqn:delta3ell0bounds}
\begin{split}
    f_{0;\;{\tt 76}} \in \left[0,\frac{5}{576}\right]\;,
    \quad
    f_{0;\;{\tt 78}} \in \left[0,\frac{25}{2304}\right]\;,
    \quad
    f_{0;\;{\tt 81}} \in \left[0,\frac{25}{1152}\right]\;,
    \\
    f_{0;\;{\tt 83}} \in \left[0,\frac{25}{1024}\right]\;,
    \quad
    f_{0;\;{\tt 85}} \in \left[\frac{81}{1152},\frac{427}{4608}\right]\;.
\end{split}
\end{align}
To further constrain the OPE coefficients one may consider correlators with non-BPS external legs and/or additional input from localisation~\cite{Chester:2021aun,Paul:2022piq}. Having external non-BPS operators is at the moment technically challenging, since the super-conformal blocks for these representations are not yet known.

\section{Conclusion}\la{sec:concl}

We present a highly efficient numerical method and its C++ implementation for studying the spectrum of non-protected observables in planar ${\cal N}=4$ supersymmetric Yang-Mills theory (SYM) using the Quantum Spectral Curve (QSC) method. This work addresses the need for a more comprehensive study of a larger set of states, due to its relevance in the conformal Bootstrap approach.

Our new method facilitates efficient initialisation of numerical iterations benefiting from already available perturbative solvers at small $g$, overcoming a key challenge in previous computations. Furthermore, our C++ implementation delivers a 8-fold performance improvement over the most efficient existing {\it Mathematica} implementation.

We used our developed algorithm to compute all states with a bare dimension $\leq 6$.
The data generated and the codes used have been made publicly available on  {\tt \href{https://GitHub.com/julius-julius/qsc}{GitHub}} for the wider scientific community.
We provide our code as open-source and kindly request that users cite this paper when using the code or any of its original components.

By combining the strong coupling analysis of our data with the recent results from conformal bootstrap we managed to resolve the degeneracy issue in some cases and give a prediction for strong coupling expansion coefficient of a structure constant of two BPS one non-BPS operator (with bare dimension $\Delta_0=4$).

Possible future applications and directions for our approach include:

\begin{itemize}
\item Combining our results with the constraints from conformal symmetry and numerical conformal bootstrap to establish narrow bounds on the structure constants or other observables beyond the spectrum.
\item Studying other non-local observables, including light-ray operators and Regge trajectories.
\item Adapting the method for use in different theories, such as ABJM.
\item Modifying the approach for use in $AdS_3$ cases.
\item Using the method to calculate the Hagedorn temperature non-perturbatively.
\end{itemize}

By making our code and data publicly accessible, we aim to aid and inspire further research in these and related systems.

\acknowledgments

We are very grateful to Luis Fernando Alday, Tobias Hansen and Joao Silva for stimulating discussions, and for sharing some of their unpublished work with us.
We thank Dionysios Anninos, Andrea Cavagli\`a, Jeremy Mann,  Sameer Murthy, Michelangelo Preti and Dmytro Volin for discussions related to some parts of this work. We thank Andrea Cavagli\`a for beta testing the GitHub repository. 
The work of  \'A.H. was supported by the NKFIH grant K134946.
The work of N.G., J.J. and N.S. was supported by the European Research Council (ERC) under the European Union’s Horizon 2020 research and innovation programme (grant agreement No. 865075) EXACTC. The work of N.G. was also partially supported by the STFC grant (ST/P000258/1).

\appendix

\newpage
\section{Spectral data for all states in planar ${\cal N} = 4$ SYM with $\Delta_0\leq 6$}
In this appendix we give the list of all the states which we studied, i.e. all states with bare dimension $\leq 6$.
In order to help with the identification of the states we provide in the table below their oscillator numbers, quantum numbers, one loop anomalous dimension and the first two strong coupling orders: sting level $\delta$ and the one loop analytic guess for the correction $d_1$. The states of the types II, III and IV, which brakes some of the discrete symmetries are paired with the states with exactly the same $\Delta(g)$ in the last column. The numerical values of the $\Delta$ can be found on \GitHub
 repository.

{
\scriptsize
\setlength\LTleft{-17mm}\setlength\LTright{-0mm}
\begin{longtable}{@{\extracolsep{\fill}}|C|C|C|C|C|C|C|C|C|C|C|@{}}
\hline
\text{{\tt St.} {\tt No.}} & \texttt{State ID} & [\ell_1\;\ell_2\;q_1\;p\;q_2] & \Delta _ 0 & \Delta _ 0\text{ \
+ $\#$ }g^2 & \delta & 2\,\sqrt{\delta } \lambda ^{1/4} - 2
 & d_1 & j_1 & \text{Type} & \text{Degs.} \\

 & &  &  &  &  & + \frac{\#}{\sqrt{\delta }}\frac{1}{\lambda ^{1/4}} & & & & \\[3pt]
\hline\hline
 1 & \text{}_ 2 \text{[0 0 1 1 1 1 0 0]}_ 1 & \text{[0 0 0 0 0]} & 2 & 12 & 1 & 2.0000000000000 & 2 & 2 & \text{I} & \text{} \\ \hline
 2 & \text{}_ 3 \text{[0 0 2 2 1 1 0 0]}_ 1 & \text{[0 0 0 1 0]} & 3 & 8 & 1 & 3.250000001 & \frac{13}{4} & 2 & \text{I} & \text{} \\ \hline
 3 & \text{}_ 4 \text{[0 0 2 2 2 2 0 0]}_ 1 & \text{[0 0 0 0 0]} & 4 & 13-\sqrt{41} & 2 & 1.9999999 & 2 & 2 & \text{I} & \text{} \\ \hline
 4 & \text{}_ 4 \text{[0 0 2 2 2 2 0 0]}_ 2 & \text{[0 0 0 0 0]} & 4 & 13+\sqrt{41} & 2 & 8.00000 & 8 & 14 & \text{I} & \text{} \\ \hline
 5 & \text{}_ 4 \text{[0 0 3 2 2 1 0 0]}_ 1 & \text{[0 0 1 0 1]} & 4 & 12 & 2 & 4.000000 & 4 & 2 & \text{I} & \text{} \\ \hline
 6 & \text{}_ 4 \text{[0 0 3 3 1 1 0 0]}_ 1 & \text{[0 0 0 2 0]} & 4 & 10+2 \sqrt{5} & 2 & 4.9999998 & 5 & 2 & \text{I} & \text{} \\
\end{longtable}}

\section{Conversion to the left-right symmetric form}
\label{app:LRconvert}
This appendix contains technical details on how to convert the perturbative results of \cite{Marboe:2017dmb, Marboe:2018ugv} into our conventions in order to be used as starting points for the numerical procedure.

The perturbative solver \cite{Marboe:2017dmb, Marboe:2018ugv} does not present $\mathbf{P}_{a}$ and $\mathbf{P}^{a}$ in the left-right symmetric form for states which have this symmetry. In order to use the perturbative data from the solver, we need to convert it to the left-right symmetric form,
{\it i.e.}~so that $\bP^a = \chi^{a\,b}\bP_b$,
by a composition of $\Lambda$- and $H$-transformations~({\it cf.} equations
\eqref{LambdaTrans}- \eqref{HtransUp}).

Let $\bP_a^{\tt MV}$ and $\bP^{a\;{\tt MV}}$ be the $\bP$-functions obtained from the perturbative QSC solver of~
\cite{Marboe:2017dmb, Marboe:2018ugv}. Then, the Ansatz looks as follows in their conventions
\begin{gather}
\mathbf{P}^{\texttt{MV}}_{a} = x^{\texttt{powP}_{a}^{\texttt{MV}}} \left(\mathbb{A}_{a} + \sum_{n=1}^{\infty} \frac{c_{a, n}^{\texttt{MV}}}{x^{n}} \right),\\
\mathbf{P}^{a, \texttt{MV}} = x^{-\texttt{powP}_{a}^{\texttt{MV}}-1} \left(\mathbb{A}^{a} + \sum_{n=1}^{\infty} \frac{c^{a, n, \ \texttt{MV}}}{x^{n}} \right).
\end{gather}

We use the $\Lambda$-symmetry ($\texttt{powP}^{\texttt{MV}}_{a} \to \texttt{powP}^{\texttt{MV}}_{a}-\Lambda = \texttt{powP}_{a}$) in order to ensure the asymptotics of the $\bP_a$ and $\bP^a$ satisfy the left-right symmetric condition~\eqref{lr}.
Since for LR symmetric states we have
\begin{align}
\begin{split}
    {\tt powP}_1 &= - {\tt powP}_4 - 1\;,\qquad {\tt powP}_2 = - {\tt powP}_3 - 1\;,\\
    {\tt powP}_3 &= - {\tt powP}_2 - 1\;,\qquad {\tt powP}_4 = - {\tt powP}_1 - 1\;,
\end{split}
\end{align}
we need to solve equations of the form
\begin{align}
    {\tt powP}_1^{\texttt{MV}}  - \Lambda = - ({\tt powP}_4^{\texttt{MV}}  - \Lambda) - 1
    \;
\end{align}
and similarly for other $\bP_a$ to get
\begin{equation}
\label{eqn:LambdaLR}
    \Lambda = \frac{1 + \texttt{powP}_1^{\texttt{MV}} + \texttt{powP}_4^{\texttt{MV}}}{2} = \frac{1 + \texttt{powP}_2^{\texttt{MV}} + \texttt{powP}_4^{\texttt{MV}}}{2}.
\end{equation}
Such a $\Lambda$ can only be found if
\begin{align}
    {\tt powP}_1^{\texttt{MV}} + {\tt powP}_4^{\texttt{MV}} = {\tt powP}_2^{\texttt{MV}} + {\tt powP}_3^{\texttt{MV}}
    \;,
\end{align}
or in terms of oscillator numbers
\begin{align}
\label{eqn:LRcondition}
    n_{{\bf f}_1} + n_{{\bf f}_4} = n_{{\bf f}_2} + n_{{\bf f}_3}\;.
\end{align}
It must be noted, however, that this is a necessary and not sufficient condition for left-right symmetry. Indeed, we can have non left-right symmetric states, whose oscillator numbers satisfy the equation~\eqref{eqn:LRcondition}.

 Let us denote the $\bP$-functions obtained after performing the $\Lambda$-transformation~\eqref{eqn:LambdaLR} by $\bP_a^{\tt Lambda}$ and $\bP^{a\;{\tt Lambda}}$.
Now that the asymptotic have been fixed, we are looking for an $H$-transformation \eq{Htrans}, \eq{HtransUp}
 \begin{gather}
 \label{eqn:HLR}
\mathbf{P}_{a}^{\texttt{LR}} = H_{a}^{b}\, \mathbf{P}_{b}^{\tt Lambda}  \\
\mathbf{P}^{a\ \texttt{LR}} = \mathbf{P}^{c\;{\tt Lambda}}\,(H^{-1})_{c}^{a} \nonumber
\end{gather}
which brings $\mathbf{P}_{a}^{\texttt{Lambda}}$ and $\mathbf{P}^{a, \texttt{Lambda}}$ to the form that manifests left-right symmetry, {\it i.e.}
\begin{equation}
\label{eqn:chiLR}
    \mathbf{P}^{\texttt{LR},\ a} = \chi^{a b} \mathbf{P}^{\texttt{LR}}_{b}.
\end{equation}
In order to ensure that the $H$-transformation does not change the asymptotic of $\bP_a^{\tt Lambda}$ and $\bP^{a\;{\tt Lambda}}$, it must be lower triangular.
Thus, we need to find 10 lower triangular matrix elements $H_{a}^{\;b}$.
Plugging~\eqref{eqn:HLR} into~\eqref{eqn:chiLR}, we get
\begin{equation}
\label{eqn:PMP}
    \mathbf{P}^{a\;{\tt Lambda}} = (H^{T}.\chi.H)^{ab} \mathbf{P}_{b}^{\tt Lambda}\;,
\end{equation}
where due to $H$ being lower triangular and the anti-symmetry of $\chi,$ $H^{T}.\chi.H$ can be parameterised only by 4 independent elements:
\begin{gather}
H^{T}.\chi.H  = \begin{pmatrix}
0 & h_1 & h_2 & h_3\\
-h_1 & 0 & h_4 & 0\\
-h_2 & -h_4 & 0 & 0\\
-h_3 & 0 & 0 & 0
\end{pmatrix}\;,
\end{gather}
where $h_i$ can be determined from \eqref{eqn:PMP} in terms of the coefficients of the $\mathbf{P}^{\texttt{MV}}$-functions:
\begin{align}
\begin{split}
    h_1 &= \frac{c^{1,\;\texttt{powP}_4 - \texttt{powP}_2\ \texttt{MV}}}{\mathbb{A}_2} = - \frac{c^{2,\;\texttt{powP}_3 - \texttt{powP}_1\ \texttt{MV}}}{\mathbb{A}_{1}}
    \;, \quad
    h_2 = \frac{c^{1,\;\texttt{powP}_4 - \texttt{powP}_3\ \texttt{MV}}}{\mathbb{A}_{3}} = - \frac{c^{3,\;\texttt{powP}_2 - \texttt{powP}_1\ \texttt{MV}}}{\mathbb{A}_{1}}\;, \\
    h_3 &= \frac{\mathbb{A}^{1}}{\mathbb{A}_{4}}=-\frac{\mathbb{A}^{4}}{\mathbb{A}_{1}}
    \;, \qquad\qquad\qquad\qquad\quad\;\;\,
    h_4 = \frac{\mathbb{A}^{2}}{\mathbb{A}_{3}}=-\frac{\mathbb{A}^{3}}{\mathbb{A}_{2}}\;.
\end{split}
\end{align}
From \eqref{eqn:PMP} the relation of $h_i$  to the matrix elements $H_{a}^{\;b}$ can be read off, as well:
\begin{align}
    h_1 &= -H_{2}^{\;2}H_{3}^{\;1}+H_{2}^{\;1}H_{3}^{\;2}-H_{1}^{\;1}H_{4}^{\;2}\;,
    \qquad
    h_2 = H_{2}^{\;1}H_{3}^{\;3}-H_{1}^{\;1}H_{4}^{\;3}\;, \\
    h_3 &= -H_{1}^{\;1}H_{4}^{\;4}\;,
    \qquad\qquad\qquad\qquad\qquad\qquad
    h_4 = H_{2}^{\;2}H_{3}^{\;3}\;.
\end{align}
Thus, we have $4$ conditions on the $10$ elements of the lower triangular matrix $H$.
We have the freedom to fix the other $6$ conditions.
We have chosen to fix $4$ gauge conditions by imposing \eqref{eqn:LRgauge},
from where it follows that $H_{2,1}=0, H_{3,1}=0, H_{3,2}=0, H_{4,1}=0$.
The remaining two conditions are fixed by imposing
 $H_{1}^{\;1} = H_{2}^{\;2} = 1$, so that  $H$ takes the final form as follows:
\begin{equation}
H = \begin{pmatrix}
1 & 0 & 0 & 0\\
0 & 1 & 0 & 0\\
0 & 0 & h_4 & 0\\
0 & -h_1 & -h_2 & h_3
\end{pmatrix}\;.
\end{equation}
In the repository we have added a {\it Mathematica} notebook which performs the conversion as described in this appendix.

\section{One-loop spectrum of $\mathfrak{sl}(2)$ sector and mode numbers}
\label{app:sl2}
The one-loop anomalous dimension of $\mathfrak{sl}(2)$ operators at weak coupling may be obtained by solving the $\mathfrak{sl}(2)$ Bethe ansatz equations (BAE)~\cite{Faddeev:1996iy,Beisert:2004ry}.
A multiplet is characterised by a number of derivatives $S$ and a length $L,$ via~\eqref{eq:Sl2}. To obtain the one-loop dimension, we need solve the following set of equations: the $\mathfrak{sl}(2)$ BAE
\begin{align}
\label{eq:Bethe}
\left( \frac{v_{k} + i/2}{v_k - i/2} \right)^{L} = \prod_{l=1, l\neq k}^{S} \frac{v_k - v_l - i}{v_k - v_l + i}\;.
\end{align}
These equations are solved by a set of Bethe roots $\{v_1,\dots,\, v_S\}$. The number of different Bethe root sets corresponds to the multiplicity of the multiplet described by $S$ and $L$, via~\eqref{eq:Sl2}.
Thus, we can associate each set of Bethe roots with a particular value of the multiplicity label \texttt{sol}. Denote such a set as $\{v_1,\dots,\, v_S\}_{\tt sol}$.

Given a set of Bethe roots, the one-loop anomalous dimension of the state can be obtained from~\cite{Faddeev:1996iy}
\begin{align}
    \Delta \simeq L + S - 2 +\sum_{v_k \in \{v_1,\dots,\, v_S\}_{\tt sol}} \frac{2\,g^2}{v_k^2 + 1/4} \;.
\end{align}
Another way of classifying solutions to the BAE is by using ``mode numbers''. We explain how in the sequel. Consider the
logarithm of the equations \eqref{eq:Bethe}.
We need to specify the branch of the logarithm, and this introduces an integer constant $n_k$, associated with the logarithm of the $k^\text{th}$ Bethe equation. We have
\begin{gather}
    L \log{\left( \frac{v_{k} + i/2}{v_k - i/2} \right)} = \sum_{l = 1, l\neq k}^{S}
    \log \bigg[
    \frac{v_k - v_l - i}{v_k - v_l + i} \bigg] + 2 \pi i\ {n}_k\;.
\end{gather}
Thus, to the solution $\{v_1,\dots,\, v_S\}_\texttt{sol}$, we can associate a set $\{n_1,\dots,\,n_S\}_\texttt{sol}$, which we call the mode numbers of this solution of the BAE, and consequently the state described by them. The mode numbers can be used as an alternative multiplicity label, in lieu of \texttt{sol}, for $\mathfrak{sl}(2)$ multiplets.

Below, in table~\ref{tab:Sl2StatesModeVectors}, we present a table of mode numbers, for  all the 27 ${\mathfrak {sl} }(2)$ states in our database.

{
\scriptsize
\setlength\LTleft{-12mm}\setlength\LTright{-0mm}
\begin{longtable}{@{\extracolsep{\fill}}|C|C|C|C|C|C|C|C|C|C|C|@{}}
\hline
\text{{\tt St.} {\tt No.}} & \texttt{State ID} & [\ell_1\;\ell_2\;q_1\;p\;q_2] & \Delta _ 0 & \Delta _ 0\text{ \
+ $\#$ }g^2 & \delta & d_1 & [S\;L\;n] & \text{Mode numbers} & \text{Type} & \text{Degs.} \\

 &  &  &  &  &  &  &  & & & \\[3pt]
\hline\hline
\rowcolor{Apricot} 1 & \text{}_ 2 \text{[0 0 1 1 1 1 0 0]}_ 1 & \text{[0 0 0 0 0]} & 2 & 12 & 1 & 2 & \text{[2 2 1]} & \{1,-1\} & \text{I} & \text{} \\
\rowcolor{Apricot} 2 & \text{}_ 3 \text{[0 0 2 2 1 1 0 0]}_ 1 & \text{[0 0 0 1 0]} & 3 & 8 & 1 & \frac{13}{4} & \text{[2 3 1]} & \{1,-1\} & \text{I} & \text{} \\
\rowcolor{Apricot} 7 & \text{}_ 4 \text{[0 0 3 3 1 1 0 0]}_ 2 & \text{[0 0 0 2 0]} & 4 & 10-2 \sqrt{5} & 1 & 5 & \text{[2 4 1]} & \{1,-1\} & \text{I} & \text{} \\
\rowcolor{SeaGreen} 6 & \text{}_ 4 \text{[0 0 3 3 1 1 0 0]}_ 1 & \text{[0 0 0 2 0]} & 4 & 10+2 \sqrt{5} & 2 & 5 & \text{[2 4 2]} & \{2,-2\} & \text{I} & \text{} \\
\rowcolor{Apricot} 23 & \text{}_ 5 \text{[0 0 4 4 1 1 0 0]}_ 1 & \text{[0 0 0 3 0]} & 5 & 4 & 1 & \frac{29}{4} & \text{[2 5 1]} & \{1,-1\} & \text{I} & \text{} \\
\rowcolor{SeaGreen} 24 & \text{}_ 5 \text{[0 0 4 4 1 1 0 0]}_ 2 & \text{[0 0 0 3 0]} & 5 & 12 & 2 & \frac{29}{4} & \text{[2 5 2]} & \{2,-2\} & \text{I} & \text{} \\
\rowcolor{Apricot} 118 & \text{}_ 6 \text{[0 0 5 5 1 1 0 0]}_ 1 & \text{[0 0 0 4 0]} & 6 & 3.012081585 & 1 & 10 & \text{[2 6 1]} & \{1,-1\} & \text{I} & \text{} \\
\rowcolor{SeaGreen} 119 & \text{}_ 6 \text{[0 0 5 5 1 1 0 0]}_ 2 & \text{[0 0 0 4 0]} & 6 & 9.780167472 & 2 & 10 & \text{[2 6 2]} & \{2,-2\} & \text{I} & \text{} \\
\rowcolor{Thistle} 120 & \text{}_ 6 \text{[0 0 5 5 1 1 0 0]}_ 3 & \text{[0 0 0 4 0]} & 6 & 15.20775094 & 3 & 9 & \text{[2 6 3]} & \{3,-3\} & \text{I} & \text{} \\
 9 & \text{}_ 4 \text{[0 1 2 2 1 1 1 0]}_ 1 & \text{[1 1 0 1 0]} & 4 & 15 & 2 & \frac{11}{2} & \text{[3 3 1]} & \{-1,-1,2\} & \text{II} & 10 \\
 10 & \text{}_ 4 \text{[0 1 2 2 1 1 1 0]}_ 2 & \text{[1 1 0 1 0]} & 4 & 15 & 2 & \frac{11}{2} & \text{[3 3 1]} & \{1,1,-2\} & \text{II} & 9 \\
 34 & \text{}_ 5 \text{[0 1 3 3 1 1 1 0]}_ 1 & \text{[1 1 0 2 0]} & 5 & 12 & 2 & \frac{29}{4} & \text{[3 4 1]} & \{-1,-1,2\} & \text{II} & 35 \\
 35 & \text{}_ 5 \text{[0 1 3 3 1 1 1 0]}_ 2 & \text{[1 1 0 2 0]} & 5 & 12 & 2 & \frac{29}{4} & \text{[3 4 1]} & \{1,1,-2\} & \text{II} & 34 \\
 173 & \text{}_ 6 \text{[0 1 4 4 1 1 1 0]}_ 3 & \text{[1 1 0 3 0]} & 6 & \frac{25}{2}-\frac{\sqrt{37}}{2} & 2 & \frac{19}{2} & \text{[3 5 1]} & \{3,-2,-1\} & \text{II} & 174 \\
 174 & \text{}_ 6 \text{[0 1 4 4 1 1 1 0]}_ 4 & \text{[1 1 0 3 0]} & 6 & \frac{25}{2}-\frac{\sqrt{37}}{2} & 2 & \frac{19}{2} & \text{[3 5 1]} & \{1,1,-2\} & \text{II} & 173 \\
 171 & \text{}_ 6 \text{[0 1 4 4 1 1 1 0]}_ 1 & \text{[1 1 0 3 0]} & 6 & \frac{25}{2}+\frac{\sqrt{37}}{2} & 3 & 10 & \text{[3 5 2]} & \{3,-2,-1\} & \text{II} & 172 \\
 172 & \text{}_ 6 \text{[0 1 4 4 1 1 1 0]}_ 2 & \text{[1 1 0 3 0]} & 6 & \frac{25}{2}+\frac{\sqrt{37}}{2} & 3 & 10 & \text{[3 5 2]} & \{1,2,-3\} & \text{II} & 171 \\
\rowcolor{Apricot} 12 & \text{}_ 4 \text{[0 2 1 1 1 1 2 0]}_ 1 & \text{[2 2 0 0 0]} & 4 & \frac{50}{3} & 2 & 6 & \text{[4 2 1]} & \{1,1,-1,-1\} & \text{I} & \text{} \\
\rowcolor{Apricot} 39 & \text{}_ 5 \text{[0 2 2 2 1 1 2 0]}_ 1 & \text{[2 2 0 1 0]} & 5 & 12 & 2 & \frac{29}{4} & \text{[4 3 1]} & \{1,1,-1,-1\} & \text{I} & \text{} \\
\rowcolor{Apricot} 206 & \text{}_ 6 \text{[0 2 3 3 1 1 2 0]}_ 3 & \text{[2 2 0 2 0]} & 6 & 8.765533583 & 2 & 9 & \text{[4 4 1]} & \{1,1,-1,-1\} & \text{I} & \text{} \\
 205 & \text{}_ 6 \text{[0 2 3 3 1 1 2 0]}_ 2 & \text{[2 2 0 2 0]} & 6 & 16.71846288 & 3 & 11 & \text{[4 4 2]} & \{1,2,-2,-1\} & \text{I} & \text{} \\
 207 & \text{}_ 6 \text{[0 2 3 3 1 1 2 0]}_ 4 & \text{[2 2 0 2 0]} & 6 & \frac{46}{3} & 3 & 11 & \text{[4 4 2]} & \{1,1,1,1\} & \text{II} & 208 \\
 208 & \text{}_ 6 \text{[0 2 3 3 1 1 2 0]}_ 5 & \text{[2 2 0 2 0]} & 6 & \frac{46}{3} & 3 & 11 & \text{[4 4 2]} & \{-1,-1,-1,-1\} & \text{II} & 207 \\
\rowcolor{SeaGreen} 204 & \text{}_ 6 \text{[0 2 3 3 1 1 2 0]}_ 1 & \text{[2 2 0 2 0]} & 6 & 23.18267020 & 4 & 12 & \text{[4 4 3]} & \{2,2,-2,-2\} & \text{I} & \text{} \\
 214 & \text{}_ 6 \text{[0 3 2 2 1 1 3 0]}_ 1 & \text{[3 3 0 1 0]} & 6 & \frac{35}{2} & 3 & 12 & \text{[5 3 1]} & \{1,1,1,-2,-1\} & \text{II} & 215 \\
 215 & \text{}_ 6 \text{[0 3 2 2 1 1 3 0]}_ 2 & \text{[3 3 0 1 0]} & 6 & \frac{35}{2} & 3 & 12 & \text{[5 3 1]} & \{1,2,-1,-1,-1\} & \text{II} & 214 \\
\rowcolor{Apricot} 219 & \text{}_ 6 \text{[0 4 1 1 1 1 4 0]}_ 1 & \text{[4 4 0 0 0]} & 6 & \frac{98}{5} & 3 & 13 & \text{[6 2 1]} & \{1,1,1,-1,-1,-1\} & \text{I} & \text{} \\
\hline
\caption{\small\it    Mode numbers for $\mathfrak{sl}(2)$ states in our database.}
\label{tab:Sl2StatesModeVectors}
\end{longtable}
}
\noindent
In table~\ref{tab:Sl2StatesModeVectors} above, consider the states in shaded rows. Entries shaded with

\begin{tabular}{l }
\fcolorbox{black}{Apricot}{\rule{0pt}{6pt}\rule{10pt}{0pt}} \quad have mode numbers $\pm 1$\;, \\
\fcolorbox{black}{SeaGreen}{\rule{0pt}{6pt}\rule{10pt}{0pt}} \quad have mode numbers $\pm 2$\;, \\
\fcolorbox{black}{Thistle}{\rule{0pt}{6pt}\rule{10pt}{0pt}} \quad have mode numbers $\pm 3$\;.  \\
\end{tabular}

\noindent

\section{Counting of states at higher $\delta$}
\label{app:higherdelta}
Here we present details of the counting of the states at strong coupling as discussed in section~\ref{sec:KKtowers}. 
In particular, we assign states in our database, with $\delta = 2$ and $\delta = 3$, to KK-towers available from formulas in the literature \cite{Alday:2023flc, AHSprivate}. In some cases, where there is an ambiguity, we use the conjecture~\eqref{eqn:d1pred}, to break this ambiguity. Therefore, we show that the subleading Casimir $j_1$, is a good classifier of KK-towers.
\subsection{States with $\delta = 2$}
\label{app:delta2states}
As we mentioned in~\ref{sec:KKtowers} there are $9$ KK-towers expected at the level $\delta = 2$.
For the convenience of the reader, we display the counting from~\cite{Alday:2023flc}, {\it i.e.}~equation~\eqref{eqn:count2} below:
\begin{align*}
    \mathtt{count}_2 =
    2\,[0\;0 ;\; 0\;0]
    +[0\;0 ;\; 2\;0]
    +[0\;0 ;\; 0\;2]
    +2\,[1\;1 ;\; 1\;0]
    +[2\;2 ;\; 0\;0]
    +[2\;0 ;\; 0\;0]
    +[0\;2 ;\; 0\;0]
    \;.
\end{align*}
First consider the case of zero spin. For the Lorentz spin labels $(\ell_1\;\ell_2) = (0\;0)$, we have the following states, presented in table~\ref{tab:delta2l00}:

{
\scriptsize
\setlength\LTleft{-6mm}\setlength\LTright{-0mm}
\begin{longtable}{@{\extracolsep{\fill}}|C|C|C|C|C|C|C|C|C|C|C|@{}}
\hline
\text{{\tt St.} {\tt No.}} & \texttt{State ID} & [\ell_1\;\ell_2\;q_1\;p\;q_2] & \Delta _ 0 & \Delta _ 0\text{ \
+ $\#$ }g^2 & \delta & 2\,\sqrt{\delta } \lambda ^{1/4} - 2
 & d_1 & j_1 & \text{Type} & \text{Degs.} \\

 & &  &  &  &  & + \frac{\#}{\sqrt{\delta }}\frac{1}{\lambda ^{1/4}} & & & & \\[3pt]
\hline\hline
\rowcolor{LimeGreen} 3 & \text{}_ 4 \text{[0 0 2 2 2 2 0 0]}_ 1 & \text{[0 0 0 0 0]} & 4 & \
13-\sqrt{41} & 2 & 1.9999999 & 2 & 2 &  \text{I} & \text{} \\
\rowcolor{LimeGreen} 4 & \text{}_ 4 \text{[0 0 2 2 2 2 0 0]}_ 2 & \text{[0 0 0 0 0]} & 4 \
& 13+\sqrt{41} & 2 & 8.0000 & 8 & 14 &  \text{I} & \text{} \\
\rowcolor{CornflowerBlue} 5 & \text{}_ 4 \text{[0 0 3 2 2 1 0 0]}_ 1 & \text{[0 0 1 0 1]} & 4 \
& 12 & 2 & 4.000000 & 4 & 2 & \text{I} & \text{} \\
\rowcolor{Goldenrod} 6 & \text{}_ 4 \text{[0 0 3 3 1 1 0 0]}_ 1 & \text{[0 0 0 2 0]} & 4 \
& 10+2 \sqrt{5} & 2 & 4.9999998 & 5 & 2 &  \text{I} & \
\text{} \\
\rowcolor{CornflowerBlue} 13 & \text{}_ 5 \text{[0 0 3 3 3 1 0 0]}_ 1 & \text{[0 0 0 0 2]} & 5 \
& 6.788897449 & 2 & 4.25 & \text{} & 2.00 & \text{III} & 15 \
\\
\rowcolor{CornflowerBlue} 15 & \text{}_ 5 \text{[0 0 4 2 2 2 0 0]}_ 1 & \text{[0 0 2 0 0]} & 5 \
& 6.788897449 & 2 & 4.25 & \text{} & 2.00 & \text{III} & 13 \
\\
\rowcolor{LimeGreen} 17 & \text{}_ 5 \text{[0 0 3 3 2 2 0 0]}_ 1 & \text{[0 0 0 1 0]} & 5 \
& 5.527864045 & 2 & 3.2500000 & \frac{13}{4} & 2 & \text{I} \
& \text{} \\
\rowcolor{LimeGreen} 18 & \text{}_ 5 \text{[0 0 3 3 2 2 0 0]}_ 2 & \text{[0 0 0 1 0]} & 5 \
& 14.47213595 & 2 & 9.249999 & \frac{37}{4} & 14 & \text{I} \
& \text{} \\
 21 & \text{}_ 5 \text{[0 0 4 3 2 1 0 0]}_ 1 & \text{[0 0 1 1 1]} & 5 \
& 10 & 2 & 5.75007 & \frac{23}{4} & 2 & \text{II} & 22 \\
 22 & \text{}_ 5 \text{[0 0 4 3 2 1 0 0]}_ 2 & \text{[0 0 1 1 1]} & 5 \
& 10 & 2 & 5.750 & \frac{23}{4} & 2 & \text{II} & 21 \\
\rowcolor{Goldenrod} 24 & \text{}_ 5 \text{[0 0 4 4 1 1 0 0]}_ 2 & \text{[0 0 0 3 0]} & 5 \
& 12 & 2 & 7.24999996 & \frac{29}{4} & 2 &  \text{I} \
& \text{} \\
\rowcolor{Goldenrod} 97 & \text{}_ 6 \text{[0 0 5 3 3 1 0 0]}_ 2 & \text{[0 0 2 0 2]} & 6 \
& 6.491188584 & 2 & 7.00000 & 7 & 2 & \text{I} & \text{} \\
\rowcolor{CornflowerBlue} 99 & \text{}_ 6 \text{[0 0 4 4 3 1 0 0]}_ 1 & \text{[0 0 0 1 2]} & 6 \
& 5.395626364 & 2 & 6.000 & 6 & 2 &  \text{III} & 102 \\
\rowcolor{CornflowerBlue} 102 & \text{}_ 6 \text{[0 0 4 4 3 1 0 0]}_ 1 & \text{[0 0 2 1 0]} & 6 \
& 5.395626364 & 2 & 6.000 & 6 & 2 &  \text{III} & 99 \\
\rowcolor{LimeGreen} 107 & \text{}_ 6 \text{[0 0 4 4 2 2 0 0]}_ 3 & \text{[0 0 0 2 0]} & \
6 & 10.67351551 & 2 & 11.000000 & 11 & 14 &  \text{I} & \text{} \\
\rowcolor{LimeGreen} 109 & \text{}_ 6 \text{[0 0 4 4 2 2 0 0]}_ 5 & \text{[0 0 0 2 0]} & \
6 & 4.524563121 & 2 & 5.00000 & 5 & 2 &  \text{I} & \text{} \\
 116 & \text{}_ 6 \text{[0 0 5 4 2 1 0 0]}_ 2 & \text{[0 0 1 2 1]} & \
6 & 8 & 2 & 8.000 & 8 & 2 &  \text{II} & 117 \\
 117 & \text{}_ 6 \text{[0 0 5 4 2 1 0 0]}_ 3 & \text{[0 0 1 2 1]} & \
6 & 8 & 2 & 8.000 & 8 & 2 &  \text{II} & 116 \\
\rowcolor{Goldenrod} 119 & \text{}_ 6 \text{[0 0 5 5 1 1 0 0]}_ 2 & \text{[0 0 0 4 0]} & \
6 & 9.780167472 & 2 & 10.00000000 & 10 & 2 &  \text{I} \
& \text{} \\
\hline
\caption{\small\it    States in our database with $\delta = 2$ and $(\ell_1\;\ell_2) = (0\;0)$. We predict that the states in rows with the same colour belong to the same KK-tower.}
\label{tab:delta2l00}
\end{longtable}
}
\noindent
We claim that in table~\ref{tab:delta2l00}, the entries shaded with

\begin{tabular}{l l}
\multicolumn{2}{l}{\fcolorbox{black}{LimeGreen}{\rule{0pt}{6pt}\rule{10pt}{0pt}} \quad belong to one of the 2 $[0\;0;\;0\;0]$\;, with the value of $j_1$ breaking the ambiguity,}
\\
\fcolorbox{black}{CornflowerBlue}{\rule{0pt}{6pt}\rule{10pt}{0pt}} \quad belong to $[0\;0;\;0\;2]$\;, &
\fcolorbox{black}{Goldenrod}{\rule{0pt}{6pt}\rule{10pt}{0pt}} \quad belong to $[0\;0;\;2\;0]$\;, \\
\multicolumn{2}{l}{\fcolorbox{black}{white}{\rule{0pt}{6pt}\rule{10pt}{0pt}} \quad belong to either $[0\ 0;\ 2\ 0]$, $[0\ 0;\ 0\ 2]$
but it can't be identified uniquely.}
\\
\end{tabular}
\noindent
In order to assign some of the $[0\;0\;0\;p\;0]$, with $p\geq 2$ states to the $[0\;0;\;2\;0]$, we also used the fact that they are all $\mathfrak{sl}(2)$ states of the form $[2,\;p+2,\;2]$, and assumed that these are in the same KK-tower.

Next, let us consider states with Lorentz spin labels $(1\; 1)$. There are 10 such states, presented in table~\ref{tab:delta2l11}. We have

{
\scriptsize
\setlength\LTleft{-4mm}\setlength\LTright{-0mm}
\begin{longtable}{@{\extracolsep{\fill}}|C|C|C|C|C|C|C|C|C|C|C|@{}}
\hline
\text{{\tt St.} {\tt No.}} & \texttt{State ID} & [\ell_1\;\ell_2\;q_1\;p\;q_2] & \Delta _ 0 & \Delta _ 0\text{ \
+ $\#$ }g^2 & \delta & 2\,\sqrt{\delta } \lambda ^{1/4} - 2
 & d_1 & j_1 & \text{Type} & \text{Degs.} \\

 & &  &  &  &  & + \frac{\#}{\sqrt{\delta }}\frac{1}{\lambda ^{1/4}} & & & & \\[3pt]
\hline\hline
\rowcolor{LimeGreen} 9 & \text{}_ 4 \text{[0 1 2 2 1 1 1 0]}_ 1 & \text{[1 1 0 1 0]} & 4 & \
15 & 2 & 5.50 & \frac{11}{2} & 8 & \text{II} & 10 \\
\rowcolor{LimeGreen} 10 & \text{}_ 4 \text{[0 1 2 2 1 1 1 0]}_ 2 & \text{[1 1 0 1 0]} & 4 \
& 15 & 2 & 5.4999 & \frac{11}{2} & 8  & \text{II} & 9 \
\\
\rowcolor{LimeGreen} 30 & \text{}_ 5 \text{[0 1 3 2 2 1 1 0]}_ 1 & \text{[1 1 1 0 1]} & 5 \
& 10 & 2 & 6.2500 & \frac{25}{4} & 8  & \text{II} & 31 \\
\rowcolor{LimeGreen} 31 & \text{}_ 5 \text{[0 1 3 2 2 1 1 0]}_ 2 & \text{[1 1 1 0 1]} & 5 \
& 10 & 2 & 6.2500 & \frac{25}{4} & 8 & \text{II} & 30 \\
\rowcolor{LimeGreen} 34 & \text{}_ 5 \text{[0 1 3 3 1 1 1 0]}_ 1 & \text{[1 1 0 2 0]} & 5 \
& 12 & 2 & 7.24998 & \frac{29}{4} & 8 & \text{II} & \
35 \\
\rowcolor{LimeGreen} 35 & \text{}_ 5 \text{[0 1 3 3 1 1 1 0]}_ 2 & \text{[1 1 0 2 0]} & 5 \
& 12 & 2 & 7.24998 & \frac{29}{4} & 8 & \text{II} & \
34 \\
\rowcolor{LimeGreen} 167 & \text{}_ 6 \text{[0 1 4 3 2 1 1 0]}_ 5 & \text{[1 1 1 1 1]} & \
6 & 7.429856598 & 2 & 7.999 & 8 & 8 & \text{II} & 168 \\
\rowcolor{LimeGreen} 168 & \text{}_ 6 \text{[0 1 4 3 2 1 1 0]}_ 6 & \text{[1 1 1 1 1]} & \
6 & 7.429856598 & 2 & 7.999 & 8 & 8 & \text{II} & 167 \\
\rowcolor{LimeGreen} 173 & \text{}_ 6 \text{[0 1 4 4 1 1 1 0]}_ 3 & \text{[1 1 0 3 0]} & \
6 & \frac{25}{2}-\frac{\sqrt{37}}{2} & 2 & 9.5000 & \frac{19}{2} & 8 \
 & \text{II} & 174 \\
\rowcolor{LimeGreen} 174 & \text{}_ 6 \text{[0 1 4 4 1 1 1 0]}_ 4 & \text{[1 1 0 3 0]} & \
6 & \frac{25}{2}-\frac{\sqrt{37}}{2} & 2 & 9.5000 & \frac{19}{2} & 8 \
 & \text{II} & 173 \\
\hline
\caption{\small\it    States in our database with $\delta = 2$ and $(\ell_1\;\ell_2) = (1\;1)$.}
\label{tab:delta2l11}
\end{longtable}
}
\noindent

In table~\ref{tab:delta2l11}, the entries shaded with

\begin{tabular}{l l}
\fcolorbox{black}{LimeGreen}{\rule{0pt}{6pt}\rule{10pt}{0pt}} \quad belong to one of the 2 $[1\;1;\;1\;0]$\;.
\end{tabular}

\noindent

Next, we consider states with Lorentz spin labels $(0\;2)$ and $(2\;0)$. There are 6 such states, presented in table~\ref{tab:delta2l0220}. We have

{
\scriptsize
\setlength\LTleft{-2mm}\setlength\LTright{-0mm}
\begin{longtable}{@{\extracolsep{\fill}}|C|C|C|C|C|C|C|C|C|C|C|@{}}
\hline
\text{{\tt St.} {\tt No.}} & \texttt{State ID} & [\ell_1\;\ell_2\;q_1\;p\;q_2] & \Delta _ 0 & \Delta _ 0\text{ \
+ $\#$ }g^2 & \delta & 2\,\sqrt{\delta } \lambda ^{1/4} - 2
 & d_1 & j_1 & \text{Type} & \text{Degs.} \\

 & &  &  &  &  & + \frac{\#}{\sqrt{\delta }}\frac{1}{\lambda ^{1/4}} & & & & \\[3pt]
\hline\hline
\rowcolor{LimeGreen} 8 & \text{}_ 4 \text{[0 0 1 1 1 1 2 0]}_ 1 & \text{[0 2 0 0 0]} & 4 & \
18 & 2 & 7.000 & 7 & 14  & \text{III} & 11 \\
\rowcolor{LimeGreen} 27 & \text{}_ 5 \text{[0 0 2 2 1 1 2 0]}_ 2 & \text{[0 2 0 1 0]} & 5 \
& 16-2 \sqrt{2} & 2 & 8.250 & \frac{33}{4} & 14 &  \text{III} & 38 \\
\rowcolor{LimeGreen} 135 & \text{}_ 6 \text{[0 0 3 3 1 1 2 0]}_ 3 & \text{[0 2 0 2 0]} & \
6 & 9.656169691 & 2 & 10.0 & 10 & 14 & \text{III} & 193 \\
\rowcolor{Apricot} 11 & \text{}_ 4 \text{[0 2 2 2 2 2 0 0]}_ 1 & \text{[2 0 0 0 0]} & 4 \
& 18 & 2 & 7.000 & 7 & 14 & \text{III} & 8 \\
\rowcolor{Apricot} 38 & \text{}_ 5 \text{[0 2 3 3 2 2 0 0]}_ 2 & \text{[2 0 0 1 0]} & 5 \
& 16-2 \sqrt{2} & 2 & 8.250 & \frac{33}{4} & 14 & \text{III} \
& 27 \\
\rowcolor{Apricot} 193 & \text{}_ 6 \text{[0 2 4 4 2 2 0 0]}_ 3 & \text{[2 0 0 2 0]} & \
6 & 9.656169691 & 2 & 10.0 & 10 & 14 & \text{III} & 135 \\
\hline
\caption{\small\it    States in our database with $\delta = 2$ and $(\ell_1\;\ell_2) = (0\;2)$ or $(\ell_1\;\ell_2) = (2\;0)$.}
\label{tab:delta2l0220}
\end{longtable}
}
\noindent

In table~\ref{tab:delta2l0220}, the entries shaded with

\begin{tabular}{l l}
\fcolorbox{black}{LimeGreen}{\rule{0pt}{6pt}\rule{10pt}{0pt}} \quad belong to $[0\;2;\;0\;0]$\;, &
\fcolorbox{black}{Apricot}{\rule{0pt}{6pt}\rule{10pt}{0pt}} \quad belong to  $[2\;0;\;0\;0]$\;.
\end{tabular}

Finally, let us consider states with Lorentz spin labels $(2\;2)$. There are 3 such states in our database, and are displayed in table~\ref{tab:delta2l22}. We have

{
\scriptsize
\setlength\LTleft{-2mm}\setlength\LTright{-0mm}
\begin{longtable}{@{\extracolsep{\fill}}|C|C|C|C|C|C|C|C|C|C|C|@{}}
\hline
\text{{\tt St.} {\tt No.}} & \texttt{State ID} & [\ell_1\;\ell_2\;q_1\;p\;q_2] & \Delta _ 0 & \Delta _ 0\text{ \
+ $\#$ }g^2 & \delta & 2\,\sqrt{\delta } \lambda ^{1/4} - 2
 & d_1 & j_1 & \text{Type} & \text{Degs.} \\

 & &  &  &  &  & + \frac{\#}{\sqrt{\delta }}\frac{1}{\lambda ^{1/4}} & & & & \\[3pt]
\hline\hline
\rowcolor{LimeGreen} 12 & \text{}_ 4 \text{[0 2 1 1 1 1 2 0]}_ 1 & \text{[2 2 0 0 0]} & 4 \
& \frac{50}{3} & 2 & 6.000000 & 6 & 14 &  \text{I} & \
\text{} \\
\rowcolor{LimeGreen} 39 & \text{}_ 5 \text{[0 2 2 2 1 1 2 0]}_ 1 & \text{[2 2 0 1 0]} & 5 \
& 12 & 2 & 7.2499996 & \frac{29}{4} & 14 & \text{I} \
& \text{} \\
\rowcolor{LimeGreen} 206 & \text{}_ 6 \text{[0 2 3 3 1 1 2 0]}_ 3 & \text{[2 2 0 2 0]} & \
6 & 8.765533583 & 2 & 9.000000 & 9 & 14 & \text{I} & \
\text{} \\
\hline
\caption{\small\it    States in our database with $\delta = 2$ and $(\ell_1\;\ell_2) = (2\;2)$.}
\label{tab:delta2l22}
\end{longtable}
}
\noindent
In table~\ref{tab:delta2l22}, the entries shaded with

\begin{tabular}{l l}
\fcolorbox{black}{LimeGreen}{\rule{0pt}{6pt}\rule{10pt}{0pt}} \quad belong to $[2\;2;\;0\;0]$\;.
\end{tabular}

\noindent
They are all $\mathfrak{sl}(2)$ states of the form $[4\;p+2\;1]$, as can be seen by comparing with section~\ref{sec:dkLit}. Their strong coupling dimensions, at first three sub-leading orders is given by equation~\eqref{eqn:sl2n1}.

\subsection{States with $\delta = 3$}
\label{app:delta3states}
The KK-towers at $\delta = 3$ are given by~\cite{AHSprivate}. We have
\begin{multline}
\label{eqn:delta3count}
    \mathtt{count}_3 =
    6\,[0\;0;0\;0] + 2\,[0\;0;1\;0] + 4\,[0\;0;0\;2] + 4\,[0\;0;2\;0] + [0\;0;2\;2] + [0\;0;0\;4] + [0\;0;4\;0]\
    \\
    + 4\,[0\;1;0\;1] + 4\,[0\;1;1\;1] + 2\,[0\;1;2\;1] + 4\,[1\;0;0\;1] + 4\,[1\;0;1\;1] + 2\,[1\;0;2\;1]
    \\ + 2\,[1\;1;0\;0] + 8\,[1\;1;1\;0] + 2\,[1\;1;0\;2] + 2\,[1\;1;1\;2] + 2\,[1\;1;3\;0]
    \\ + 3\,[0\;2;0\;0] + [0\;2;1\;0] + 2\,[0\;2;0\;2] + [0\;2;2\;0] + 3\,[2\;0;0\;0] + [2\;0;1\;0] + 2\,[2\;0;0\;2] + [2\;0;2\;0]
    \\
    + 4\,[1\;2;0\;1] + 2\,[1\;2;1\;1] + 4\,[2\;1;0\;1] + 2\,[2\;1;1\;1] + 4\,[2\;2;0\;0] + [2\;2;0\;2] + 3\,[2\;2;2\;0]
    \\
    + 2\,[1\;3;1\;0] + 2\,[3\;1;1\;0] + 2\,[2\;3;0\;1] + 2\,[3\;2;0\;1] + 2\,[3\;3;1\;0]
    \\
    + [0\;4;0\;0] + [4\;0;0\;0] + [2\;4;0\;0] + [4\;2;0\;0] + [4\;4;0\;0].
\end{multline}
There are 118 states in our database with $\delta = 3$. Out of them, 27 have Lorentz spin $(0\;0)$. They are

{\scriptsize
\setlength\LTleft{-8mm}\setlength\LTright{-0mm}
\begin{longtable}{@{\extracolsep{\fill}}|C|C|C|C|C|C|C|C|C|C|C|@{}}
\hline
\text{{\tt St.} {\tt No.}} & \texttt{State ID} & [\ell_1\;\ell_2\;q_1\;p\;q_2] & \Delta _ 0 & \Delta _ 0\text{ \
+ $\#$ }g^2 & \delta & 2\,\sqrt{\delta } \lambda ^{1/4} - 2
 & d_1 & j_1 & \text{Type} & \text{Degs.} \\

 & &  &  &  &  & + \frac{\#}{\sqrt{\delta }}\frac{1}{\lambda ^{1/4}} & & & & \\[3pt]
\hline\hline
\rowcolor{CornflowerBlue} 14 & \text{}_ 5 \text{[0 0 3 3 3 1 0 0]}_ 2 & \text{[0 0 0 0 2]} & 5 \
& 21.21110255 & 3 & 9.9 & \text{} & 13.2  &  \text{III} & 16 \\
\rowcolor{CornflowerBlue} 16 & \text{}_ 5 \text{[0 0 4 2 2 2 0 0]}_ 2 & \text{[0 0 2 0 0]} & 5 \
& 21.21110255 & 3 & 9.9 & \text{} & 13.2  &  \text{III} & 14 \\
\rowcolor{LimeGreen} 19 & \text{}_ 5 \text{[0 0 3 3 2 2 0 0]}_ 3 & \text{[0 0 0 1 0]} & 5 \
 & 20 & 3 & 8.9 & \text{} & 13.3  &  \text{III} & 20 \\
\rowcolor{LimeGreen} 20 & \text{}_ 5 \text{[0 0 3 3 2 2 0 0]}_ 4 & \text{[0 0 0 1 0]} & 5 \
& 20 & 3 & 8.9 & \text{} & 13.3  &  \text{III} & 19 \\
\rowcolor{LimeGreen} 76 & \text{}_ 6 \text{[0 0 2 2 2 2 1 1]}_ 1 & \text{[0 0 0 0 0]} & 6 \
& 13.67544468 & 3 & 7.749 & \frac{31}{4} & \frac{27}{2} &  \text{III} & 85 \\
\rowcolor{LimeGreen} 78 & \text{}_ 6 \text{[0 0 3 3 3 3 0 0]}_ 1 & \text{[0 0 0 0 0]} & 6 \
& 4.808186901 & 3 & 1.0000 & 1 & 0 &  \text{I} & \text{} \\
\end{longtable}
\noindent
}

In table~\ref{tab:delta3l00}, the entries shaded with

\begin{tabular}{l l}
\fcolorbox{black}{LimeGreen}{\rule{0pt}{6pt}\rule{10pt}{0pt}} \quad belong to one of the 6 $[0\;0;\;0\;0]$\;, &
\fcolorbox{black}{Apricot}{\rule{0pt}{6pt}\rule{10pt}{0pt}} \quad belong to one of the 2 $[0\;0;\;1\;0]$\;, \\
\fcolorbox{black}{CornflowerBlue}{\rule{0pt}{6pt}\rule{10pt}{0pt}} \quad belong to one of the 4 $[0\;0;\;0\;2]$\;, &
\fcolorbox{black}{Goldenrod}{\rule{0pt}{6pt}\rule{10pt}{0pt}} \quad belong to one of the 4 $[0\;0;\;2\;0]$\;, \\
\fcolorbox{black}{WildStrawberry}{\rule{0pt}{6pt}\rule{10pt}{0pt}} \quad belong to  $[0\;0;\;2\;2]$\;, &
\fcolorbox{black}{Orchid}{\rule{0pt}{6pt}\rule{10pt}{0pt}} \quad belong to $[0\;0;\;0\;4]$\;, \\
\fcolorbox{black}{GreenYellow}{\rule{0pt}{6pt}\rule{10pt}{0pt}} \quad belong to $[0\;0;\;4\;0]$\;
\end{tabular}

\noindent

The 10 states with Lorentz spin labels $(0\;1)$ and 10 states with Lorentz spin labels $(1\;0)$ are given below, in table~\ref{tab:delta3l01}:
{
\scriptsize
\setlength\LTleft{-8mm}\setlength\LTright{-0mm}
\begin{longtable}{@{\extracolsep{\fill}}|C|C|C|C|C|C|C|C|C|C|C|@{}}
\hline
\text{{\tt St.} {\tt No.}} & \texttt{State ID} & [\ell_1\;\ell_2\;q_1\;p\;q_2] & \Delta _ 0 & \Delta _ 0\text{ \
+ $\#$ }g^2 & \delta & 2\,\sqrt{\delta } \lambda ^{1/4} - 2
 & d_1 & j_1 & \text{Type} & \text{Degs.} \\

 & &  &  &  &  & + \frac{\#}{\sqrt{\delta }}\frac{1}{\lambda ^{1/4}} & & & & \\[3pt]
\hline\hline
\rowcolor{LimeGreen} 40 & \text{}_ {\frac{11}{2}} \text{[0 0 3 2 2 2 1 0]}_ 1 & \text{[0 1 \
1 0 0]} & \frac{11}{2} & 18.61803399 & 3 & 9.4 & \text{} & 15.5  & \text{IV} & \text{41, 50, 51} \\
\rowcolor{LimeGreen} 41 & \text{}_ {\frac{11}{2}} \text{[0 0 3 2 2 2 1 0]}_ 2 & \text{[0 \
1 1 0 0]} & \frac{11}{2} & 18.61803399 & 3 & 9.4 & \text{} & 15.5  & \text{IV} & \text{40, 50, 51} \\
\rowcolor{LimeGreen} 42 & \text{}_ {\frac{11}{2}} \text{[0 0 3 2 2 2 1 0]}_ 3 & \text{[0 \
1 1 0 0]} & \frac{11}{2} & 16.38196601 & 3 & 0.4 & \text{} & 0.7  & \text{IV} & \text{43, 52, 53} \\
\rowcolor{LimeGreen} 43 & \text{}_ {\frac{11}{2}} \text{[0 0 3 2 2 2 1 0]}_ 4 & \text{[0 \
1 1 0 0]} & \frac{11}{2} & 16.38196601 & 3 & 0.4 & \text{} & 0.7  & \text{IV} & \text{42, 52, 53} \\
\rowcolor{Apricot} 44 & \text{}_ {\frac{11}{2}} \text{[0 0 3 3 2 1 1 0]}_ 1 & \text{[0 \
1 0 1 1]} & \frac{11}{2} & 18.82842712 & 3 & 12.0 & \text{} & 0.7  & \text{IV} & \text{45, 54, 55} \\
\rowcolor{Apricot} 45 & \text{}_ {\frac{11}{2}} \text{[0 0 3 3 2 1 1 0]}_ 2 & \text{[0 \
1 0 1 1]} & \frac{11}{2} & 18.82842712 & 3 & 12.0 & \text{} & 0.7  & \text{IV} & \text{44, 54, 55} \\
\end{longtable}}
\noindent
In table~\ref{tab:delta3l01}, the entries shaded with

\begin{tabular}{l l}
\fcolorbox{black}{LimeGreen}{\rule{0pt}{6pt}\rule{10pt}{0pt}} \quad belong to one of the 4 $[1\;0;\;0\;1]$\;, &
\fcolorbox{black}{Apricot}{\rule{0pt}{6pt}\rule{10pt}{0pt}} \quad belong to one of the 4 $[1\;0;\;1\;1]$\;, \\
\fcolorbox{black}{CornflowerBlue}{\rule{0pt}{6pt}\rule{10pt}{0pt}} \quad belong to one of the 2 $[1\;0;\;2\;1]$\;, &
\fcolorbox{black}{Goldenrod}{\rule{0pt}{6pt}\rule{10pt}{0pt}} \quad belong to one of the 4 $[0\;1;\;0\;1]$\;, \\
\fcolorbox{black}{WildStrawberry}{\rule{0pt}{6pt}\rule{10pt}{0pt}} \quad belong to one of the 4 $[0\;1;\;1\;1]$\;, &
\fcolorbox{black}{Orchid}{\rule{0pt}{6pt}\rule{10pt}{0pt}} \quad belong to one of the 4 $[0\;1;\;2\;1]$\;.
\end{tabular}

\noindent The 20 states with Lorentz spin labels $(1\;1)$ are given below in table~\ref{tab:delta3l11}, they are
{
\scriptsize
\setlength\LTleft{-10mm}\setlength\LTright{-0mm}
\begin{longtable}{@{\extracolsep{\fill}}|C|C|C|C|C|C|C|C|C|C|C|@{}}
\hline
\text{{\tt St.} {\tt No.}} & \texttt{State ID} & [\ell_1\;\ell_2\;q_1\;p\;q_2] & \Delta _ 0 & \Delta _ 0\text{ \
+ $\#$ }g^2 & \delta & 2\,\sqrt{\delta } \lambda ^{1/4} - 2
 & d_1 & j_1 & \text{Type} & \text{Degs.} \\

 & &  &  &  &  & + \frac{\#}{\sqrt{\delta }}\frac{1}{\lambda ^{1/4}} & & & & \\[3pt]
\hline\hline
\rowcolor{LimeGreen} 28 & \text{}_ 5 \text{[0 1 2 2 2 2 1 0]}_ 1 & \text{[1 1 0 0 0]} & 5 \
& 18 & 3 & 7.0 & \text{} & 13.5 &  \text{IV} & 29 \\
\rowcolor{LimeGreen} 29 & \text{}_ 5 \text{[0 1 2 2 2 2 1 0]}_ 2 & \text{[1 1 0 0 0]} & 5 \
& 18 & 3 & 6.9 & \text{} & 13.3 &  \text{IV} & 28 \\
\rowcolor{Apricot} 32 & \text{}_ 5 \text{[0 1 3 2 2 1 1 0]}_ 3 & \text{[1 1 1 0 1]} & 5 \
& 20 & 3 & 0.4 & \text{} & 9.0 &  \text{IV} & 33 \\
\rowcolor{Apricot} 33 & \text{}_ 5 \text{[0 1 3 2 2 1 1 0]}_ 4 & \text{[1 1 1 0 1]} & 5 \
& 20 & 3 & 0.4 & \text{} & 9.0 &  \text{IV} & 32 \\
 147 & \text{}_ 6 \text{[0 1 3 3 2 2 1 0]}_ 1 & \text{[1 1 0 1 0]} & \
6 & 19.61100719 & 3 & 15.0 & 15 & 27 &  \text{II} & 148 \\
 148 & \text{}_ 6 \text{[0 1 3 3 2 2 1 0]}_ 2 & \text{[1 1 0 1 0]} & \
6 & 19.61100719 & 3 & 15.0 & 15 & 27 &  \text{II} & 147 \\
\end{longtable}}
\noindent
In table~\ref{tab:delta3l11}, the entries shaded with

\begin{tabular}{l l}
\fcolorbox{black}{LimeGreen}{\rule{0pt}{6pt}\rule{10pt}{0pt}} \quad belong to one of the 2 $[1\;1;\;0\;0]$\;, &
\fcolorbox{black}{Apricot}{\rule{0pt}{6pt}\rule{10pt}{0pt}} \quad belong to one of the 8 $[1\;1;\;1\;0]$\;, \\
\fcolorbox{black}{WildStrawberry}{\rule{0pt}{6pt}\rule{10pt}{0pt}} \quad belong to one of the 2 $[1\;1;\;1\;2]$\;, &
\fcolorbox{black}{Orchid}{\rule{0pt}{6pt}\rule{10pt}{0pt}} \quad belong to one of the 2 $[1\;1;\;3\;0]$\;, \\
\fcolorbox{black}{White}{\rule{0pt}{6pt}\rule{10pt}{0pt}} \quad belong either to $[1\;1;\;1\;0]$ or $[1\;1;\;0\;2]$ \;.
\end{tabular}

\noindent The 11 states with Lorentz spin labels $(0\;2)$ and 11 states with Lorentz spin labels $(2\;0)$, are displayed in table~\ref{tab:delta3l0220}.
{
\scriptsize
\setlength\LTleft{-11mm}\setlength\LTright{-0mm}
\begin{longtable}{@{\extracolsep{\fill}}|C|C|C|C|C|C|C|C|C|C|C|@{}}
\hline
\text{{\tt St.} {\tt No.}} & \texttt{State ID} & [\ell_1\;\ell_2\;q_1\;p\;q_2] & \Delta _ 0 & \Delta _ 0\text{ \
+ $\#$ }g^2 & \delta & 2\,\sqrt{\delta } \lambda ^{1/4} - 2
 & d_1 & j_1 & \text{Type} & \text{Degs.} \\

 & &  &  &  &  & + \frac{\#}{\sqrt{\delta }}\frac{1}{\lambda ^{1/4}} & & & & \\[3pt]
\hline\hline
\rowcolor{CornflowerBlue} 25 & \text{}_ 5 \text{[0 0 3 1 1 1 2 0]}_ 1 & \text{[0 2 2 0 0]} & 5 \
& 3 & 3 & 8.86 & \text{} & 13.22 &  \text{III} & 36 \\
\rowcolor{LimeGreen} 26 & \text{}_ 5 \text{[0 0 2 2 1 1 2 0]}_ 1 & \text{[0 2 0 1 0]} & 5 \
& 16+2 \sqrt{2} & 3 & 7.9 & \text{} & 13.3 &  \text{III} & \
37 \\
\rowcolor{LimeGreen} 122 & \text{}_ 6 \text{[0 0 2 2 2 2 2 0]}_ 2 & \text{[0 2 0 0 0]} & \
6 & 11.34832950 & 3 & 9.00 & 9 & 18 &  \text{III} & 182 \\
\rowcolor{LimeGreen} 124 & \text{}_ 6 \text{[1 1 1 1 1 1 2 0]}_ 1 & \text{[0 2 0 0 0]} & \
6 & 11.54748008 & 3 & 6.750 & \frac{27}{4} & \frac{27}{2} &  \
\text{III} & 179 \\
\rowcolor{LimeGreen} 125 & \text{}_ 6 \text{[1 1 1 1 1 1 2 0]}_ 2 & \text{[0 2 0 0 0]} & \
6 & 23.78585325 & 3 & 17.999 & 18 & 36 &  \text{III} & 180 \
\\
\rowcolor{Apricot} 127 & \text{}_ 6 \text{[0 0 3 2 2 1 2 0]}_ 2 & \text{[0 2 1 0 1]} & \
6 & 14.87689437 & 3 & 11.00 & 11 & 18 &  \text{III} & 185 \\
\end{longtable}}
\noindent
In table~\ref{tab:delta3l0220}, the entries shaded with

\begin{tabular}{l l}
\fcolorbox{black}{LimeGreen}{\rule{0pt}{6pt}\rule{10pt}{0pt}} \quad belong to one of the 3 $[0\;2;\;0\;0]$\;, &
\fcolorbox{black}{Apricot}{\rule{0pt}{6pt}\rule{10pt}{0pt}} \quad belong  $[0\;2;\;0\;2]$\;, \\
\fcolorbox{black}{CornflowerBlue}{\rule{0pt}{6pt}\rule{10pt}{0pt}} \quad belong to one of the 2 $[0\;2;\;0\;2]$\;, &
\fcolorbox{black}{Goldenrod}{\rule{0pt}{6pt}\rule{10pt}{0pt}} \quad belong to  $[0\;2;\;2\;0]$\;  \\
\fcolorbox{black}{WildStrawberry}{\rule{0pt}{6pt}\rule{10pt}{0pt}} \quad belong to one of the 3 $[2\;0;\;0\;2]$\;, &
\fcolorbox{black}{Orchid}{\rule{0pt}{6pt}\rule{10pt}{0pt}} \quad belong to  $[2\;0;\;0\;2]$\; \\
\fcolorbox{black}{GreenYellow}{\rule{0pt}{6pt}\rule{10pt}{0pt}} \quad belong to one of the 2 $[2\;0;\;0\;2]$\;, &
\fcolorbox{black}{BurntOrange}{\rule{0pt}{6pt}\rule{10pt}{0pt}} \quad belong to  $[2\;0;\;2\;0]$\;
\end{tabular}

\noindent The 6 states with Lorentz spin labels $(1\;2)$ and 6 states with Lorentz spin labels $(2\;1)$, are given in table~\ref{tab:delta3l1221}. They are
{
\scriptsize
\setlength\LTleft{-10mm}\setlength\LTright{-0mm}
\begin{longtable}{@{\extracolsep{\fill}}|C|C|C|C|C|C|C|C|C|C|C|@{}}
\hline
\text{{\tt St.} {\tt No.}} & \texttt{State ID} & [\ell_1\;\ell_2\;q_1\;p\;q_2] & \Delta _ 0 & \Delta _ 0\text{ \
+ $\#$ }g^2 & \delta & 2\,\sqrt{\delta } \lambda ^{1/4} - 2
 & d_1 & j_1 & \text{Type} & \text{Degs.} \\

 & &  &  &  &  & + \frac{\#}{\sqrt{\delta }}\frac{1}{\lambda ^{1/4}} & & & & \\[3pt]
\hline\hline
\rowcolor{LimeGreen} 60 & \text{}_ {\frac{11}{2}} \text{[0 1 2 2 2 1 2 0]}_ 1 & \text{[1 2 \
0 0 1]} & \frac{11}{2} & 15.45861873 & 3 & 0.4 & \text{} & 0.8 &  \text{IV} & \text{61, 66, 67} \\
\rowcolor{LimeGreen} 61 & \text{}_ {\frac{11}{2}} \text{[0 1 2 2 2 1 2 0]}_ 2 & \text{[1 \
2 0 0 1]} & \frac{11}{2} & 15.45861873 & 3 & 0.4 & \text{} & 0.8 &  \text{IV} & \text{60, 66, 67} \\
\rowcolor{LimeGreen} 62 & \text{}_ {\frac{11}{2}} \text{[0 1 2 2 2 1 2 0]}_ 3 & \text{[1 \
2 0 0 1]} & \frac{11}{2} & 21.54138127 & 3 & 14.9 & \text{} & 1.2 & \text{IV} & \text{63, 68, 69} \\
\rowcolor{LimeGreen} 63 & \text{}_ {\frac{11}{2}} \text{[0 1 2 2 2 1 2 0]}_ 4 & \text{[1 \
2 0 0 1]} & \frac{11}{2} & 21.54138127 & 3 & 14.9 & \text{} & 1.2 &  \text{IV} & \text{62, 68, 69} \\
\rowcolor{Apricot} 64 & \text{}_ {\frac{11}{2}} \text{[0 1 3 2 1 1 2 0]}_ 1 & \text{[1 \
2 1 1 0]} & \frac{11}{2} & 18 & 3 & 10.9 & \text{} & 17.9 &  \
\text{IV} & \text{65, 70, 71} \\
\rowcolor{Apricot} 65 & \text{}_ {\frac{11}{2}} \text{[0 1 3 2 1 1 2 0]}_ 2 & \text{[1 \
2 1 1 0]} & \frac{11}{2} & 18 & 3 & 10.9 & \text{} & 17.9 &  \
\text{IV} & \text{64, 70, 71} \\
\end{longtable}}
\noindent
In table~\ref{tab:delta3l1221}, the entries shaded with

\begin{tabular}{l l}
\fcolorbox{black}{LimeGreen}{\rule{0pt}{6pt}\rule{10pt}{0pt}} \quad belong to one of the 4 $[1\;2;\;0\;1]$\;, &
\fcolorbox{black}{Apricot}{\rule{0pt}{6pt}\rule{10pt}{0pt}} \quad belong to one of the 2 $[1\;2;\;1\;1]$\;, \\
\fcolorbox{black}{CornflowerBlue}{\rule{0pt}{6pt}\rule{10pt}{0pt}} \quad belong to one of the 4 $[2\;1;\;0\;1]$\;, &
\fcolorbox{black}{Goldenrod}{\rule{0pt}{6pt}\rule{10pt}{0pt}} \quad belong to one of the 2 $[2\;1;\;1\;1]$\;.
\end{tabular}

\noindent
The 8 states with Lorentz spin labels $(2\;2)$, are given in table~\ref{tab:delta3l22}. They are
{
\scriptsize
\setlength\LTleft{-6mm}\setlength\LTright{-0mm}
\begin{longtable}{@{\extracolsep{\fill}}|C|C|C|C|C|C|C|C|C|C|C|C|@{}}
\hline
\text{{\tt St.} {\tt No.}} & \texttt{State ID} & [\ell_1\;\ell_2\;q_1\;p\;q_2] & \Delta _ 0 & \Delta _ 0\text{ \
+ $\#$ }g^2 & \delta & 2\,\sqrt{\delta } \lambda ^{1/4} - 2
 & d_1 & j_1 & \text{Type} & \text{Degs.} \\

 & &  &  &  &  & + \frac{\#}{\sqrt{\delta }}\frac{1}{\lambda ^{1/4}} & & & & \\[3pt]
\hline\hline
\rowcolor{LimeGreen} 196 & \text{}_ 6 \text{[0 2 2 2 2 2 2 0]}_ 2 & \text{[2 2 0 0 0]} & 6 \
& 9.804773390 & 3 & 8.0000 & 8 & 18 &  \text{I} & \text{} \\
\rowcolor{LimeGreen} 197 & \text{}_ 6 \text{[0 2 2 2 2 2 2 0]}_ 3 & \text{[2 2 0 0 0]} & \
6 & 23.10773903 & 3 & 17.0000 & 17 & 36 &  \text{I} & \text{} \\
\rowcolor{LimeGreen} 198 & \text{}_ 6 \text{[0 2 2 2 2 2 2 0]}_ 4 & \text{[2 2 0 0 0]} & \
6 & \frac{58}{3} & 3 & 12.999 & 13 & 28 &  \text{II} & 199 \
\\
\rowcolor{LimeGreen} 199 & \text{}_ 6 \text{[0 2 2 2 2 2 2 0]}_ 5 & \text{[2 2 0 0 0]} & \
6 & \frac{58}{3} & 3 & 12.999 & 13 & 28 &  \text{II} & 198 \
\\
\rowcolor{Apricot} 203 & \text{}_ 6 \text{[0 2 3 2 2 1 2 0]}_ 2 & \text{[2 2 1 0 1]} & \
6 & \frac{55}{3}-\frac{\sqrt{145}}{3} & 3 & 10.0000 & 10 & 18 &  \text{I} & \text{} \\
\rowcolor{CornflowerBlue} 205 & \text{}_ 6 \text{[0 2 3 3 1 1 2 0]}_ 2 & \text{[2 2 0 2 0]} & \
6 & 16.71846288 & 3 & 11.0000 & 11 & 18 &  \text{I} & \text{} \\
\rowcolor{CornflowerBlue} 207 & \text{}_ 6 \text{[0 2 3 3 1 1 2 0]}_ 4 & \text{[2 2 0 2 0]} & \
6 & \frac{46}{3} & 3 & 11.000 & 11 & 18 &  \text{II} & 208 \\
\rowcolor{CornflowerBlue} 208 & \text{}_ 6 \text{[0 2 3 3 1 1 2 0]}_ 5 & \text{[2 2 0 2 0]} & \
6 & \frac{46}{3} & 3 & 11.000 & 11 & 18 &  \text{II} & 207 \\
\hline
\caption{\small\it    States in our database with $\delta = 3$ and $(\ell_1\;\ell_2) = (2\;2)$.}
\label{tab:delta3l22}
\end{longtable}}
\noindent
In table~\ref{tab:delta3l22}, the entries shaded with

\begin{tabular}{l l}
\fcolorbox{black}{LimeGreen}{\rule{0pt}{6pt}\rule{10pt}{0pt}} \quad belong to one of the 4 $[2\;2;\;0\;0]$\;, &
\fcolorbox{black}{Apricot}{\rule{0pt}{6pt}\rule{10pt}{0pt}} \quad belong to $[2\;2;\;0\;2]$\;, \\
\fcolorbox{black}{CornflowerBlue}{\rule{0pt}{6pt}\rule{10pt}{0pt}} \quad belong to one of the 3 $[2\;2;\;2\;0]$\;.
\end{tabular}

\noindent
We didn't find any states with Lorentz spin labels $(0\;3)$ and $(3\;0)$, exactly as expected by~\cite{AHSprivate} through~\eqref{eqn:delta3count}.
The 2 states with Lorentz spin labels $(1\;3)$ and 2 states with Lorentz spin labels $(3\;1)$, are given in table~\ref{tab:delta3l1331}:
{
\scriptsize
\setlength\LTleft{-10mm}\setlength\LTright{-0mm}
\begin{longtable}{@{\extracolsep{\fill}}|C|C|C|C|C|C|C|C|C|C|C|C|@{}}
\hline
\text{{\tt St.} {\tt No.}} & \texttt{State ID} & [\ell_1\;\ell_2\;q_1\;p\;q_2] & \Delta _ 0 & \Delta _ 0\text{ \
+ $\#$ }g^2 & \delta & 2\,\sqrt{\delta } \lambda ^{1/4} - 2
 & d_1 & j_1 & \text{Type} & \text{Degs.} \\

 & &  &  &  &  & + \frac{\#}{\sqrt{\delta }}\frac{1}{\lambda ^{1/4}} & & & & \\[3pt]
\hline\hline
\rowcolor{LimeGreen} 175 & \text{}_ 6 \text{[0 1 2 2 1 1 3 0]}_ 1 & \text{[1 3 0 1 0]} & \
6 & 18.48338852 & 3 & 12.0 & \text{} & 23.5 &  \text{IV} & \text{176, 210, 211} \\
\rowcolor{LimeGreen} 176 & \text{}_ 6 \text{[0 1 2 2 1 1 3 0]}_ 2 & \text{[1 3 0 1 0]} & \
6 & 18.48338852 & 3 & 12.0 & \text{} & 23.5 &  \text{IV} & \text{175, 210, 211} \\
\rowcolor{Apricot} 210 & \text{}_ 6 \text{[0 3 3 3 2 2 1 0]}_ 1 & \text{[3 1 0 1 0]} & 6 \
& 18.48338852 & 3 & 12.0 & \text{} & 23.5 &  \text{IV} & \text{175, 176, 211} \\
\rowcolor{Apricot} 211 & \text{}_ 6 \text{[0 3 3 3 2 2 1 0]}_ 2 & \text{[3 1 0 1 0]} & \
6 & 18.48338852 & 3 & 12.0 & \text{} & 23.5 &  \text{IV} & \text{175, 176, 210} \\
\hline
\caption{\small\it    States in our database with $\delta = 3$ and $(\ell_1\;\ell_2) = (1\;3)$ or $(\ell_1\;\ell_2) = (3\;1)$.}
\label{tab:delta3l1331}
\end{longtable}}
\noindent
In table~\ref{tab:delta3l1331}, the entries shaded with

\begin{tabular}{l l}
\fcolorbox{black}{LimeGreen}{\rule{0pt}{6pt}\rule{10pt}{0pt}} \quad belong to one of the 2 $[1\;3;\;1\;0]$\;, &
\fcolorbox{black}{Apricot}{\rule{0pt}{6pt}\rule{10pt}{0pt}} \quad belong to one of the 2 $[3\;1;\;1\;0]$\;.
\end{tabular}

\noindent
The 2 states with Lorentz spin labels $(2\;3)$ and 2 states with Lorentz spin labels $(3\;2)$, are
{
\scriptsize
\setlength\LTleft{-10mm}\setlength\LTright{-0mm}
\begin{longtable}{@{\extracolsep{\fill}}|C|C|C|C|C|C|C|C|C|C|C|@{}}
\hline
\text{{\tt St.} {\tt No.}} & \texttt{State ID} & [\ell_1\;\ell_2\;q_1\;p\;q_2] & \Delta _ 0 & \Delta _ 0\text{ \
+ $\#$ }g^2 & \delta & 2\,\sqrt{\delta } \lambda ^{1/4} - 2
 & d_1 & j_1 & \text{Type} & \text{Degs.} \\

 & &  &  &  &  & + \frac{\#}{\sqrt{\delta }}\frac{1}{\lambda ^{1/4}} & & & & \\[3pt]
\hline\hline
\rowcolor{LimeGreen} 72 & \text{}_ {\frac{11}{2}} \text{[0 2 2 1 1 1 3 0]}_ 1 & \text{[2 3 \
1 0 0]} & \frac{11}{2} & 20 & 3 & 12.9 & \text{} & 28.2 &  \text{IV} & \text{73, 74, 75} \\
\rowcolor{LimeGreen} 73 & \text{}_ {\frac{11}{2}} \text{[0 2 2 1 1 1 3 0]}_ 2 & \text{[2 \
3 1 0 0]} & \frac{11}{2} & 20 & 3 & 12.9 & \text{} & 28.2 &
 \text{IV} & \text{72, 74, 75} \\
\rowcolor{Apricot} 74 & \text{}_ {\frac{11}{2}} \text{[0 3 2 2 2 1 2 0]}_ 1 & \text{[3 \
2 0 0 1]} & \frac{11}{2} & 20 & 3 & 12.9 & \text{} & 28.2 &
\text{IV} & \text{72, 73, 75} \\
\rowcolor{Apricot} 75 & \text{}_ {\frac{11}{2}} \text{[0 3 2 2 2 1 2 0]}_ 2 & \text{[3 \
2 0 0 1]} & \frac{11}{2} & 20 & 3 & 12.9 & \text{} & 28.2 &
\text{IV} & \text{72, 73, 74} \\
\hline
\caption{\small\it    States in our database with $\delta = 3$ and $(\ell_1\;\ell_2) = (2\;3)$ or $(\ell_1\;\ell_2) = (3\;2)$.}
\label{tab:delta3l2332}
\end{longtable}}
\noindent
In table~\ref{tab:delta3l2332}, the entries shaded with

\begin{tabular}{l l}
\fcolorbox{black}{LimeGreen}{\rule{0pt}{6pt}\rule{10pt}{0pt}} \quad belong to one of the 2 $[2\;3;\;0\;1]$\;, &
\fcolorbox{black}{Apricot}{\rule{0pt}{6pt}\rule{10pt}{0pt}} \quad belong to one of the 2 $[3\;2;\;0\;1]$\;.
\end{tabular}

\noindent
The 2 states with Lorentz spin labels $(3\;3)$, are displayed in table~\ref{tab:delta3l33}. They are
{
\scriptsize
\setlength\LTleft{-8mm}\setlength\LTright{-0mm}
\begin{longtable}{@{\extracolsep{\fill}}|C|C|C|C|C|C|C|C|C|C|C|@{}}
\hline
\text{{\tt St.} {\tt No.}} & \texttt{State ID} & [\ell_1\;\ell_2\;q_1\;p\;q_2] & \Delta _ 0 & \Delta _ 0\text{ \
+ $\#$ }g^2 & \delta & 2\,\sqrt{\delta } \lambda ^{1/4} - 2
 & d_1 & j_1 & \text{Type} & \text{Degs.} \\

 & &  &  &  &  & + \frac{\#}{\sqrt{\delta }}\frac{1}{\lambda ^{1/4}} & & & & \\[3pt]
\hline\hline
\rowcolor{LimeGreen} 214 & \text{}_ 6 \text{[0 3 2 2 1 1 3 0]}_ 1 & \text{[3 3 0 1 0]} & \
6 & \frac{35}{2} & 3 & 12.00 & 12 & 27 & \text{II} & 215 \\
\rowcolor{LimeGreen} 215 & \text{}_ 6 \text{[0 3 2 2 1 1 3 0]}_ 2 & \text{[3 3 0 1 0]} & \
6 & \frac{35}{2} & 3 & 12.00 & 12 & 27 &  \text{II} & 214 \\
\hline
\caption{\small\it    States in our database with $\delta = 3$ and $(\ell_1\;\ell_2) = (3\;3)$.}
\label{tab:delta3l33}
\end{longtable}}
\noindent
In table~\ref{tab:delta3l33}, the entries shaded with

\begin{tabular}{l l}
\fcolorbox{black}{LimeGreen}{\rule{0pt}{6pt}\rule{10pt}{0pt}} \quad belong to one of the 2 $[3\;3;\;1\;0]$\;.
\end{tabular}

\noindent
The 1 state with Lorentz spin labels $(0\;4)$ and 1 state with Lorentz spin labels $(4\;0)$ are dislayed in table~\ref{tab:delta3l0440}. They are
{
\scriptsize
\setlength\LTleft{-8mm}\setlength\LTright{-0mm}
\begin{longtable}{@{\extracolsep{\fill}}|C|C|C|C|C|C|C|C|C|C|C|@{}}
\hline
\text{{\tt St.} {\tt No.}} & \texttt{State ID} & [\ell_1\;\ell_2\;q_1\;p\;q_2] & \Delta _ 0 & \Delta _ 0\text{ \
+ $\#$ }g^2 & \delta & 2\,\sqrt{\delta } \lambda ^{1/4} - 2
 & d_1 & j_1 & \text{Type} & \text{Degs.} \\

 & &  &  &  &  & + \frac{\#}{\sqrt{\delta }}\frac{1}{\lambda ^{1/4}} & & & & \\[3pt]
\hline\hline
\rowcolor{LimeGreen} 138 & \text{}_ 6 \text{[1 1 1 1 1 1 4 0]}_ 1 & \text{[0 4 0 0 0]} & 6 \
& 22 & 3 & 16.0000 & 16 & 36 &  \text{III} & 216 \\
\hline
\rowcolor{Apricot} 216 & \text{}_ 6 \text{[0 4 2 2 2 2 1 1]}_ 1 & \text{[4 0 0 0 0]} & 6 \
& 22 & 3 & 16.0000 & 16 & 36 &  \text{III} & 138 \\
\hline
\caption{\small\it    States in our database with $\delta = 3$ and $(\ell_1\;\ell_2) = (0\;4)$ or $(\ell_1\;\ell_2) = (4\;0)$.}
\label{tab:delta3l0440}
\end{longtable}}
\noindent
In table~\ref{tab:delta3l0440}, the entries shaded with

\begin{tabular}{l l}
\fcolorbox{black}{LimeGreen}{\rule{0pt}{6pt}\rule{10pt}{0pt}} \quad belong to $[0\;4;\;0\;0]$\;, &
\fcolorbox{black}{Apricot}{\rule{0pt}{6pt}\rule{10pt}{0pt}} \quad belong to $[4\;0;\;0\;0]$\;.
\end{tabular}

\noindent
We didn't find any states with Lorentz spin labels $(1\;4)$ and $(4\;1)$, exactly as expected by~\cite{AHSprivate} through~\eqref{eqn:delta3count}.
The 1 state with Lorentz spin labels $(2\;4)$ and 1 state with Lorentz spin labels $(4\;2)$, are given in table~\ref{tab:delta3l2442}, they are
{
\scriptsize
\setlength\LTleft{-8mm}\setlength\LTright{-0mm}
\begin{longtable}{@{\extracolsep{\fill}}|C|C|C|C|C|C|C|C|C|C|C|C|@{}}
\hline
\text{{\tt St.} {\tt No.}} & \texttt{State ID} & [\ell_1\;\ell_2\;q_1\;p\;q_2] & \Delta _ 0 & \Delta _ 0\text{ \
+ $\#$ }g^2 & \delta & 2\,\sqrt{\delta } \lambda ^{1/4} - 2
 & d_1 & j_1 & \text{Type} & \text{Degs.} \\

 & &  &  &  &  & + \frac{\#}{\sqrt{\delta }}\frac{1}{\lambda ^{1/4}} & & & & \\[3pt]
\hline\hline
\rowcolor{LimeGreen} 209 & \text{}_ 6 \text{[0 2 1 1 1 1 4 0]}_ 1 & \text{[2 4 0 0 0]} & \
6 & \frac{64}{3} & 3 & 15.000 & 15 & 36  & \text{III} & 218 \ \\
\hline
\rowcolor{Apricot} 218 & \text{}_ 6 \text{[0 4 2 2 2 2 2 0]}_ 1 & \text{[4 2 0 0 0]} & \
6 & \frac{64}{3} & 3 & 15.000 & 15 & 36  & \text{III} & 209 \ \\
\hline
\caption{\small\it    States in our database with $\delta = 3$ and $(\ell_1\;\ell_2) = (2\;4)$ or $(\ell_1\;\ell_2) = (4\;2)$.}
\label{tab:delta3l2442}
\end{longtable}}
\noindent
In table~\ref{tab:delta3l2442}, the entries shaded with

\begin{tabular}{l l}
\fcolorbox{black}{LimeGreen}{\rule{0pt}{6pt}\rule{10pt}{0pt}} \quad belong to $[2\;4;\;0\;0]$\;, &
\fcolorbox{black}{Apricot}{\rule{0pt}{6pt}\rule{10pt}{0pt}} \quad belong to $[4\;2;\;0\;0]$\;.
\end{tabular}

\noindent
We didn't find any states with Lorentz spin labels $(3\;4)$ and $(4\;3)$, exactly as expected by~\cite{AHSprivate} through~\eqref{eqn:delta3count}.
The 1 state with Lorentz spin labels $(4\;4)$ is given in table~\ref{tab:delta3l44}. We have
{
\scriptsize
\setlength\LTleft{-4mm}\setlength\LTright{-0mm}
\begin{longtable}{@{\extracolsep{\fill}}|C|C|C|C|C|C|C|C|C|C|C|@{}}
\hline
\text{{\tt St.} {\tt No.}} & \texttt{State ID} & [\ell_1\;\ell_2\;q_1\;p\;q_2] & \Delta _ 0 & \Delta _ 0\text{ \
+ $\#$ }g^2 & \delta & 2\,\sqrt{\delta } \lambda ^{1/4} - 2
 & d_1 & j_1 & \text{Type} & \text{Degs.} \\

 & &  &  &  &  & + \frac{\#}{\sqrt{\delta }}\frac{1}{\lambda ^{1/4}} & & & & \\[3pt]
\hline\hline
\rowcolor{LimeGreen} 219 & \text{}_ 6 \text{[0 4 1 1 1 1 4 0]}_ 1 & \text{[4 4 0 0 0]} & 6 \
& \frac{98}{5} & 3 & 13.0000 & 13 & 36 &  \text{I} & \
\text{} \\
\hline
\caption{\small\it    The states in our database with $\delta = 3$ and $(\ell_1\;\ell_2) = (4\;4)$.}
\label{tab:delta3l44}
\end{longtable}}
\noindent
In table~\ref{tab:delta3l44}, the entries shaded with

\begin{tabular}{l l}
\fcolorbox{black}{LimeGreen}{\rule{0pt}{6pt}\rule{10pt}{0pt}} \quad belong to $[4\;4;\;0\;0]$\;.
\end{tabular}

\noindent

To conclude, we observe matching with the counting at strong coupling for $\delta=2$ and $\delta=3$. In addition, we break degeneracies of KK-towers using the sub-leading Casimir $j_1$ which as we notice is a good classifier.

\bibliographystyle{JHEP.bst}
\bibliography{references}

\end{document}